\documentclass[aps,prd,groupaddress,floatfix,tighten,nofootinbib,showpacs,amsfonts,superscriptaddress]{revtex4-2}
\usepackage{url}
\usepackage{xcolor}
\usepackage{hyperref}
\usepackage[utf8]{inputenc}
\usepackage{appendix}
\colorlet{shadecolor}{gray!15}
\definecolor{greenLinks}{rgb}{0, 0.6, 0}
\definecolor{blueLinks}{rgb}{0, 0, 0.6}
\definecolor{redLinks}{rgb}{0.6, 0, 0}
\definecolor{tempText}{rgb}{0.55, 0.10,0.67}
\definecolor{eprintLinks}{rgb}{0.4, 0.4, 0.4}
\definecolor{journalLinks}{rgb}{0.6, 0, 0}
\usepackage{amssymb}
\usepackage{amsmath,color}
\usepackage{epsfig}
\usepackage{graphicx,epsf,epsfig}
\usepackage{footnote}
\usepackage{multirow}
\usepackage{float}
\usepackage{enumitem}
\usepackage{color}
\def\slc#1{\setbox0=\hbox{$#1$}                  
    \dimen0=\wd0                                 
    \setbox1=\hbox{/} \dimen1=\wd1               
    \ifdim\dimen0>\dimen1                        
       \rlap{\hbox to \dimen0{\hfil/\hfil}}      
       #1                                        
    \else                                        
       \rlap{\hbox to \dimen1{\hfil$#1$\hfil}}   
       /                                         
    \fi}

\def\be{\begin{equation}}
\def\ee{\end{equation}}
\def\gs{\mathrel{
   \rlap{\raise 0.511ex \hbox{$>$}}{\lower 0.511ex \hbox{$\sim$}}}}
\def\ls{\mathrel{
   \rlap{\raise 0.511ex \hbox{$<$}}{\lower 0.511ex \hbox{$\sim$}}}}

\newcommand{\ba}{\begin{array}{c}}
\newcommand{\baz}{\begin{array}{cc}}
\newcommand{\barrr}{\begin{array}{rrr}}
\newcommand{\bad}{\begin{array}{ccc}}
\newcommand{\bav}{\begin{array}{cccc}}
\newcommand{\baf}{\begin{array}{ccccc}}
\newcommand{\bea}{\begin{equation} \begin{array}{c}}
\newcommand{\eea}{\end{array} \end{equation}}
\newcommand{\ea}{\end{array}}

\usepackage[normalem]{ulem}

\def\21{$\mathrm{SU(2)_L \otimes U(1)_Y}$ }

\newcommand {\ignore}[1]{}

\newcommand{\vt}{\vert}
\newcommand{\nn}{\nonumber}

\newcommand{\mc}{\mathcal}
\allowdisplaybreaks \allowdisplaybreaks[2]
\newcommand{\AddrIFUNAM}{
Instituto~de F{\'{\i}}sica, 
Universidad~Nacional Aut\'onoma de M\'exico, \\
Apdo. Postal 20-364, CDMX 01000, M\'exico.}

\newcommand{\AddrCONA}{
C\'atedras Conahcyt - Instituto~de F{\'{\i}}sica, 
Universidad~Nacional Aut\'onoma de M\'exico, \\
Apdo. Postal 20-364, CDMX 01000, M\'exico.}

\newcommand{\AddrCECYT}{Centro de Estudios Cient\'ificos y Tecnol\'ogicos No 16, Instituto Polit\'ecnico Nacional, Pachuca: Ciudad del Conocimiento y la Cultura, Carretera Pachuca Actopan km 1+500, San Agust\'in Tlaxiaca, Hidalgo, M\'exico.}

\begin{document}
\title{Inverse See-Saw Mechanism with $\mathbf{S}_{3}$ flavor symmetry} 
%
\author{Juan Carlos Gómez-Izquierdo}
\email{cizquierdo@ipn.mx}
\affiliation{\AddrCECYT}

\author{Catalina Espinoza}
\email{m.catalina@fisica.unam.mx}
\affiliation{\AddrCONA}

\author{Lucia E. Gutiérrez Luna}
\email{lgutierrez@estudiantes.fisica.unam.mx}
\affiliation{\AddrIFUNAM}

\author{Myriam Mondrag\'on}
\email{myriam@fisica.unam.mx}
\affiliation{\AddrIFUNAM}

%

%

\date{\bf \today} 

\begin{abstract}\vspace{2cm}
The current neutrino experiments provide an opportunity for testing the inverse see-saw mechanism through charged lepton flavor violating processes and neutrinoless double beta decay.  Motivated by this, in this paper we study the $\mathbf{S}_{3}\otimes \mathbf{Z}_{2}$ discrete symmetry in the $B-L$ gauge model where the active light neutrino mass matrix comes from the aforementioned mechanism. In this framework, the effect of complex vacuum expectation values of the Higgs doublets on the fermion masses is explored and, under certain assumptions on the Yukawa couplings, we find that the neutrino mixing is controlled by the Cobimaximal pattern, but a sizeable deviation from the charged lepton sector breaks the well known predictions on the atmospheric angle ($45^{\circ}$) and the Dirac CP-violating phase ($-90^{\circ}$). In addition, due to the presence of heavy neutrinos at the $TeV$ scale,   charged lepton flavor violation (CLFV) and neutrinoless double beta decay get notable contributions. Analytical formulae for these observables are obtained, and then a numerical calculation allows to fit quite well the lepton mixing for the normal and inverted hierarchies, however, the branching ratios decay values for CLFV disfavors the latter one. Along with this, the region of parameter space for the $m_{ee}$ effective neutrino mass lies below the GERDA bounds for both the normal and inverted hierarchies. On the other hand, with a particular benchmark, the quark mass matrices are found to have textures that allow to fit with great accuracy the CKM mixing matrix. 
\end{abstract}

\begin{flushright}
CECyT 16-24-I
\end{flushright}

\maketitle
%

\section{Introduction}
Neutrino oscillations confirmed that those neutral particles have a tiny but non negligible mass in contrast to the charged lepton and quark ones. The type I see-saw mechanism~\cite{Minkowski:1977sc,GellMann:1980vs,Yanagida:1979as,Mohapatra:1980yp,Mohapatra:1979ia,Schechter:1980gr,Schechter:1981cv} provides a possible explanation to generate small neutrino masses, however, it is not testable in the current experiments. Moreover, the mixing pattern remains as an outstanding issue. Once the neutrino masses and mixing angles are added to the well established quark sector, the flavor puzzle \cite{Feruglio:2015jfa,Abbas:2023ivi,Nilles:2023shk} remains as one of the prominent problems to be solved.  In particular, a compelling solution that explains the very different masses and mixing patters in the quark and lepton sectors is still lacking.

Besides  the type I see-saw mechanism, other mechanisms have been proposed in the past~\cite{Wyler:1982dd,Akhmedov:1995vm,Barr:2005ss,Romao:2007ny,Xing:2009in} to give mass to  the active neutrinos. Among these, the inverse see-saw  mechanism~\cite{Mohapatra:1986aw,Mohapatra:1986bd,Bernabeu:1987gr} (ISSM), is a falsifiable scheme where the lepton masses and mixing may find a reasonable explanation. In its minimal version,  three heavy right-handed neutrinos (RHN's) $N_{R}$, three SM gauge singlets neutrinos  $S_{R}$, and one Higgs SM gauge singlet are added to the SM content such that a lepton number violating mass term ($\bar{S}^{c}S$) can be written. As a result,  the active neutrino mass matrix has a double suppression due to a large and small pseudo-Dirac ($\bar{N}^{c}S$) and Majorana ($\bar{S}^{c}S$) mass respectively, while the Dirac neutrino mass is at the electroweak scale \cite{Xing:2009hx,Malinsky:2009df,Abada:2014vea,Abada:2014zra,Boucenna:2014zba,Abada:2014nwa,Abada:2016awd}. 
Remarkably, the ISSM has been implemented in the $B-L$ gauge model\cite{Khalil:2010iu,Abdallah:2011ew,Abdallah:2019svm} where six extra neutral fermions are required due to the gauge symmetry; in this scheme, the sterile neutrino mass is generated dynamically. In these different realizations, the ISSM gives interesting signatures like CLFV processes, contributions to the neutrinoless double beta decay, and so forth~\cite{Ibarra:2010xw,Ibarra:2011xn,Abada:2016awd,Abada:2014nwa,Pinheiro:2021mps}.

On the other hand, multi-Higgs models \cite{Barroso:2005da,Mantry:2007ar,Mantry:2007ar,Ferreira:2008zy,Botella:2009pq,Ferreira:2010xe,Blechman:2010cs,Diaz-Cruz:2014pla,Ivanov:2017dad} have been attractive scenarios to look for extra sources of CP violation beyond the Standard Model (SM), which turn out  to be relevant to explain baryogenesis. These scalar extensions possess distinctive features, as for example,  possible effects on the SM Higgs boson phenomenology. Nonetheless, the mentioned extensions have a drawback in the scalar potential as well as in the Yukawa sector, where the proliferation of free parameters is not under control. This may be alleviated by including a discrete symmetry~\cite{Ishimori:2010au, Grimus:2011fk, Altarelli:2012bn, Altarelli:2012ss, King:2013eh,  King:2015aea,Fonseca:2014lfa,Chauhan:2022gkz}, at the same time, scalar extensions combined with flavor symmetries might be a route to find out the underlying physics 
related to some issues on the flavor problem. In this line of thought, motivated by the number of families in the quark and lepton sector, appealing models with three Higgs doublets together with the $\mathbf{S}_{3}$ discrete symmetry (3HDM-S3) have been explored since many years ago. Some works are focused on the masses and mixings~\cite{Kubo:2003iw,Chen:2004rr, Mondragon:2007af,Mondragon:2007nk,Meloni:2010aw,Canales:2012dr,Canales:2013cga}, some others are interested in studying the phenomenology of the scalar potential~\cite{Pakvasa:1977in,Kubo:2004ps,EmmanuelCosta:2007zz, Beltran:2009zz,Teshima:2012cg,Barradas-Guevara:2014yoa, Das:2014fea, Gomez-Bock:2021uyu}, which may be tested in future experiments. Due to the discovery of the Higgs boson at the LHC in 2012~\cite{ATLAS:2012yve,CMS:2012qbp}, the multi-Higgs models have increased in relevance and the 3HDM-S3 scenario is not the exception. As a result, detailed studies on the scalar potential and its minimization have been released in the last years with particular attention to the structure of the vacuum expectation values (vev's) that are allowed by the discrete symmetry~\cite{Das:2014fea,Gomez-Bock:2021uyu,Kuncinas:2020wrn,Khater:2021wcx,Kuncinas:2023ycz}. Besides this, some scalar alignments may play an important role in addressing the dark matter problem~\cite{Khater:2021wcx, Espinoza:2018itz}. Certainly, in general,  discrete symmetries have inspired a lot of models which can be tested experimentally~\cite{Gehrlein:2022nss}. In particular,  although extensive work has been realized  in the 3HDM-S3 scenario also,  there is  so far not one complete model that can describe all the observables. 

Therefore, in this letter, the main aim is to address the fermion masses and mixing on the alternative B-L model~\cite{Khalil:2010iu,Abdallah:2011ew,Abdallah:2019svm}, where the ISSM gives rise to the effective neutrino masses at the $TeV$ scale. In addition, the inclusion of the $\mathbf{S}_{3}\otimes \mathbf{Z}_{2}$ discrete symmetry is needed to control the Yukawa couplings, and is therefore crucial in shaping the fermion mass and mixing matrices. Despite the fact that there are already many flavored models with ISSM in the literature \cite{Hirsch:2009mx, Karmakar:2016cvb, Gautam:2019pce,Thapa:2023fxu, Garnica:2023ccx,Duy:2024lbd}, our paper is different from the aforementioned models since it has peculiar features. In this approach, three Higgs doublets and three singlets scalars are added to the $B-L$ matter content, and under the flavor symmetry, the scalar and quark sectors are not treated on the  same footing as compared to the leptons. 
This distinctive feature allows to obtain, under certain assumptions on the Yukawa couplings and considering complex Higgs vev's~\cite{Emmanuel-Costa:2016vej,Kuncinas:2020wrn,Khater:2021wcx,Kuncinas:2023ycz}, quark mass textures that fit the CKM mixing with very good accuracy. On the other hand, despite the large number of scalar fields, we do identify the Cobimaximal pattern~\cite{Fukuura:1999ze,Miura:2000sx,Ma:2002ce,Grimus:2003yn,Chen:2014wxa,Ma:2015fpa,Joshipura:2015dsa,Li:2015rtz,He:2015xha,Chen:2015siy,Ma:2016nkf,Damanik:2017jar,Ma:2017trv,Grimus:2017itg,CarcamoHernandez:2017owh,CarcamoHernandez:2018hst,Ma:2019iwj,Hernandez:2021kju,CarcamoHernandez:2024ycd} in the effective neutrino mass matrix while the charged lepton sector breaks the well known predictions on the atmospheric ($\pi/4$) and Dirac CP-violating phase ($3\pi/2$), effectively giving a deviation of the Cobimaximal scenario and making the results compatible with experimental data. At the same time, due to the active neutrino field having a contribution from the heavy neutrinos ones, their impact on the neutrinoless double decay and CLFV processes are taken into account. We ought to comment that the current work is an extension of a previous paper~\cite{Gomez-Izquierdo:2023mph} where the lepton sector was studied in the type I see-saw mechanism. Similar results are obtained in the PMNS matrix, nonetheless CLFV and neutrinoless double beta decay make a notable difference between both works  as we will show here. In addition, in order to have a complete model the quark sector was added, so we end up having a richer model from the phenomenological point of view.

The paper is organized as follows.  In section II, we describe briefly the $B-L$ gauge model~\cite{Khalil:2010iu,Abdallah:2011ew,Abdallah:2019svm} and the main features are pointed out. Then, the $\mathbf{S}_{3}\otimes \mathbf{Z}_{2}$ symmetry is implemented in the model where a minimal scalar extension is realized to take into account the fermion masses and mixings. Then, we obtain analytical expressions for the CKM and PMNS mixing matrices to perform a numerical study on the mixing angles by  scanning the allowed region for the free parameters in the model. In a low scale see-saw mechanism as the ISSM, the neutrinoless double beta decay and CLFV processes get an extra contribution, apart from the active neutrinos,  due to the heavy sterile neutrinos which can modify notable the observables. Therefore, in section III, these issues are addressed. We end up giving our conclusions in section IV.

\section{$B-L$ model with $\mathbf{S}_{3} \otimes \mathbf{Z}_{2}$ symmetry}
\subsection{Inverse See-Saw Mechanism in the $B-L$ gauge model}
One of the most popular mechanism to explain the smallness of neutrino mass is the type I see-saw mechanism~\cite{Minkowski:1977sc,GellMann:1980vs,Yanagida:1979as,Mohapatra:1980yp,Mohapatra:1979ia,Schechter:1980gr,Schechter:1981cv}, where the inclusion of right-handed neutrinos is necessary. In that scheme, the aforementioned particles must have a large mass ($\mathcal{O}({10^{15}})~GeV$) in order to get active neutrino masses at  the $eV$ scale. This scenario is beyond the accessible energy that future experiments can reach,  thus not possible to test. In this line of thought, testable mechanisms have been proposed in the literature such as the type II see-saw, radiative, linear, and inverse see-saw mechanisms~\cite{Khan:2012zw}. Each one has its particular features and matter content. As we already commented, in this paper, we focus on the ISSM in the $B-L$ gauge model~\cite{Khalil:2010iu, Abdallah:2011ew, Abdallah:2019svm}. In this framework, in addition to the SM particles, three right-handed ($N_{R}$) and six sterile ($s^{i=1,2,3}$ and $S^{i=1,2,3}$) neutrinos are added to the matter content. Along with these, a $\phi$ Higgs singlet field  that breaks $B-L$ number is added, to provide mass to the pseudo Dirac neutrinos. Under the $B-L$ gauge model, the complete matter content has the following charges. 

\begin{table}[ht]
	\begin{center}
		\begin{tabular}{|c|c|c|c|c|c|c|c|c|c|c|c|c|c|c|c}
			\hline \hline
			Matter & Quarks & Leptons & Sterile Neutrino ($s$)& Sterile Neutrino ($S$)& Higgs & $\phi$	\\ \hline
			B-L &	$1/3$ & $-1$ & $-2$ & $2$ &$0$ & $-1$ \\\hline \hline
		\end{tabular}\caption{Matter content in the $B-L$ model}\label{TB1}
	\end{center}
\end{table}

The Yukawa mass term in this framework is given as\footnote{In the original paper, there is an extra $\mathbb{Z}_{2}$ symmetry that forbids that the $s$ ($\mathbb{Z}=-1$) sterile neutrino couples to the rest of the particles. Also, the $\mathbf{M}_{(1,2)}$ sterile neutrino masses are assumed to be the same. Their masses may be generated dynamically by means the  scalar field $\phi$~\cite{Abdallah:2019svm}.}
\begin{equation}
-\mathcal{L}=y^{d}\bar{Q} H d_{R}+y^{u}\bar{Q}\tilde{H} u_{R}+y^{l}\bar{L} H l_{R}+y^{D}\bar{L}\tilde{H}N_{R}+y^{R}\overline{(N_{R})^{c}}\phi S
+\frac{1}{2}\mathbf{M}_{1} \overline{(s)^{c}}
s+\frac{1}{2}\mathbf{M}_{2} \overline{(S)^{c}}
S+h.c.,
\end{equation} 
where $Q=\left(u_{L}, d_{L}\right)^{T}$, $L=\left(\nu_{L}, l_{L}\right)^{T}$ and $H=\left(H^{+}, H^{0}\right)^{T}$ denote the quark, lepton, and Higgs doublet under $ \mathbf{SU(2)}_{L}$, respectively;  with $\tilde{H}_{i}=i\sigma_{2}H^{\ast}_{i}$ and $\sigma_{2}$  the second Pauli matrix. Focusing in the neutrino sector, after the spontaneous symmetry breaking, the mass term is written as
\begin{eqnarray}
-\mathcal{L}&=&\frac{1}{2}\left( \bar{\nu}_{L}\hspace{2mm} \overline{(N_{R})^{c}}\hspace{2mm} \overline{(S)^{c}}\right)\overbrace{\begin{pmatrix}
0	& \mathbf{M}_{D} & 0 \\
\mathbf{M}^{T}_{D} & 0 & \mathbf{M}_{R} \\
0	& \mathbf{M}^{T}_{R} & \mathbf{M}_{2}
\end{pmatrix}}^{\mathcal{M}_{\nu}}  \begin{pmatrix}
(\nu_{L})^{c}\\
N_{R}\\
S
\end{pmatrix}+h.c.;\nn\\
&=&\frac{1}{2}\left( \bar{\nu}_{L}\hspace{2mm} \overline{(n_{R})^{c}}\right)\mathcal{M}_{\nu} \begin{pmatrix}
	(\nu_{L})^{c}\\
	n_{R}
\end{pmatrix}+h.c.,
\end{eqnarray}
where we have defined $n^{T}_{R}=\left( N_{R}\hspace{5mm} S \right)$, and 
\begin{equation}
\mathcal{M}_{\nu}=\begin{pmatrix}
	0& \mathcal{M}_{D} \\
	\mathcal{M}^{T}_{D} & \mathcal{M}_{R}
\end{pmatrix},\qquad \mathcal{M}_{D}=\begin{pmatrix}
\mathbf{M}_{D} & 0
\end{pmatrix}, \qquad \mathcal{M}_{R}= \begin{pmatrix}
0& \mathbf{M}_{R} \\
\mathbf{M}^{T}_{R} & \mathbf{M}_{2}
\end{pmatrix} ~.
\end{equation}
Here, $\mathbf{M}_{D}$, $\mathbf{M}_{R}$, and $\mathbf{M}_{2}$  stand for the Dirac, pseudo Dirac, and Majorana mass matrices, respectively. In flavor space, they are $3\times 3$ matrices. As one can notice, strictly speaking, there is a missing mass matrix, $\mathbf{M}_{1}$ which is decoupled from the rest of the neutrino sector; in the minimal version of this model~\cite{Abdallah:2019svm}, $s$ is a candidate to dark matter. In the current work, we will just focus on the ISSM and its implications on the masses and mixings, as well as the related phenomenology on neutrinoless double beta decay and CLFV processes.

As it is well known, $\mathcal{M}_{\nu}$ is diagonalized by $\mathcal{U}_{\nu}$ such that $\mathcal{U}^{\dagger}_{\nu} \mathcal{M}_{\nu} \mathcal{U}^{\ast}_{\nu}=\hat{\mathcal{M}_{\nu}}$ with $\hat{\mathcal{M}_{\nu}}=\textrm{Diag.} \left(\mathbf{U}_{\nu} \hat{\mathbf{M}}_{\nu}\mathbf{U}^{T}_{\nu}, \mathbf{V}_{R} \hat{\mathbf{M}}_{R} \mathbf{V}^{T}_{R} \right)$ where $\mathbf{U}_{\nu}$ and $\mathbf{V }_{R}$ are $3\times 3$ and $6\times 6$ matrices~\cite{Hettmansperger:2011bt}. Also, $\hat{\mathbf{M}}_{\nu}$ is a diagonal matrix where the active physical neutrino masses are contained; in addition to this, there are six heavy neutrino masses in $\hat{\mathbf{M}}_{R}$. Assuming a hierarchy among the mass matrices, that is, $\mathbf{M}_{R}>\mathbf{M}_{D}>\mathbf{M}_{2}$, in the standard framework, we have
\begin{equation}
\mathcal{U}_{\nu}\approx\begin{pmatrix}
\left(\mathbf{1}_{3\times 3}-\frac{1}{2}\mathcal{A}\mathcal{A}^{\dagger}\right)	&  \mathcal{A} \\
-\mathcal{A}^{\dagger}	&  \left(\mathbf{1}_{6\times 6}-\frac{1}{2}\mathcal{A}^{\dagger}\mathcal{A} \right)	
\end{pmatrix} \begin{pmatrix}
\mathbf{U}_{\nu} & 0 \\
0 & \mathbf{V}_{R}
\end{pmatrix},
\end{equation}
where $\mathcal{A}=\mathcal{M}_{D}(\mathcal{M}_{R})^{-1}$ is a $3\times 6$ matrix and  stands for the mixing between the light and heavy neutrino sector. Explicitly, this is given as $\mathcal{A}=\left( \mathbf{M}_{D} (\mathbf{M}^{T}_{R})^{-1}\mathbf{M}_{2}\mathbf{M}^{-1}_{R}\hspace{5mm}  \mathbf{M}_{D}(\mathbf{M}^{T}_{R})^{-1}\right)\approx \left(0\hspace{5mm} \mathbf{A}\right)$ with $\mathbf{A}=\mathbf{M}_{D}(\mathbf{M}^{T}_{R})^{-1}$ being a $3\times 3$ matrix.

In addition, the $\mathbf{M}_{\nu}$ effective mass matrix is given by
\begin{equation}\label{EffNu}
\mathbf{M}_{\nu}=\mathbf{M}_{D}\left(\mathbf{M}^{T}_{R}\right)^{-1}\mathbf{M}_{2} \left(\mathbf{M}^{-1}_{R}\right)\mathbf{M}^{T}_{D}.
\end{equation}
The above expression is known as the inverse see-saw mechanism~\cite{Mohapatra:1986aw,Mohapatra:1986bd,Bernabeu:1987gr}, and it has a double suppression due to the $\mathbf{M}_{R}$ as well as the $\mathbf{M}_{2}$ small mass that breaks lepton number, which  is restored in the limit $\mathbf{M}_{2}\rightarrow 0$~\cite{tHooft:1979rat}.

On the other hand, the weak charge-current Lagrangian in this extension is given as
\begin{equation}
-\mathcal{L}=\frac{g_{L}}{\sqrt{2}}W^{\mu}_{L}\bar{l}_{L}\gamma_{\mu}\nu_{L}+h.c.
\end{equation}
After diagonalizing the $\mathcal{M}_{\nu}$ neutrino mass matrix, in the mass basis, the above expression is replaced by
\begin{equation}
	-\mathcal{L}=\frac{g_{L}}{\sqrt{2}}W^{\mu}_{L}\bar{\tilde{l}}_{L}\gamma_{\mu}\left[\mathbf{U}\tilde{\nu}_{L}+\mathcal{K}(\tilde{n}_{R})^{c}\right]+h.c.,
\end{equation}
where there is an extra contribution to the PMNS matrix due to the $\mathbf{\eta}=\mathcal{A}\mathcal{A}^{\dagger}/2$ mixing between the active and sterile neutrinos so that $\mathbf{U}=\mathbf{U}^{\dagger}_{l}\left(1-\mathbf{\eta}\right)\mathbf{U}_{\nu}$. Along with this, the neutrinoless double beta decay and the charged lepton flavor violation processes will have also an extra term due to the heavy neutrinos, in the standard notation, $\mathcal{K}=\mathbf{U}^{\dagger}_{l}\mathcal{A}\mathbf{V}_{R}$ which is a $3\times 6$ matrix.

As one can notice, the $\mathbf{V}_{R}$ matrix diagonalizes  the $\mathcal{M}_{R}$ matrix such that $\mathbf{V}^{\dagger}_{R} \mathcal{M}_{R} \mathbf{V}^{\ast}_{R}=\hat{\mathbf{M}}_{R}$. Explicitly, we have
\begin{equation}
	\mathbf{V}^{\dagger}_{R} \mathcal{M}_{R} \mathbf{V}^{\ast}_{R}=\hat{\mathbf{M}}_{R}\approx\begin{pmatrix}
	\mathbf{V}^{\dagger}_{1}\left[-\mathbf{M}_{R}+\frac{1}{2}\mathbf{M}_{2}\right] \mathbf{V}^{\ast}_{1}	&  0  \\
	0	&  \mathbf{V}^{\dagger}_{2}\left[\mathbf{M}_{R}+\frac{1}{2}\mathbf{M}_{2}\right] \mathbf{V}^{\ast}_{2}
	\end{pmatrix}~,
\end{equation}
where $\mathbf{M}_{R}$ has been assumed to be symmetric. As it was shown in~\cite{Karmakar:2016cvb}, $\mathbf{V}_{R}$ is given as
\begin{equation}
\mathbf{V}_{R}\approx\frac{1}{\sqrt{2}}\begin{pmatrix}
\mathbf{1}+\frac{\mathbf{M_{2}}\mathbf{M}^{-1}_{R}}{4}	& \mathbf{1}-\frac{\mathbf{M_{2}}\mathbf{M}^{-1}_{R}}{4} \\
-\mathbf{1}+\frac{\mathbf{M_{2}}\mathbf{M}^{-1}_{R}}{4} & \mathbf{1}+\frac{\mathbf{M_{2}}\mathbf{M}^{-1}_{R}}{4}
\end{pmatrix} \begin{pmatrix}
\mathbf{V}_{1} & 0 \\
0  & \mathbf{V}_{2} 
\end{pmatrix}.
\end{equation}

\subsection{Flavored $B-L$ model}
Having described briefly the $B-L$ gauge model~\cite{Khalil:2010iu,Abdallah:2011ew,Abdallah:2019svm}, where the ISSM
gives rise to the active neutrino masses at the $TeV$ scale, we focus on the mixing patterns.  
To do this, we consider the $\mathbf{S}_{3}$ symmetry to drive mainly the mixing. As it is well known, the mentioned non-abelian discrete group has three irreducible representations: two singlets $\mathbf{1}_{S}$ and  $\mathbf{1}_{A}$, and one doublet $\mathbf{2}$~\cite{Ishimori:2010au}. Thus, there are two reducible representations $\mathbf{3}_{S}=\mathbf{2}\oplus \mathbf{1}_{S}$ and $\mathbf{3}_{A}=\mathbf{2}\oplus \mathbf{1}_{A}$ which are useful to assign three families of particles, in this work,  we will use the former assignment.

Then, the scalar and fermion sectors are augmented by three Higgs doublets ($H$) and singlets ($\phi$); three right-handed ($N_{R}$) and six sterile ($s$ and $S$) neutrinos are added to the fermion sector. In this case, under the flavor symmetry,  the scalars and fermion fields are assigned as follows. Focusing on the quarks and scalars,  the first and second family are put together within a $\mathbf{2}$,  the third one belongs to $\mathbf{1}_{S}$. In the  scalar sector we use the same assignment, and the main motivation has to do with the results on the $\mathbf{S}_{3}$ scalar potential with three Higgs (3HD-S3) where this assignment is used with a particular vev alignment, giving a viable SM Higgs and several exotic scalars with possibilities of detection in future experiments  \cite{Das:2014fea,Gomez-Bock:2021uyu}. Moreover, in these exhaustive studies~\cite{Emmanuel-Costa:2016vej, Kuncinas:2020wrn,Khater:2021wcx, Kuncinas:2023ycz}, one can find several more vev alignments which turn out to be useful in shaping the fermion masses. On the other hand, in the lepton sector, the first family is assigned to $\mathbf{1}_{S}$ singlet; the second and third families live in a $\mathbf{2}$ doublet ($\mathbf{3}_{S}= \mathbf{1}_{S}\oplus \mathbf{2}$). Such assignment allows the identification of the $\mu \leftrightarrow \tau$~\cite{Xing:2015fdg} symmetry or the Cobimaximal pattern~\cite{Fukuura:1999ze,Miura:2000sx,Ma:2002ce,Grimus:2003yn,Chen:2014wxa,Ma:2015fpa,Joshipura:2015dsa,Li:2015rtz,He:2015xha,Chen:2015siy,Ma:2016nkf,Damanik:2017jar,Ma:2017trv,Grimus:2017itg,CarcamoHernandez:2017owh,CarcamoHernandez:2018hst,Ma:2019iwj}. 

Besides the $\mathbf{S}_{3}$ symmetry, the model is supplemented by an extra symmetry, $\mathbf{Z}_{2}$. This forbids some Yukawa couplings in the lepton sector, and as a result the charged lepton and Dirac mass matrices are almost diagonal, as we will show. In short, 
the explicit assignment for the matter content is displayed in Table \ref{TAB2}. 
\begin{table}[ht]
	\begin{center}
		\begin{tabular}{|c|c|c|c|c|c|c|c|c|c|c|c|c|c|c|}
			\hline \hline	
			{\footnotesize Matter} & {\footnotesize $Q_{I},d_{I R},u_{I, R}, H_{I}, L_{J}, e_{J R}, N_{J, R}, s_{J L}$} & {\footnotesize $L_{1}, e_{1 R}, N_{1 R}, s_{1 L}$} & {\footnotesize $Q_{3}, d_{3 R}, u_{3 R},H_{3}, \phi_{3}$} & {\footnotesize $\phi_{I}$}   \\ \hline
			{\footnotesize \bf $\mathbf{S}_{3}$} &  {\footnotesize \bf $2$} & {\footnotesize \bf $1_{S}$}   & {\footnotesize \bf $1_{S}$} & {\footnotesize \bf $2$} \\ \hline
			{\footnotesize \bf $\mathbf{Z}_{2}$} & {\footnotesize $1$} & {\footnotesize $-1$}  &  {\footnotesize $1$} & {\footnotesize $-1$} \\ \hline \hline
		\end{tabular}\caption{Flavored $B-L$ model. Here, $I=1,2$ and $J=2,3$.}\label{TAB2}
	\end{center}
\end{table}
Thereby, we have the following allowed Lagrangian for the lepton sector
\begin{align}
-\mathcal{L}_{\ell}&=y^{e}_{1}\bar{L}_{1}H_{3}e_{1 R}+y^{e}_{2}\left[(\bar{L}_{2}H_{2}+\bar{L}_{3}H_{1})e_{2 R}+(\bar{L}_{2}H_{1}-\bar{L}_{3}H_{2})e_{3 R} \right]+y^{e}_{3}\left[\bar{L}_{2}H_{3}e_{2 R}+\bar{L}_{3}H_{3}e_{3 R}\right]+y^{D}_{1}\bar{L}_{1}\tilde{H}_{3}N_{1 R}\nn\\&+
y^{D}_{2}\left[(\bar{L}_{2}\tilde{H}_{2}+\bar{L}_{3}\tilde{H}_{1})N_{2 R}+(\bar{L}_{2}\tilde{H}_{1}-\bar{L}_{3}\tilde{H}_{2})N_{3 R} \right]	
+ y^{D}_{3}\left[\bar{L}_{2}\tilde{H}_{3}N_{2 R}+\bar{L}_{3}\tilde{H}_{3}N_{3 R}\right]
+y^{R}_{1}\overline{(N_{1 R})^{c}}\phi_{3}S_{1}\nn\\&+y^{R}_{2}\overline{(N_{1 R})^{c}}[\phi_{1}S_{2}+\phi_{2}S_{3}]+y^{R}_{3}[\overline{(N_{2 R})^{c}}\phi_{1}+\overline{(N_{3 R})^{c}}\phi_{2}]S_{1}+y^{R}_{4}\left[\overline{(N_{2 R})^{c}}\phi_{3 }S_{2}+\overline{(N_{3 R})^{c}}\phi_{3}S_{3}\right]+M_{1}\overline{S^{c}_{1}}S_{1}\nn\\&+M_{2}\left[\overline{S^{c}_{2}}S_{2}+\overline{S^{c}_{3}}S_{3}\right]+h.c.\label{EQ4}
\end{align}
and for the quark sector
\begin{eqnarray}
-\mathcal{L}_{q}&=&y^{d}_{1}\left[\bar{Q}_{1 L}\left(H_{1} d_{2 R}+H_{2} d_{1 R}\right)+\bar{Q}_{2 L}\left(H_{1} d_{1 R}-H_{2}d_{2 R}\right)\right]+y^{d}_{2}\left[\bar{Q}_{1 L}H_{3}d_{1 R}+\bar{Q}_{2 L}H_{3} d_{2 R}\right]+y^{d}_{3}\left[\bar{Q}_{1 L}H_{1}+\bar{Q}_{2 L}H_{2} \right]d_{3 R}\nn\\&&+y^{d}_{4}\bar{Q}_{3 L}\left[H_{1}d_{1 R}+H_{2} d_{2 R}\right]+y^{d}_{5}\bar{Q}_{3 L}H_{3}d_{3 R}+
y^{u}_{1}\left[\bar{Q}_{1 L}\left(\tilde{H}_{1} u_{2 R}+\tilde{H}_{2} u_{1 R}\right)+\bar{Q}_{2 L}\left(\tilde{H}_{1} u_{1 R}-\tilde{H}_{2} u_{2 R}\right)\right]\nn\\&&+y^{u}_{2}\left[\bar{Q}_{1 L} \tilde{H}_{3} u_{1 R}+\bar{Q}_{2 L} \tilde{H}_{3} u_{2 R}\right]+y^{u}_{3}\left[\bar{Q}_{1 L} \tilde{H}_{1}+\bar{Q}_{2 L} \tilde{H}_{2} \right]u_{3 R}+y^{u}_{4}\bar{Q}_{3 L}\left[\tilde{H}_{1} u_{1 R}+\tilde{H}_{2} u_{2 R}\right]\nn\\&&+y^{u}_{5}\bar{Q}_{3 L} \tilde{H}_{3} u_{3 R}+h.c.
\end{eqnarray}

For simplicity, the sterile neutrino mass term was added by hand. There has been also a focused effort in providing a dynamical mechanism to generate the sterile neutrino mass, see for instance~\cite{Abdallah:2011ew}

Once the spontaneous symmetry breaking happens, the lepton mass matrices are given as
\begin{equation}
{\bf M}_{\ell}=\begin{pmatrix}
y^{\ell}_{1}\langle H_{3}\rangle & 0 & 0 \\ 
0 & y^{\ell}_{3}\langle H_{3}\rangle+y^{\ell}_{2}\langle H_{2}\rangle & y^{\ell}_{2}\langle H_{1}\rangle \\ 
0 & y^{\ell}_{2}\langle H_{1}\rangle & y^{\ell}_{3}\langle H_{3}\rangle-y^{\ell}_{2}\langle H_{2}\rangle \end{pmatrix},\quad
{\bf M}_{R}=\begin{pmatrix}
	y^{R}_{1} \langle \phi_{3}\rangle & y^{R}_{2} \langle \phi_{1}\rangle & y^{R}_{2} \langle \phi_{2}\rangle \\ 
	y^{R}_{3} \langle \phi_{1}\rangle & y^{R}_{4}\langle \phi_{3}\rangle & 0 \\ 
	y^{R}_{3} \langle \phi_{2}\rangle & 0 & y^{R}_{4}\langle \phi_{3}\rangle, 
\end{pmatrix};
\end{equation}
with $\mathbf{M}_{2}=\textrm{Diag.}\left(M_{1}, M_{2}, M_{2}\right)$. Besides this, in the quark sector we have
\begin{equation}
{\bf M}_{q}=\begin{pmatrix}
	y^{q}_{2}\langle H_{3}\rangle+y^{q}_{1}\langle H_{2}\rangle & y^{q}_{1}\langle H_{1}\rangle & y^{d}_{3}\langle H_{1}\rangle \\ 
	y^{q}_{1}\langle H_{1}\rangle & y^{q}_{2}\langle H_{3}\rangle-y^{q}_{1}\langle H_{2}\rangle  & y^{q}_{3}\langle H_{2}\rangle \\ 
	y^{q}_{4}\langle H_{1}\rangle & y^{q}_{4}\langle H_{2}\rangle & y^{q}_{3}\langle H_{3}\rangle
\end{pmatrix}.\label{EQ6} 
\end{equation}
Notice that  $q=u,d$ stands for the up and down quark sector; $\ell=D, e$ denotes the Dirac neutrinos and charged leptons. In addition, we have to keep in mind that for the up sector, $H=i\sigma_{2}H^{\ast}$. 

On the other hand, the scalar potential is crucial to get  viable mass textures which will provide the mixing matrices. The general potential of the model is based on the following structure,
\begin{equation} \label{pot1}
	V= V(H) + V(\phi) + V(H, \phi),
\end{equation}
where $V(H)$ corresponds to the most general potential with three  Higgs doublets allowed by the symmetry group $\mathbf{S}_{3}$, and is structured as shown below, 
\begin{equation} \label{001}
	V(H) = M^2 ( H^{\dagger} H) + \dfrac{a}{2} ( H^{\dagger} H )^2 .
\end{equation}
Notice that the three Higgs doublets scalar potential with the $\mathbf{S}_{3}$ symmetry has extensively been studied (3HD-S3) with the usual assignment: $(H_{1},H_{2})\sim \mathbf{2} $ and $H_{3}\sim \mathbf{1}_{S}$~\cite{Pakvasa:1977in, Derman:1978rx, Pakvasa:1978tx, Derman:1979nf, Kubo:2003iw,Das:2014fea,Barradas-Guevara:2014yoa,Gomez-Bock:2021uyu}. Recently, a study of all possible alignments that are allowed by the imposed discrete symmetry, allowing for complex vev's, was released~\cite{Emmanuel-Costa:2016vej, Kuncinas:2020wrn, Khater:2021wcx, Kuncinas:2023ycz}, making it  necessary to explore their effects on the mass matrices as well as the mixings.

With respect to the rest of the potential, we have the following,
\begin{equation} \label{potencialescalar}
	\begin{aligned}
		V(\phi) & = \mu^2_{BL} ( \phi^{\dagger} \phi) + \dfrac{\lambda}{2} (\phi^{\dagger} \phi )^2 \\
		V(H, \phi) & = - L \left( H^{\dagger} H \right)  \left( \phi^{\dagger} \phi \right) ~,
	\end{aligned}
\end{equation}
where $V(H, \phi)$ is an interaction between the electroweak sector and the fields $\phi$ sector, allowed by the flavour symmetry. However, in this work we will consider the scenario where both sectors are decoupled, so we set $L =0$.

Henceforth, we will concentrate on  $V(\phi)$, which is responsible for providing mass to the heavy right handed neutrinos. It is important to note that, since in the potential the sector of the $\phi$ fields and the Higgs doublets are decoupled, the following calculations do not change any results in the low energy Higgs sector.
In terms of complex fields, we express the singlet fields $\phi_i$, in the following way,
\begin{equation} \label{componentes}
	\begin{aligned}
		\phi_1 & = \frac{1}{\sqrt{2}} (x_1 + iy_1) \hspace{13mm}
		\phi_2 = \frac{1}{\sqrt{2}} (x_2 + iy_2) \hspace{13mm}
		\phi_3 = \frac{1}{\sqrt{2}} (x_3 + iy_3) .
	\end{aligned}
\end{equation} 
Using the natural choices of conservation of electric charge and CP invariance, we considered the spontaneous symmetry breaking, implying that only the real part of the neutral fields will acquire the vacuum expectation value, 
\begin{equation}   \label{vevs}
\begin{aligned}
& \langle x_1 \rangle = w_1 \hspace{5mm} \langle x_2 \rangle = w_2 \hspace{5mm}  \langle x_3 \rangle = w_3. 
\end{aligned}
\end{equation}
By requiring $\partial V(\phi)/  \partial x_i = 0$, we can minimize the potential, then solving the tree level tadpole equations with the requirement $w_1, w_2 \neq 0$, the following result is found,
\begin{equation} \label{alineamiento}
	w_1 = w_2 \hspace{5mm} \text{or} \hspace{5mm} \langle \phi_{2}\rangle=\langle \phi_{1}\rangle.
\end{equation}
The calculations that follow will include this alignment as a fact.
Two massless scalar bosons appear when we examine the mass spectrum generated by the potential $V(\phi)$ invariant under $\mathbf{S}_{3} \otimes \mathbf{Z}_{2}$ symmetry with this particular solution.  One massless scalar will give mass to an extra gauge boson associated to the $U(1)_{B-L}$ symmetry, but we still have an extra one due to the presence of a continuous symmetry, as found in \cite{Kubo:2003iw,Beltran:2009zz}. To avoid this problem we introduce a breaking term of the $\mathbf{S}_{3}$ symmetry, which leaves the alignment invariant. More information about the potential is provided in Appendix \ref{potencial}.

\subsubsection{Quark sector}
Once the spontaneous symmetry breaking is realized, the quark mass matrices are given 
\begin{equation}
{\bf M}_{q}=\begin{pmatrix}
	y^{q}_{2}\langle H_{3}\rangle+y^{q}_{1}\langle H_{2}\rangle & y^{q}_{1}\langle H_{1}\rangle & y^{d}_{3}\langle H_{1}\rangle \\ 
	y^{q}_{1}\langle H_{1}\rangle & y^{q}_{2}\langle H_{3}\rangle-y^{q}_{1}\langle H_{2}\rangle  & y^{q}_{3}\langle H_{2}\rangle \\ 
	y^{q}_{4}\langle H_{1}\rangle & y^{q}_{4}\langle H_{2}\rangle & y^{q}_{3}\langle H_{3}\rangle
\end{pmatrix}
\end{equation}
The above mass matrix has been studied exhaustively in the 3HD-S3 framework~\cite{Canales:2013cga}. In this paper, a different approach is taken which is completely different to the cited work. Now, we are interested in exploring the alignment $\langle H_{1}\rangle=v_{1}$, $\langle H_{2}\rangle=iv_{2}$ and $\langle H_{3}\rangle=v_{3}$ which may be relevant for the masses. This alignment comes from the 3HD-S3 framework~\cite{Emmanuel-Costa:2016vej, Kuncinas:2020wrn, Khater:2021wcx, Kuncinas:2023ycz}, although to our knowledge it has never been used  explicitly before  to calculate quark masses. In this scenario, we end up having
\begin{equation}
	{\bf M}_{d}=\frac{1}{\sqrt{2}}\begin{pmatrix}
		y^{d}_{2} v_{3}+iy^{d}_{1} v_{2} & y^{d}_{1} v_{1} & y^{d}_{3} v_{1} \\ 
		y^{d}_{1}v_{1}  & y^{d}_{2}v_{3}-iy^{d}_{1}v_{2}   & iy^{d}_{3} v_{2}  \\ 
		y^{d}_{4}v_{1}  & iy^{d}_{4}v_{2}  & y^{d}_{5} v_{3}
	\end{pmatrix},\,
	{\bf M}_{u}=\frac{1}{\sqrt{2}}\begin{pmatrix}
		y^{u}_{2}v_{3}-iy^{u}_{1} v_{2} & y^{u}_{1} v_{1} & y^{u}_{3} v_{1} \\ 
		y^{u}_{1}v_{1} & y^{u}_{2}v_{3}+iy^{u}_{1} v_{2} & -iy^{u}_{3} v_{2} \\ 
		y^{u}_{4} v_{1} & -iy^{u}_{4}v_{2} & y^{u}_{5} v_{3}
	\end{pmatrix}.
\end{equation}

In this case, both mass matrices have many free parameters as one can see. We ought to point out that the general mass matrix will not be diagonalized. Instead of doing that, we will work in the following benchmark:

\begin{itemize}
	\item The following hierarchies $\vert \langle H_{1}\rangle\vert<\vert \langle H_{2}\rangle\vert<\vert \langle H_{3}\rangle\vert$ and $y^{q}_{3},y^{q}_{4}\ll y^{q}_{1}$ are assumed such that the $(\mathbf{M}_{q})_{13}\approx 0$. As it is well known, this could have been realized by means a transformation on the quarks field however we decided to make the mentioned approximation. 
	
	\item In this B-L model, there is no right-handed currents as consequence a suitable rotation may be realized  on these fields. Thus,  the quark mass matrices come out being hermitian~\cite{Fritzsch:1999ee}. However, we will just assume that  $(\mathbf{M}_{q})_{23} \approx (\mathbf{M}_{q})_{32}$.
\end{itemize}

With these assumptions, the quark mass matrices are parametrized as

\begin{equation}
	{\bf M}_{d}=\begin{pmatrix}
		A_{d} & b_{d} & 0 \\ 
		b_{d} &  B_{d} & C_{d} \\ 
		0 & C_{d} & h_{d}
	\end{pmatrix},\qquad
	{\bf M}_{u}=\begin{pmatrix}
		A_{u} & b_{u} & 0 \\ 
		b_{u} &  B_{u} & -C_{u} \\ 
		0 & -C_{u} & h_{u}
	\end{pmatrix}.
\end{equation}

Both matrices are written in the polar form in order to factorize the CP-violating phases and diagonalize them. This is  $\hat{\bf M}_{q}=\textrm{Diag.}\left(\tilde{m}_{q_{1}}, \tilde{m}_{q_{2}}, 1\right)=\mathbf{U}^{\dagger}_{q L}\mathbf{M}_{q}\mathbf{U}_{q R}$ with $\mathbf{M}_{q}=\mathbf{P}_{q}\bar{\mathbf{M}}_{q}\mathbf{P}_{q}$ where $\mathbf{P}_{q}=\textrm{Diag.}\left(e^{i\eta_{q_{1}}}, e^{i\eta_{q_{2}}}, e^{i\eta_{q_{3}}} \right)$. As noticed, we have normalized the quark masses by the heaviest one, $m_{q_{3}}$, so $\tilde{m}_{q_{i}}=m_{q_{i}}/m_{q_{3}}$;  this is done for simplicity.

For up and down sector, these phases must satisfy the following conditions
\begin{eqnarray}
	\eta_{d}&=&\textrm{arg.} (A_{d})/2, \quad \eta_{s}=\textrm{arg}.(B_{d})/2, \quad \eta_{b}=\textrm{arg.} (h_{d})/2,\quad \eta_{d}+\eta_{s}=2\textrm{arg}. (b_{d}),\quad \quad \eta_{s}+\eta_{b}=2\textrm{arg}.(C_{d});\nn\\
	\eta_{u}&=&\textrm{arg.} (A_{u})/2, \quad \eta_{c}=\textrm{arg}.(B_{u})/2, \quad \eta_{t}=\textrm{arg.} (h_{u})/2,\quad \eta_{u}+\eta_{c}=2\textrm{arg}.(b_{u}),\quad \eta_{c}+\eta_{t}=2\left[\textrm{arg}.(C_{u})+\pi \right].
\end{eqnarray}

along with this,
\begin{equation}
	\bar{{\bf M}}_{q}=\begin{pmatrix}
		\vert \tilde{A}_{q}\vert  & \vert \tilde{b}_{q}\vert & 0 \\ 
		\vert \tilde{b}_{q}\vert & \vert \tilde{B}_{q}\vert & \vert \tilde{C}_{q}\vert \\ 
		0 & \vert \tilde{C}_{q}\vert & \vert \tilde{h}_{q}\vert
	\end{pmatrix}.\label{bmq}
\end{equation}

Then, $\mathbf{U}_{q L}=\mathbf{P}_{q}\mathbf{O}_{q}$ and $\mathbf{U}_{q R}=\mathbf{P}^{\dagger}_{q}\mathbf{O}_{q}$ so that 
$\hat{{\bf M}}_{q}={\bf O}^{T}_{q}\bar{\mathbf{M}}_{q}{\bf O}_{q}$. As one can notice, $\bar{\mathbf{M}}_{q}$ has six free parameters which three of them can be fixed in terms of the physical masses, this is
\begin{eqnarray}
\vert \tilde{b}_{q}\vert&=&\sqrt{\frac{\left(1-\vert \tilde{A}_{q}\vert\right)\left(\tilde{m}_{q_{2}}-\vert \tilde{A}_{q}\vert\right)\left(\vert \tilde{A}_{q}\vert+\vert \tilde{m}_{q_{1}}\vert \right)}{\vert \tilde{h}_{q}\vert-\vert \tilde{A}_{q}\vert}};\nn\\
\vert \tilde{B}_{q}\vert&=& 1+\tilde{m}_{q_{2}}-\vert \tilde{m}_{q_{1}}\vert-\vert \tilde{h}_{q}\vert-\vert \tilde{A}_{q}\vert;\nn\\ 
\vert \tilde{c}_{q}\vert&=&\sqrt{\frac{\left(1-\vert \tilde{h}_{q}\vert\right)\left(\vert \tilde{h}_{q}\vert-\tilde{m}_{q_{2}}\right)\left(\vert \tilde{h}_{q}\vert+\vert \tilde{m}_{q_{1}}\vert \right)}{\vert \tilde{h}_{q}\vert-\vert \tilde{A}_{q}\vert}},
\end{eqnarray}
where $\tilde{m}_{q_{1}}=-\vert m_{q_{1}}\vert/m_{q_{3}}$ and there is a hierarchy among the free parameters namely $1>\vert \tilde{h}_{q}\vert>\tilde{m}_{q_{2}}> \vert \tilde{m}_{q_{1}}\vert>\vert \tilde{A}_{q}\vert$.

Having realized that, the ${\bf O}_{q}$ orthogonal matrix  is given as
{\small
\begin{equation}
	\mathbf{O}_{q}=\begin{pmatrix}
		-\sqrt{\frac{\left(1-\vert \tilde{A}_{q}\vert\right)\left(\tilde{m}_{q_{2}}-\vert \tilde{A}_{q}\vert\right)\left(\vert \tilde{h}_{q}\vert+\vert \tilde{m}_{q_{1}}\vert \right)}{D_{q_{1}}}} & \sqrt{\frac{\left(1-\vert \tilde{A}_{q}\vert\right)\left(\vert \tilde{h}_{q}\vert-\tilde{m}_{q_{2}}\right)\left(\vert \tilde{m}_{q_{1}}\vert+\vert \tilde{A}_{q}\vert \right)}{D_{q_{2}}}}  & \sqrt{\frac{\left(1-\vert \tilde{h}_{q}\vert\right)\left(\tilde{m}_{q_{2}}-\vert \tilde{A}_{q}\vert\right)\left(\vert \tilde{m}_{q_{1}}\vert+\vert \tilde{A}_{q}\vert \right)}{D_{q_{3}}}} \\
		\sqrt{\frac{\left(\vert \tilde{h}_{q}\vert-\vert \tilde{A}_{q}\vert\right)\left(\vert \tilde{h}_{q}\vert+\vert \tilde{m}_{q_{1}}\vert \right)\left(\vert \tilde{m}_{q_{1}}\vert+\vert \tilde{A}_{q}\vert\right)}{D_{q_{1}}}} & \sqrt{\frac{\left(\vert \tilde{h}_{q}\vert-\vert \tilde{A}_{q}\vert\right)\left(\vert \tilde{h}_{q}\vert- \tilde{m}_{q_{2}} \right)\left(\tilde{m}_{q_{2}}-\vert \tilde{A}_{q}\vert\right)}{D_{q_{2}}}} & \sqrt{\frac{\left(1-\vert \tilde{h}_{q}\vert\right)\left(1-\vert \tilde{A}_{q}\vert\right)\left(\vert \tilde{h}_{q}\vert-\vert \tilde{A}_{q}\vert \right)}{D_{q_{3}}}} \\
		-\sqrt{\frac{\left(1-\vert \tilde{h}_{q}\vert\right)\left(\vert \tilde{h}_{q}\vert-\tilde{m}_{q_{2}}\right)\left(\vert \tilde{m}_{q_{1}}\vert +\vert \tilde{A}_{q}\vert\right)}{D_{q_{1}}}} & -\sqrt{\frac{\left(1-\vert \tilde{h}_{q}\vert\right)\left(\vert \tilde{h}_{q}\vert+\vert \tilde{m}_{q_{1}}\vert\right)\left(\tilde{m}_{q_{2}} -\vert \tilde{A}_{q}\vert\right)}{D_{q_{2}}}} & \sqrt{\frac{\left(1-\vert \tilde{A}_{q}\vert\right)\left(\vert \tilde{h}_{q}\vert-\tilde{m}_{q_{2}}\right)\left(\vert \tilde{h}_{q}\vert+\vert \tilde{m}_{q_{1}}\vert \right)}{D_{q_{3}}}}
	\end{pmatrix}.
\end{equation}}
Notice that 
\begin{eqnarray}
D_{q_{1}}&=&\left(1+\vert \tilde{m}_{q_{1}}\vert\right)\left(\tilde{m}_{q_{2}}+\vert \tilde{m}_{q_{1}}\vert\right)\left(\vert \tilde{h}_{q}\vert-\vert \tilde{A}_{q}\vert\right);\nn\\ D_{q_{2}}&=&\left(1-\tilde{m}_{q_{2}}\right)\left(\tilde{m}_{q_{2}}+\vert \tilde{m}_{q_{1}}\vert\right)\left(\vert \tilde{h}_{q}\vert-\vert \tilde{A}_{q}\vert\right);\nn\\ D_{q_{3}}&=&\left(1+\vert \tilde{m}_{q_{1}}\vert\right)\left(1-\tilde{m}_{q_{2}}\right)\left(\vert \tilde{h}_{q}\vert-\vert \tilde{A}_{q}\vert\right).
\end{eqnarray}
Consequently, the CKM matrix ($\mathbf{V}=\mathbf{U}^{\dagger}_{u L} \mathbf{U}_{d L}$) is written as $\mathbf{V}=\mathbf{O}^{T}_{u} \bar{\mathbf{P}}_{q} \mathbf{O}_{d}$ with $\bar{\mathbf{P}}_{q}=\mathbf{P}^{\dagger}_{u}\mathbf{P}_{d}\equiv\textrm{Diag}. \left(e^{i\alpha_{q}},e^{i\beta_{q}}, e^{i\gamma_{q}}\right)$. As notices, three phases take place in the CKM matrix however two of them are relevant.
\begin{eqnarray}
\vert \mathbf{V}_{ud}\vert =\big|\left(\mathbf{O}_{u}\right)_{11}\left(\mathbf{O}_{d}\right)_{11}+\left(\mathbf{O}_{u}\right)_{21}\left(\mathbf{O}_{d}\right)_{21}e^{i\bar{\beta}}+\left(\mathbf{O}_{u}\right)_{31}\left(\mathbf{O}_{d}\right)_{31}e^{i\bar{\gamma}} \big|;\nn\\
\vert	\mathbf{V}_{us}\vert =\big| \left(\mathbf{O}_{u}\right)_{11}\left(\mathbf{O}_{d}\right)_{12}+\left(\mathbf{O}_{u}\right)_{21}\left(\mathbf{O}_{d}\right)_{22}e^{i\bar{\beta}}+\left(\mathbf{O}_{u}\right)_{31}\left(\mathbf{O}_{d}\right)_{32}e^{i\bar{\gamma}}\big|;\nn\\
\vert	\mathbf{V}_{ub}\vert =\big| \left(\mathbf{O}_{u}\right)_{11}\left(\mathbf{O}_{d}\right)_{13}+\left(\mathbf{O}_{u}\right)_{21}\left(\mathbf{O}_{d}\right)_{23}e^{i\bar{\beta}}+\left(\mathbf{O}_{u}\right)_{31}\left(\mathbf{O}_{d}\right)_{33}e^{i\bar{\gamma}}\big|;\nn\\
\vert 	\mathbf{V}_{cd}\vert =\big|\left(\mathbf{O}_{u}\right)_{12}\left(\mathbf{O}_{d}\right)_{11}+\left(\mathbf{O}_{u}\right)_{22}\left(\mathbf{O}_{d}\right)_{21}e^{i\bar{\beta}}+\left(\mathbf{O}_{u}\right)_{32}\left(\mathbf{O}_{d}\right)_{31}e^{i\bar{\gamma}} \big|;\nn\\
\vert	\mathbf{V}_{cs}\vert=\big|\left(\mathbf{O}_{u}\right)_{12}\left(\mathbf{O}_{d}\right)_{12}+\left(\mathbf{O}_{u}\right)_{22}\left(\mathbf{O}_{d}\right)_{22}e^{i\bar{\beta}}+\left(\mathbf{O}_{u}\right)_{32}\left(\mathbf{O}_{d}\right)_{32}e^{i\bar{\gamma}} \big|;\nn\\
\vert \mathbf{V}_{cb}\vert=\big|\left(\mathbf{O}_{u}\right)_{12}\left(\mathbf{O}_{d}\right)_{13}+\left(\mathbf{O}_{u}\right)_{22}\left(\mathbf{O}_{d}\right)_{23}e^{i\bar{\beta}}+\left(\mathbf{O}_{u}\right)_{32}\left(\mathbf{O}_{d}\right)_{33}e^{i\bar{\gamma}} \big|;\nn\\
\vert 	\mathbf{V}_{td}\vert=\big|\left(\mathbf{O}_{u}\right)_{13}\left(\mathbf{O}_{d}\right)_{11}+\left(\mathbf{O}_{u}\right)_{23}\left(\mathbf{O}_{d}\right)_{21}e^{i\bar{\beta}}+\left(\mathbf{O}_{u}\right)_{33}\left(\mathbf{O}_{d}\right)_{31}e^{i\bar{\gamma}} \big|;\nn\\
\vert	\mathbf{V}_{ts}\vert=\big|\left(\mathbf{O}_{u}\right)_{13}\left(\mathbf{O}_{d}\right)_{12}+\left(\mathbf{O}_{u}\right)_{23}\left(\mathbf{O}_{d}\right)_{22}e^{i\bar{\beta}}+\left(\mathbf{O}_{u}\right)_{33}\left(\mathbf{O}_{d}\right)_{32}e^{i\bar{\gamma}} \big|;\nn\\
\vert	\mathbf{V}_{tb}\vert =\big|\left(\mathbf{O}_{u}\right)_{13}\left(\mathbf{O}_{d}\right)_{13}+\left(\mathbf{O}_{u}\right)_{23}\left(\mathbf{O}_{d}\right)_{23}e^{i\bar{\beta}}+\left(\mathbf{O}_{u}\right)_{33}\left(\mathbf{O}_{d}\right)_{33}e^{i\bar{\gamma}} \big|,
\end{eqnarray} 
with $\bar{\beta}=\beta_{q}-\alpha_{q}$ and $\bar{\gamma}=\gamma_{q}-\alpha_{q}$ being two relative phases. Thus, in this benchmark there are six parameters namely $\vert h_{q}\vert$, $\vert A_{q}\vert$ ($q=u,d$) and two relative CP-violating phases. Eventually, the quark mixing angles are given by
\begin{equation}
	\sin^{q}{\theta_{13}}= \vert \left(\mathbf{V}\right)_{ub}\vert;\qquad
	\sin^{q}{\theta_{23}}= \frac{\vert \left(\mathbf{V}\right)_{cb}\vert}{\sqrt{1-\vert \left(\mathbf{V}\right)_{ub}\vert^{2}}};\qquad
	\sin^{q}{\theta_{12}}=\frac{\vert \left(\mathbf{V}\right)_{us}\vert}{\sqrt{1-\vert \left(\mathbf{V}\right)_{ub}\vert^{2}}}.
\end{equation}

In addition, a brief analytical study is realized in order to show that the Gatto-Sartori-Tonin relations are obtained as a limiting case. To do so, let us consider that $\vert \tilde{h}_{q}\vert\approx 1-\tilde{m}_{q_{2}}$ and $\vert \tilde{A}_{q}\vert\approx 0$ \footnote{ The quark mass matrix, given in  Eqn. (\ref{bmq}), with $\vert A_{q}\vert=0$ was studied in \cite{Garcia-Aguilar:2022gfw}, and this kind of matrix fits quite well the CKM one.}. Therefore,

\begin{equation}
	\mathbf{O}_{q}\approx \begin{pmatrix}
		-\sqrt{1-\bar{\tilde{m}}_{q_{1}}} & \sqrt{\bar{\tilde{m}}_{q_{1}}}  & \tilde{m}_{q_{2}} \sqrt{\tilde{m}_{q_{1}}} \\
		\sqrt{\bar{m}_{q_{1}}\left(1-\tilde{m}_{q_{2}}\right)} & \sqrt{1-\tilde{m}_{q_{2}}-\bar{\tilde{m}}_{q_{1}}} & \sqrt{\tilde{m}_{q_{2}}\left(1-\tilde{m}_{q_{1}}\right)} \\
		-\sqrt{\tilde{m}_{q_{1}}\left(1-\tilde{m}_{q_{2}}\right)} & -\sqrt{\tilde{m}_{q_{2}}} & \sqrt{1-\tilde{m}_{q_{2}}}
	\end{pmatrix},
\end{equation}
where $\bar{\tilde{m}}_{q_{1}}=m_{q_{1}}/m_{q_{2}}$.
As a result of this, 
\begin{eqnarray}
	\vert \mathbf{V}_{us} \vert &\approx& \left| 
	\sqrt{\frac{m_{d}}{m_{s}}}-\sqrt{\frac{m_{u}}{m_{c}}}e^{i\bar{\beta}}\right|;\nn\\
	\vert \mathbf{V}_{cb}\vert &\approx& \left|  \sqrt{\frac{m_{s}}{m_{b}}}e^{i\bar{\beta}}-\sqrt{\frac{m_{c}}{m_{t}}}e^{i\bar{\gamma}}\right|;\nn\\
	\vert \mathbf{V}_{ub}\vert &\approx&\left|\frac{m_{s}}{m_{b}}\sqrt{\frac{m_{d}}{m_{s}}}+ \sqrt{\frac{m_{u}}{m_{c}}\frac{m_{s}}{m_{b}}}e^{i\bar{\beta}}-\sqrt{\frac{m_{u}}{m_{t}}}e^{i\bar{\gamma}}\right|;\nn\\
	\vert \mathbf{V}_{td}\vert &\approx&\left|\frac{m_{c}}{m_{b}}\sqrt{\frac{m_{u}}{m_{b}}}+ \sqrt{\frac{m_{u}}{m_{t}}\frac{m_{d}}{m_{s}}}e^{i\bar{\beta}}-\sqrt{\frac{m_{d}}{m_{b}}}e^{i\bar{\gamma}}\right| .
\end{eqnarray}

Thus, we do identify the above expression with the Gatto-Sartori-Toni relations. Hence, we would expect to fit the CKM mixing matrix through  a numerical study, this also will help us to find out the set of values for the free parameters. We have to keep in mind the quark observables depend on six free parameters as it was already commented, and the dependence of the mixing angles and Jarlskog invariant on those is given  by
\begin{eqnarray}
	\sin^{q}{\theta_{13}}&=&\sin^{q}{\theta_{13}}\left(\vert h_{u}\vert, \vert h_{d}\vert, \vert A_{u}\vert,\vert A_{d}\vert, \bar{\beta}_{q}, \bar{\gamma}_{q} \right);\nn\\
	\sin^{q}{\theta_{12}}&=&\sin^{q}{\theta_{12}}\left(\vert h_{u}\vert, \vert h_{d}\vert, \vert A_{u}\vert,\vert A_{d}\vert, \bar{\beta}_{q}, \bar{\gamma}_{q} \right);\nn\\
	\sin^{q}{\theta_{23}}&=&\sin^{q}{\theta_{23}}\left(\vert h_{u}\vert, \vert h_{d}\vert, \vert A_{u}\vert,\vert A_{d}\vert, \bar{\beta}_{q}, \bar{\gamma}_{q} \right);\nn\\
	\mathcal{J}_{q}&=&\mathcal{J}_{q}\left(\vert h_{u}\vert, \vert h_{d}\vert, \vert A_{u}\vert,\vert A_{d}\vert, \bar{\beta}_{q}, \bar{\gamma}_{q} \right),\nn\\
\end{eqnarray}
with $1>\vert \tilde{h}_{q}\vert>\tilde{m}_{q_{2}}> \vert \tilde{m}_{q_{1}}\vert>\vert \tilde{A}_{q}\vert$ and $2\pi\geq \bar{\beta},\bar{\gamma}\geq 0$. The quark masses will be considered as inputs, at the top mass scale, these are give as~\cite{Garces:2018nar}
\begin{eqnarray}
	\tilde{m}_{u} &=& (1.33 \pm 0.73) \times 10^{-5}, \qquad \tilde{m}_{c}= (3.91 \pm 0.42) \times 10^{-3};\nn\\
	\tilde{m}_{d} &=& (1.49 \pm 0.39) \times 10^{-3},\qquad \tilde{m}_{s}= (2.19 \pm 0.53) \times 10^{-2}.
\end{eqnarray}
In addition, the quark mixing angles and the Jarlskog invariant have the following experimental values~\cite{ParticleDataGroup:2022pth}
\begin{eqnarray}
	\sin^{e q}{\theta_{13}}&=& 0.00369 \pm 0.00011;\nn\\
	\sin^{e q}{\theta_{12}}&=& 0.22500 \pm 0.00067;\nn\\
	\sin^{eq}{\theta_{23}}&=& 0.04182^{+0.00085}_{-0.00074};\nn\\
	\mathcal{J}^{e q} &=& \left( 3.08^{+0.15}_{-0.13}\right)\times 10^{-5}.
\end{eqnarray}

\begin{figure}[ht]\centering
			\includegraphics[scale=0.45]{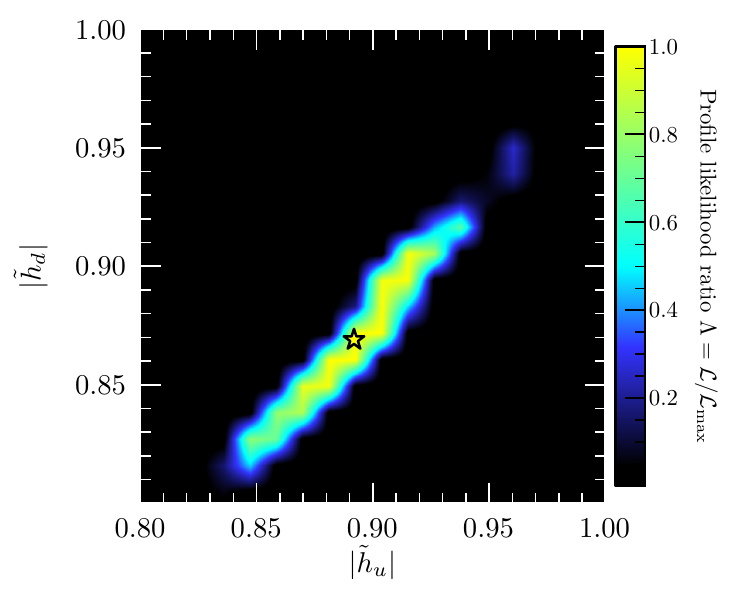}
			\hspace{1mm}\includegraphics[scale=0.45]{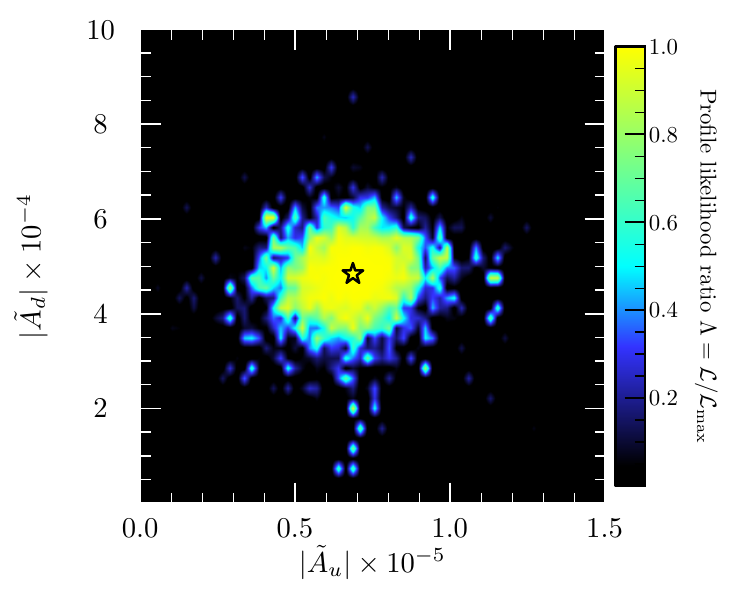}
			\hspace{1mm}\includegraphics[scale=0.45]{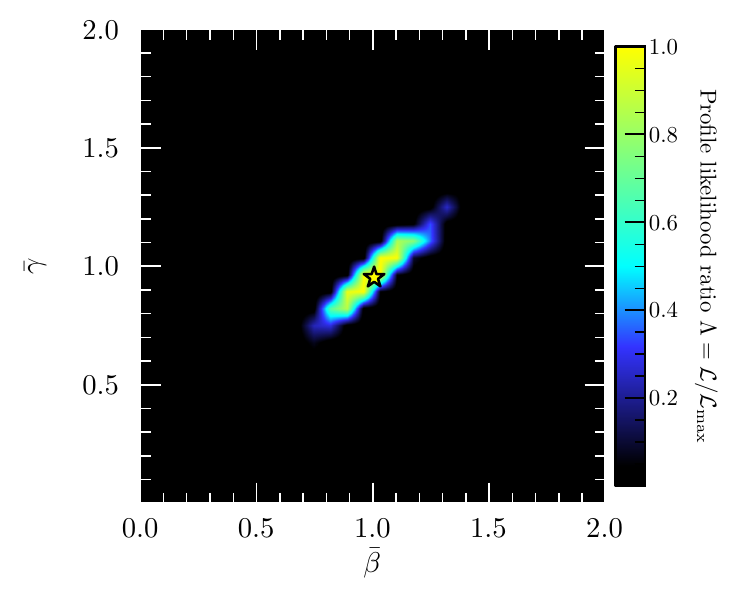}
			\caption{\label{fig:Qfits1} 
				Regions of the free parameter space where the model can fit accurately the experimental observations regarding the CKM matrix. Dark regions are not compatible with observations at all, while the best fit point (BFP) is depicted with a star. }
		\end{figure}

In order to fit to the experimental values we define Gaussian likelihood functions for each CKM matrix element in absolute value $v_{ij} \equiv |V_{ij}^{\mathrm{CKM}}|$:
		
		\begin{equation}
			{\cal L}_{ij} = \frac{1}{\sqrt{2\pi} \sigma} \exp{ \left\{  -\frac{1}{2} \frac{( v_{ij} - v_{ij}^{\mathrm{exp}} )^2}{\sigma^2} \right\} }
		\end{equation}
		the $v_{ij}$ are functions of the free parameters of the model, whilst $v_{ij}^{\mathrm{exp}}$ and $\sigma$ are the corresponding experimental value and its associated error respectively. Maximizing the log-likelihood function $\log{ {\cal L} } \equiv \sum_{i,j} \log{ {\cal L}_{ij} } $ allows us to determine chi-squared intervals $ \Delta \chi^2 \equiv \chi^2 - \chi^2_{\mathrm{min}} $ by the relation:
		
		\begin{equation}
			\Delta \chi^2 = -2 \log { \left( {\cal L} / {\cal L}_{\mathrm{max}} \right) }
		\end{equation}
		where $ {\cal L}_{\mathrm{max}} $ is the maximum of the likelihood function attained at the best fit point (BFP) or the point in parameter space wherein ${\cal L}$ has its global maximum. In addition we treat the values of the quark mass ratios as nuisance parameters and are allowed to vary within their respective experimental interval. To perform the numerical calculations we use a publicly available optimizer~\cite{Martinez:2017lzg,DarkMachinesHighDimensionalSamplingGroup:2021wkt,Scott:2012qh}, we find the BFP to have coordinates in parameter space given by:
		
\begin{eqnarray}
			\vert \tilde{h}_{u}\vert &=& 0.8919159, \qquad \vert \tilde{A}_{u}\vert = 6.85834 \times 10^{-6}, \qquad \bar{\beta} = 1.005637;\nn\\
			\vert \tilde{h}_{d}\vert &=& 0.8693808, \qquad \vert \tilde{A}_{d}\vert = 4.85842 \times 10^{-4}, \qquad \bar{\gamma} =0.956388.
		\end{eqnarray}
		In figure (\ref{fig:Qfits1}) and appendix \ref{appFits} we present heat maps for the visualization of the regions of parameter space where the model can fit accurately the experimental observations. We can conclude that, in order to fit the observations, $\vert \tilde{h}_{u}\vert$ and $\vert \tilde{h}_{d}\vert$ must be higher than $\sim 0.85$, while $\vert \tilde{A}_{u}\vert$ and $\vert \tilde{A}_{d}\vert$ have to be smaller than $\sim 10^{-5}$ and $\sim 6\times 10^{-4}$ respectively. Also, the phases $\bar{\beta}$ and $\bar{\gamma}$ have to be close to 1 radian.

\subsubsection{Lepton sector}\label{LSEC} 
According to the previous section, the alignment
$\langle H_{1}\rangle=v_{1}$, $\langle H_{2}\rangle=iv_{2}$ and $\langle H_{3}\rangle=v_{3}$ was chosen then the charged leptons and Dirac neutrinos get their masses. In addition, we consider the  following alignment
$\langle \phi_{2}\rangle=\langle \phi_{1}\rangle$, which is also a solution of the scalar potential as was shown in~\ref{potencial}, to pseudo Dirac neutrinos. So that
\begin{equation}
		{\bf M}_{e}=\begin{pmatrix}
			a_{e} & 0 & 0 \\ 
			0 & c_{e}+i b_{e} & d_{e} \\ 
			0 & d_{e} & c_{e}-i b_{e}
		\end{pmatrix},\qquad 	{\bf M}_{D}=\begin{pmatrix}
		a_{D} & 0 & 0 \\ 
		0 & c_{D}-i b_{D} & d_{D} \\ 
		0 & d_{D} & c_{D}+i b_{D}
	\end{pmatrix},\qquad {\bf M}_{R}=\begin{pmatrix}
	a_{R} & b_{R} & b_{R} \\ 
	b^{\prime}_{R} & c_{R} & 0 \\ 
	b^{\prime}_{R} & 0 & c_{R}
\end{pmatrix};
\end{equation}
where
\begin{align}
&a_{e}=y^{e}_{1} v_{3},\quad c_{e}=y^{e}_{3} v_{3},\quad b_{e}=y^{e}_{2} v_{2},\quad d_{e}=y^{e}_{2} v_{1}, \nn\\&
a_{D}=y^{D}_{1}v_{3},\quad c_{D}=y^{D}_{3}{v}_{3},\quad b_{D}=y^{D}_{2}v_{2},\quad d_{D}=y^{D}_{2} v_{1},\nn\\ 	
&a_{R}=y^{R}_{1} \langle \phi_{3}\rangle,\quad b_{R}=y^{R}_{2} \langle \phi_{1}\rangle,\quad b^{\prime}_{R}=y^{R}_{3} \langle \phi_{1}\rangle,\quad c_{R}=y^{R}_{4}\langle \phi_{3}\rangle .
\label{eq9}
\end{align}
Besides this, $\mathbf{M}_{2}=\textrm{Diag}.(M_{1 }, M_{2}, M_{2} )$. 

Here, we ought to remark that the charged Yukawa couplings are assumed to be complex. Then, $\mathbf{M}_{e}$  is parametrized as
\begin{equation}
	{\bf M}_{e}=\begin{pmatrix}
		a_{e} & 0 & 0 \\ 
		0 & A_{e} & d_{e} \\ 
		0 & d_{e} & B_{e}
	\end{pmatrix},
\end{equation}
with $A_{e}=c_{e}+ib_{e}$ and $B_{e}=c_{e}-ib_{e}$. It follows that $\hat{{\bf M}}_{e}=\textrm{Diag.}\left(m_{e}, m_{\mu}, m_{\tau}\right)=\mathbf{U}^{\dagger}_{e L} \mathbf{M}_{e} \mathbf{U}_{e R}$ with $\mathbf{U}_{eL}=\mathbf{P}_{e}\mathbf{O}_{e}$ and $\mathbf{U}_{eR}=\mathbf{P}^{\dagger}_{e}\mathbf{O}_{e}$ where $\mathbf{P}_{e}$ is a diagonal matrix that contains phases, and $\mathbf{O}_{e}$ is an orthogonal matrix. As one can notice, there is one phase in the former matrix that takes place in the PMNS one. To see this, we build the bi-lineal $\hat{{\bf M}}_{e}\hat{{\bf M}}^{\dagger}_{e}=\mathbf{U}^{\dagger}_{e L} \mathbf{M}_{e}\mathbf{M}^{\dagger}_{e} \mathbf{U}_{e L}$, then $\mathbf{M}_{e}\mathbf{M}^{\dagger}_{e}$ contains one phase which is associated to the entry $23$. For this reason, $\mathbf{U}_{e L}=\mathbf{P}_{e}\mathbf{O}_{e}$ such that $\mathbf{P}_{e}=\textrm{Diag}.(1,1, e ^{i\eta_{\ell}})$, and $\mathbf{O}_{e}$ is performed explicitly by means of $\hat{{\bf M}}_{e}=\textrm{Diag.}\left(m_{e}, m_{\mu}, m_{\tau}\right)=\mathbf{U}^{\dagger}_{e L} \mathbf{M}_{e} \mathbf{U}_{e R}$. In short, we have
\begin{equation}
		{\bf O}_{e}=\begin{pmatrix}
			1 & 0 & 0 \\ 
			0 & \cos{\theta}_{e} & \sin{\theta}_{e} \\ 
			0 & -\sin{\theta}_{e} & \cos{\theta}_{e}
		\end{pmatrix},\label{EL2}
	\end{equation}
The $\theta_{e}$ free parameter is related to $\vert A_{e}\vert $, and it is constrained by  
\begin{equation}
\tan{\theta_{e}}=\sqrt{\frac{\vert A_{e}\vert- m_{\mu}}{m_{\tau}-\vert A_{e}\vert}}.\label{EL3}
\end{equation}
in consequence $ m_{\tau}>\vt A_{e}\vt> m_{\mu}$.

On the other hand, in the neutrino sector, we obtain
\begin{equation}
 	{\bf M}^{-1}_{R}=\begin{pmatrix}
	\mc{X}& -\mc{Y}_{1} & -\mc{Y}_{1} \\ 
	-\mc{Y}_{2} & \mc{W} & \mc{Z} \\ 
	-\mc{Y}_{2} & \mc{Z} & \mc{W} 
\end{pmatrix}, 
\end{equation}
with
\begin{equation}
\mc{X}=\frac{c^{2}_{R}}{\vert \mathbf{M}_{R}\vert};\quad \mc{Y}_{1}=\frac{b_{R}c_{R}}{\vert \mathbf{M}_{R}\vert};\quad \mc{Y}_{2}=\frac{b^{\prime}_{R}c_{R}}{\vert \mathbf{M}_{R}\vert};\quad \mc{W}=\frac{a_{R}c_{R}-b_{R}b^{\prime}_{R}}{\vert \mathbf{M}_{R}\vert},\quad \mc{Z}=\frac{b_{R}b^{\prime}_{R}}{\vert \mathbf{M}_{R}\vert},
\end{equation}
where  $\vert \mathbf{M}_{R}\vert$ denotes the determinant of $\mathbf{M}_{R}$. In consequence, one performs $\mathbf{A}=\mathbf{M}_{D}(\mathbf{M}^{T}_{R})^{-1}$
\begin{equation}\label{matA}
\mathbf{A}=\begin{pmatrix}
a_{D} \mc{X} & -a_{D} \mc{Y}_{2}  & -a_{D} \mc{Y}_{2} \\
\left(-c_{D}-d_{D}+ib_{D}\right)\mc{Y}_{1} & d_{D}\mc{Z}+\left(c_{D}-i b_{D}\right)\mc{W} & d_{D}\mc{W}+\left(c_{D}-i b_{D}\right)\mc{Z} \\
\left(-c_{D}-d_{D}-ib_{D}\right)\mc{Y}_{1} &  d_{D}\mc{W}+\left(c_{D}+i b_{D}\right)\mc{Z} & d_{D}\mc{Z}+\left(c_{D}+i b_{D}\right)\mc{W}
\end{pmatrix}.
\end{equation}

Finally, the effective neutrino mass matrix, that comes from the ISSM $\mathbf{M}_{\nu}=\mathbf{M}_{D}(\mathbf{M}^{T})^{-1}_{R}\mathbf{M}_{2}\mathbf{M}^{-1}_{R}\mathbf{M}^{T}_{D}$, is given by
\begin{equation}
    \mathbf{M}_{\nu}=  \begin{pmatrix}
		A_{\nu} &  B_{\nu}  &  B^{\prime}_{\nu} \\ 
		B_{\nu} & C^{\prime}_{\nu} & D_{\nu}  \\ 
		B^{\prime}_{\nu} & D_{\nu}  & C_{\nu}
		\end{pmatrix}.
\end{equation}
with the following matrix elements
\begin{eqnarray}
A_{\nu}&=&
a^{2}_{D}\left[M_{1}\mc{X}^{2}+2 M_{2} \mc{Y}^{2}_{2}\right];\nn\\
B_{\nu}&=& a_{D}\left(-c_{D}-d_{D}+ib_{D}\right)\left[M_{1}\mc{X}\mc{Y}_{1}+M_{2}\mc{Y}_{2}\left(\mc{W}+\mc{Z}\right) \right];\nn\\
B^{\prime}_{\nu}&=& a_{D}\left(-c_{D}-d_{D}-ib_{D}\right)\left[M_{1}\mc{X}\mc{Y}_{1}+M_{2}\mc{Y}_{2}\left(\mc{W}+\mc{Z}\right) \right];\nn\\
 C_{\nu}&=&\left[d^{2}_{D}+\left(c_{D}-ib_{D}\right)^{2}\right] \left[M_{1}\mc{Y}^{2}_{1}+M_{2}\left(\mc{W}^{2}+\mc{Z}^{2}\right)\right]+2d_{D}\left(c_{D}-ib_{D}\right)\left[M_{1}\mc{Y}^{2}_{1}+2 M_{2}\mc{W}\mc{Z} \right];\nn\\
C^{\prime}_{\nu}&=&\left[d^{2}_{D}+\left(c_{D}+ib_{D}\right)^{2}\right] \left[M_{1}\mc{Y}^{2}_{1}+M_{2}\left(\mc{W}^{2}+\mc{Z}^{2}\right)\right]+2d_{D}\left(c_{D}+ib_{D}\right)\left[M_{1}\mc{Y}^{2}_{1}+2 M_{2}\mc{W}\mc{Z} \right];\nn\\
D_{\nu}&=&2 c_{D}d_{D} \left[M_{1}\mc{Y}^{2}_{1}+M_{2}\left(\mc{W}^{2}+\mc{Z}^{2}\right) \right]+\left(b^{2}_{D}+c^{2}_{D}+d^{2}_{D}\right)\left[M_{1}\mc{Y}^{2}_{1}+2 M_{2}\mc{W}\mc{Z} \right].
\end{eqnarray}
We have to highlight that the number of parameters in $\mathbf{A}$ and $\mathbf{M}_{\nu}$ might reduce notably. In addition to this, the latter one might exhibit a peculiar feature under certain considerations on the Dirac, pseudo-Dirac and sterile mass matrices. If the Yukawa couplings are real, $\mathbf{A}$,  $\mathbf{\eta}$ and $\mathbf{M}_{\nu}$ are parametrized as
 \begin{equation}
{\bf 	A}\equiv \begin{pmatrix}
		a_{1} &  -a_{2}  &  -a_{2} \\ 
		a_{3} & a^{\ast}_{4} & a^{\ast}_{5}  \\ 
		a^{\ast}_{3} & a_{5}  & a_{4}
		\end{pmatrix}, \qquad \mathbf{\eta}=\begin{pmatrix}
		a_{\nu} &  b_{\nu}  &  b^{\ast}_{\nu} \\ 
		b^{\ast}_{\nu} & c_{\nu} & d_{\nu}  \\ 
		b_{\nu} & d^{\ast}_{\nu}  & c_{\nu}
		\end{pmatrix},\qquad \mathbf{M}_{\nu}=  \begin{pmatrix}
		A_{\nu} &  B_{\nu}  &  B^{\ast}_{\nu} \\ 
		B_{\nu} & C^{\ast}_{\nu} & D_{\nu}  \\ 
		B^{\ast}_{\nu} & D_{\nu}  & C_{\nu}
		\end{pmatrix}.
\end{equation}
Under the aforementioned assumption the following parameters are real: $a_{1,2}$, $a_{\nu}$, $c_{\nu}$; $A_{\nu}$ and $D_{\nu}$. In this benchmark, we point out that $\mathbf{M}_{\nu}$ is identified with the Cobimaximal matrix ~\cite{Fukuura:1999ze,Miura:2000sx,Ma:2002ce,Grimus:2003yn,Chen:2014wxa,Ma:2015fpa,Joshipura:2015dsa,Li:2015rtz,He:2015xha,Chen:2015siy,Ma:2016nkf,Damanik:2017jar,Ma:2017trv,Grimus:2017itg,CarcamoHernandez:2017owh,CarcamoHernandez:2018hst,Ma:2019iwj}. At the same time, the model predicts that the $\mathbf{\eta}$ matrix has a peculiar features, this is, the entries $\mathbf{\eta}_{12}$ ($\mathbf{\eta}_{22}$) and $\mathbf{\eta}_{13}$ ($\mathbf{\eta}_{33}$) have the same magnitude. This fact goes against the current bounds~\cite{Fernandez-Martinez:2016lgt,Blennow:2023mqx} as it is shown below
\begin{equation}
\vert \eta \vert<\begin{pmatrix}
2.5\times 10^{-3}	& 2.4 \times 10^{-5}  & 2.7\times 10^{-3} \\
2.4 \times 10^{-5}	& 4.0 \times 10^{-4} & 1.2 \times 10^{-3} \\
2.7\times 10^{-3}	& 1.2 \times 10^{-3} & 5.6 \times 10^{-3}
\end{pmatrix}.
\end{equation}

Going back to the light neutrino sector, as it has been shown, 
$\mathbf{M}_{\nu}$ is diagonalized by $\mathbf{U}_{\nu}=\mathbf{U}_{\alpha}\mathbf{O}_{23}\mathbf{O}_{13}\mathbf{O}_{12}\mathbf{U}_{\beta}$ such that $\hat{\mathbf{M}}_{\nu}=\textrm{Diag.}\left(\vert m_{1}\vert, \vert m_{2}\vert, \vert m_{3}\vert\right)= \mathbf{U}^{\dagger}_{\nu}\mathbf{M}_{\nu}\mathbf{U}^{\ast}_{\nu}$  with $\mathbf{U}_{\alpha}=\textrm{Diag.}\left(e^{i \alpha_{1}}, e^{i \alpha_{2}}, e^{i \alpha_{3}}\right)$ and $\mathbf{U}_{\beta}=\textrm{Diag.}\left(1, e^{i \beta_{1}}, e^{i \beta_{2}}\right)$~\cite{Grimus:2003yn}. These stand for unphysical and Majorana phases, respectively. In addition, we have explicitly
\begin{equation}
\mathbf{O}_{23}=\begin{pmatrix}
1	& 0 & 0 \\
0	& \cos{\rho_{23}} & \sin{\rho_{23}} \\
0	& -\sin{\rho_{23}}  & \cos{\rho_{23}}
\end{pmatrix},\quad 	\mathbf{O}_{13}=\begin{pmatrix}
\cos{\rho_{13}}	& 0 & \sin{\rho_{13}} e^{-i \delta} \\
0	& 1 & 0 \\
-\sin{\rho_{13}} e^{i \delta} 	& 0  & \cos{\rho_{13}}
\end{pmatrix},\quad \mathbf{O}_{12}=\begin{pmatrix}
\cos{\rho_{12}}	& \sin{\rho_{12}} &  0 \\
-\sin{\rho_{12}}	& \cos{\rho_{12}} & 0 \\
0 	& 0  & 1
\end{pmatrix}.
\end{equation}

By considering $\rho_{23}=\pi/4$, $\delta=-\pi/2$, $\alpha_{1}=0=\alpha_{3}$ and $\alpha_{2}=\pi$; also,  $\beta_{1}=0$ and $\beta_{2}=\pi/2$. With all of this, one obtains
\begin{equation}
	\mathbf{U}_{\nu}=\begin{pmatrix}
		\cos{\rho_{13}}\cos{\rho_{12}}	& \cos{\rho_{13}}\sin{\rho_{12}} & -\sin{\rho_{13}} \\
		\frac{1}{\sqrt{2}}\left(\sin{\rho_{12}}-i\cos{\rho_{12}}\sin{\rho_{13}}\right)	& -\frac{1}{\sqrt{2}}\left(\cos{\rho_{12}}+i\sin{\rho_{12}}\sin{\rho_{13}}\right) & -i\frac{\cos{\rho_{13}}}{\sqrt{2}} \\
		\frac{1}{\sqrt{2}}\left(\sin{\rho_{12}}+i\cos{\rho_{12}}\sin{\rho_{13}}\right)	& -\frac{1}{\sqrt{2}}\left(\cos{\rho_{12}}-i\sin{\rho_{12}}\sin{\rho_{13}}\right) & i\frac{\cos{\rho_{13}}}{\sqrt{2}}
	\end{pmatrix}.\label{Une}
\end{equation}
In flavored models where the charged lepton mass matrix is diagonal, and the active neutrino sector is controlled by the Cobimaximal pattern, the PMNS mixing comes from the aforementioned pattern whose predictions on the atmospheric, reactor and the Dirac CP-violating phase are $\pi/4$, $\neq 0$ and $3\pi/2$, respectively. These values are in tension withe current experimental data as one can see in~\cite{deSalas:2020pgw,Esteban:2020cvm,Gonzalez-Garcia:2021dve}.

Once the Cobimaximal mixing is built, the $\mathbf{M}_{\nu}=\mathbf{U}_{\nu}\hat{{\bf M}}_{\nu}\mathbf{U}^{T}_{\nu}$ matrix elements can be rewritten in terms of the physical neutrino masses as
\begin{eqnarray}
A_{\nu}&=&\vert m_{3}\vert \sin^{2}{\rho_{13}} + \cos^{2}{\rho_{13}}\left[\vert m_{1}\vert \cos^{2}{\rho_{12}}+\vert m_{2}\vert \sin^{2}{\rho_{12}}\right];\nn\\
B_{\nu}&=&\frac{\cos{\rho_{13}}}{\sqrt{2}}\left[\left(\vert m_{1}\vert-\vert m_{2}\vert\right)\cos{\rho_{12}}\sin{\rho_{12}}+i\sin{\rho_{13}}\left(\vert m_{3}\vert-\vert m_{1}\vert \cos^{2}{\rho_{12}}- \vert m_{2}\vert \sin^{2}{\rho_{12}}\right)\right];\nn\\
C_{\nu}&=&\frac{1}{2}\left[\vert m_{2}\vert\left(\cos{\rho_{12}}-i \sin{\rho_{13}}\sin{\rho_{12}}\right)^{2}+\vert m_{1}\vert\left(\sin{\rho_{12}}+i\cos{\rho_{12}}\sin{\rho_{13}}\right)^{2}-\vert m_{3}\vert\cos^{2}{\rho_{13}} \right];\nn\\
D_{\nu}&=& \frac{1}{2}\left[\vert m_{2}\vert\left(\cos^{2}{\rho_{12}}+ \sin^{2}{\rho_{13}}\sin^{2}{\rho_{12}}\right)+\vert m_{1}\vert\left(\sin^{2}{\rho_{12}}+\cos^{2}{\rho_{12}}\sin^{2}{\rho_{13}}\right)+\vert m_{3}\vert\cos^{2}{\rho_{13}}\right].
\end{eqnarray}

On the other hand, in the current framework, the PMNS matrix is given by $\mathbf{U}=\mathbf{U}^{\dagger}_{l}\left(\mathbf{1}-\mathbf{\eta}\right)\mathbf{U}_{\nu}$ where the involved matrices have been already performed. Let us point out that the PMNS matrix is not unitary due to the $\mathbf{\eta}$ contribution but it is irrelevant. Nonetheless, the charged lepton modifies the Cobimaximal predictions on the atmospheric and Dirac CP-violating phase, as we already commented. Speaking roughly, $\mathbf{U}\approx \mathbf{U}^{\dagger}_{e}\mathbf{U}_{\nu}$ with matrix elements
\begin{eqnarray}
\mathbf{U}_{11}&=& ({\bf {U}_{\nu}})_{11};\nn\\
\mathbf{U}_{12}&=&({\bf {U}_{\nu}})_{12};\nn\\
\mathbf{U}_{13}&=&({\bf {U}_{\nu}})_{13};\nn\\
\mathbf{U}_{21}&=& \cos{\theta_{e}}({\bf {U}_{\nu}})_{21}- \sin{\theta_{e}}({\bf {U}_{\nu}})_{31}~e^{-i\eta_{\ell}};\nn\\
\mathbf{U}_{22}&=& \cos{\theta_{e}}({\bf {U}_{\nu}})_{22}- \sin{\theta_{e}}({\bf {U}_{\nu}})_{32}~e^{-i\eta_{\ell}};\nn\\
\mathbf{U}_{23}&=&\cos{\theta_{e}}({\bf {U}_{\nu}})_{23}- \sin{\theta_{e}}({\bf {U}_{\nu}})_{33}~e^{-i\eta_{\ell}};\nn\\
\mathbf{U}_{31}&=& \sin{\theta_{e}}({\bf {U}_{\nu}})_{21} +\cos{\theta_{e}}({\bf {U}_{\nu}})_{31}~e^{-i\eta_{\ell}};\nn\\
\mathbf{U}_{32}&=&\sin{\theta_{e}}({\bf {U}_{\nu}})_{22} +\cos{\theta_{e}}({\bf {U}_{\nu}})_{32}~e^{-i\eta_{\ell}};\nn\\
\mathbf{U}_{33}&=& \sin{\theta_{e}}({\bf {U}_{\nu}})_{23} +\cos{\theta_{e}}({\bf {U}_{\nu}})_{33}~e^{-i\eta_{\ell}}.\label{UPMNS}
\end{eqnarray}
Then, the mixing angles are obtained by 
comparing our theoretical formula with the  
standard parametrization of the PMNS.
\begin{eqnarray}
\sin^{2}{\theta}_{13}&=&\vert ({\bf U})_{13}\vert^{2} =\sin^{2}{\rho_{13}};\nn\\
\sin^{2}{\theta}_{23}&=&\dfrac{\vert ({\bf U})_{23}\vert^{2}}{1-\vert {\bf U}_{13}\vert^{2}}=\frac{1}{2}\left[1+\sin{2\theta_{e}}\cos{\eta_{\ell}}\right];\nn\\
\sin^{2}{\theta_{12}}&=&\dfrac{\vert ({\bf U})_{12}\vert^{2}}{1-\vert {\bf U}_{13}\vert^{2}}=\sin^{2}{\rho_{12}}
.\label{mixang}
\end{eqnarray} 

Also, the Jarlskog invariant can be performed analytically. As one can verify straightforward, using Eqns.(\ref{Une})-(\ref{mixang}), we obtain 
\begin{eqnarray}
\sin{\delta_{CP}} &=& \frac{\textrm{Im}\left[(\mathbf{U})_{23}(\mathbf{U})^{\ast}_{13}(\mathbf{U})_{12}(\mathbf{U})^{\ast}_{22}\right]}{\frac{1}{8}\sin{2\theta_{12}}\sin{2\theta_{23}}\sin{2\theta_{13}}\cos{\theta_{13}}};\nn\\	
\sin{\delta_{CP}}&=&-\frac{\cos{2\theta_{e}}}{\sqrt{1-\sin^{2}{2\theta_{e}}\cos^{2}{\eta_{\ell}}}}.\label{IJE}
\end{eqnarray}
Some comments are added in order: the reactor and solar angles are associated directly to the $\rho_{13}=\theta_{13}$ and $\rho_{12}=\theta_{12}$ parameters, respectively. Besides this, as one notices, the atmospheric angle and Dirac CP-violating phase are deviated from $\pi/4$ and $3\pi/2$, respectively. The $\theta_{e}$ (or $\vert A_{e}\vert$) and $\eta_{\ell}$ free parameters control such deviations which can be computed by a numerical study.

To make this, the charged lepton masses~\cite{Xing:2022uax} will be considered like inputs as well as the reactor and Dirac phase. At the electroweak scale, we have
\begin{equation}
m_{e}=0.483 07 \pm 0.000 45~\textrm{MeV},\quad m_{\mu}=101.766 \pm 0.023~\textrm{MeV},  \quad m_{\tau}= 1728.56 \pm 0.28~\textrm{MeV}.
\end{equation}
Along with this, the  experimental neutrino data are given as follows~\cite{deSalas:2020pgw}
\begin{eqnarray}
\sin^{2}{\theta_{23}}= 0.434-0.610, \qquad \delta_{CP}/^{\circ}= 128-359;\qquad \textrm{Normal ordering}\nn\\
\sin^{2}{\theta_{23}}= 0.433-0.608, \qquad \delta_{CP}/^{\circ}= 200-353.\qquad \textrm{Inverted  ordering}\label{OND}
\end{eqnarray}
 
From Eqns.(\ref{IJE}) and (\ref{mixang}), we vary arbitrarily the free parameters in their allowed range such that these fit the observables at $3\sigma$.
\begin{eqnarray}
\delta_{CP}\left(\vert A_{e}\vert, \eta_{\ell}\right)&=& \arcsin[-\frac{\cos{2\theta_{e}}}{\sqrt{1-\sin^{2}{2\theta_{e}}\cos^{2}{\eta_{\ell}}}}],\nn\\
\sin^{2}\theta_{23}\left(\vert A_{e}\vert, \eta_{\ell}\right)&=& \frac{1}{2}\left[1+\sin{2\theta_{e}}\cos{\eta_{\ell}}\right].
\end{eqnarray}

\section{Phenomenology aspects}

\subsection{Charged Lepton Flavor Violation decays}
The $\ell_{\alpha}\rightarrow \ell_{\beta}\gamma$ charged lepton violation decays~\cite{Petcov:1976ff,Bilenky:1977du,Cheng:1976uq, Cheng:1980tp,Ilakovac:1994kj,Deppisch:2004fa,Bernstein:2013hba} are  interesting processes where the model may be tested. Here, the light and heavy neutrinos can mediate the aforementioned decays in which the former one contribution is suppressed by the active neutrino masses. Nonetheless, the latter ones can enhance the lepton violation decays~\cite{Deppisch:2004fa}. 

Then, for the active and heavy neutrinos, the branching ratio of $\ell_{\alpha}\rightarrow \ell_{\beta}\gamma$~\cite{Ilakovac:1994kj} is given by
\begin{equation}\label{BRs}
BR\left(\ell_{\alpha}\rightarrow \ell_{\beta}\gamma\right)\approx\frac{\alpha^{3}_{W}\sin^{2}{\theta_{W}}}{256\pi^{2}}\frac{m^{4}_{\ell_{\alpha}}}{m^{4}_{W}}\frac{m_{\ell_{\alpha}}}{\varGamma_{\ell_{\alpha}}}\bigg|\sum^{3}_{i=1} \mathbf{U}_{\alpha i} \mathbf{U}^{\ast}_{\beta i}G_{\gamma}\left(\frac{m^{2}_{i }}{m^{2}_{W}}\right)+
\sum^{6}_{j=1} \mathcal{K}_{\alpha j} \mathcal{K}^{\ast}_{\beta j} G_{\gamma}\left(\frac{m^{2}_{j R}}{m^{2}_{W}}\right)\bigg|^{2}
\end{equation}
where $\varGamma_{\ell_{\alpha}}$ and $m_{\ell_{\alpha}}$ stand for the total width and the mass of the $\ell_{\alpha}$ decaying lepton. Along with this, $\mathbf{U}$ and $\mathcal{K}$ are the PMNS and mixing matrix between the heavy and light neutrinos, respectively. Along with this, the function $G_{\gamma}(x)$ has the following form
\begin{equation}
G_{\gamma}(x)=-\frac{2x^{3}-5x^{2}-x}{4(1-x)^{3}}-\frac{3x^{3}}{2(1-x)^{4}}\ln{x}
\end{equation}
with $x_{(i,j)}=(m^{2}_{i},m^{2}_{j R})/m^{2}_{W}$.

For the numerical calculations we compute equation (\ref{BRs}) directly.  At this point, it is convenient to rewrite the matrix ${\bf M}_{D}$ in terms of the eigenvalues of ${\bf M}_{D} {\bf M}_{D}^{\dagger}$, we find for the squared elements:
\begin{eqnarray}
		a_{D}^2&=&m_{1D}^2;\nn\\
		c_{D}^2&=&\frac{1}{4}( -4b_{D}^2 + (m_{2D} - m_{3D})^2 );\nn\\
		d_{D}^2&=&\frac{1}{4}(m_{2D} + m_{3D})^2.
\end{eqnarray}
In this manner, we can obtain numerically the matrix $\mathcal{K}$ as well as the predicted active neutrino masses which are obtained from the effective neutrino mass matrix (\ref{EffNu}).  Using the results from appendix \ref{AppA} we can now substitute all expressions for the free parameters in (\ref{matA}) to find the matrix $\mathbf{A}$ with its elements written as functions of the mass parameters. Since $\mathbf{U}_{l}$ depends on the lepton free parameters, we see that the branching ratios are functions of the set of free parameters $\{ |A_{e}|, \eta_{\ell}, m_{jR}, m_{iD}, b_{R}, b_{D} \}, (j=1-4,i=1-3)$, so that they are correlated with the observables discussed in the lepton sector section. To take into account these correlations, we define a log likelihood function as:
\begin{equation}\label{likeli}
		\log {\cal L} = \log {\cal L}_{\delta_{CP}} + \log {\cal L}_{\theta_{23}} + \log {\cal L}_{\Delta m^2_{21}} + \log {\cal L}_{\Delta m^2_{31}} + \log {\cal L}_{m_{\mathrm{tot}}} + \log {\cal L}_{\mu\rightarrow e\gamma} + \log {\cal L}_{\tau\rightarrow e\gamma} + \log {\cal L}_{\tau\rightarrow \mu\gamma}
\end{equation}
We take the first five as Gaussian likelihoods centered at their experimental values with widths equal to their corresponding experimental errors, here $m_{\mathrm{tot}}$ refers to the sum of the active neutrino masses which are obtained from the effective neutrino mass matrix (\ref{EffNu}), and we take for this observable the interval $0.06 \,\textrm{eV} < m_{\mathrm{tot}} < 0.12 \,\textrm{eV}$ reported in~\cite{ParticleDataGroup:2022pth} based on current experiments and cosmological observations. On the other hand, since we have experimental upper limits on the decaying branching ratios, the likelihoods for the lepton flavor violation observables are defined as:
	\begin{equation}
		{\cal L}_{\mu\rightarrow e\gamma} =\left\{
		\begin{array}{ll}
			\frac{1}{\sqrt{2\pi} \sigma_{\mu\rightarrow e\gamma}},&\mathrm{if}\, BR\left(\mu\rightarrow e\gamma\right) < BR^{\mathrm{exp}}\left(\mu\rightarrow e\gamma\right) \\
			\frac{1}{\sqrt{2\pi} \sigma_{\mu\rightarrow e\gamma}} \exp{\left[ -\frac{1}{2} \frac{(BR_{\mu\rightarrow e\gamma} - BR^{\mathrm{exp}}_{\mu\rightarrow e\gamma})^2}{\sigma^2_{\mu\rightarrow e\gamma}}\right]},&\mathrm{if}\, BR\left(\mu\rightarrow e\gamma\right) > BR^{\mathrm{exp}}\left(\mu\rightarrow e\gamma\right).
		\end{array}	
	\right.
	\end{equation}
	with similar expressions for the rest of the CLFV observables. 
	Our results for the values of the free parameters and some observables at the best fit point are shown in table \ref{tab1}.
	
	\begin{table}[ht!]
		\centering
	\begin{tabular}{ |p{3cm}|p{3cm}|p{3cm}|  }
		\hline
		Parameter & Normal ordering & Inverted ordering \\
		\hline
		$|A_e|$     & $0.1879305 \textrm{GeV}$               & $0.1075106 \textrm{GeV}$             \\      
		$\eta_\ell$ & $1.472408  \textrm{deg}$               & $5.065276\textrm{deg}$                  \\
		$M_1$       & $1.48956   \textrm{GeV}$               & $1.307211 \textrm{GeV}$  \\
    $M_{2}$     & $0.08534454 \textrm{GeV}$              & $0.07308043 \textrm{GeV}$  \\
		$m_{1R}$    & $8478.9329\times 10^{6} \textrm{GeV}$  & $506.9580\times 10^{3} \textrm{GeV}$  \\
		$m_{2R}$    & $916.4922 \textrm{GeV}$                & $865.4204 \textrm{GeV}$  \\
		$m_{4R}$    & $8478.933\times 10^{6} \textrm{GeV}$   & $506.9589\times 10^{3} \textrm{GeV}$  \\
		$m_{1D}$    & $7.860081 \textrm{GeV}$                & $0.1260918 \textrm{GeV}$  \\
		$m_{2D}$    & $9.345837 \textrm{GeV}$                & $1.53258 \textrm{GeV}$  \\
    $m_{3D}$    & $6.32177 \textrm{GeV}$                 & $8.889641 \textrm{GeV}$  \\
		$b_{R}$     & $-2484.761\times 10^{6} \textrm{GeV}$  & $-3429.706\times 10^{6} \textrm{GeV}$  \\
		$b_{D}$     & $0.5637148 \textrm{GeV}$               & $3.023898 \textrm{GeV}$  \\
		\hline
    \hline
    $\sin^2 \theta_{23}$ &  $0.522$                     &  $0.5205012$               \\
    $\delta_{CP}$        &  $243.5\textrm{deg}$         &  $276.4\textrm{deg}$       \\
    $\theta_e$           &  $13.3\textrm{deg}$          &  $3.4\textrm{deg}$         \\
    $m_{\nu_1}$             &  $0.01749147\textrm{eV}$     &  $0.04912271\textrm{eV}$   \\
    $m_{\nu_2}$             &  $0.01949462\textrm{eV}$     &  $0.04986874\textrm{eV}$   \\
    $m_{\nu_3}$             &  $0.05301888\textrm{eV}$     &  $10^{-13}\textrm{eV}$     \\
    $BR\left(\mu\rightarrow e\gamma\right)$      &  $1.155\times  10^{-25}$  & $2.492\times  10^{-45}$   \\
    $BR\left(\tau\rightarrow e\gamma\right)$     &  $2.626\times 10^{-26}$  & $4.216\times 10^{-46}$   \\
    $BR\left(\tau\rightarrow \mu\gamma\right)$   &  $4.444\times 10^{-27}$  & $2.485\times 10^{-23}$   \\
    \hline
	\end{tabular}
\caption{\label{tab1}Numerical values of the free and derived parameters, and some observables at the best fit point.}
\end{table}
	\begin{figure}[ht]\centering
		\includegraphics[scale=0.7]{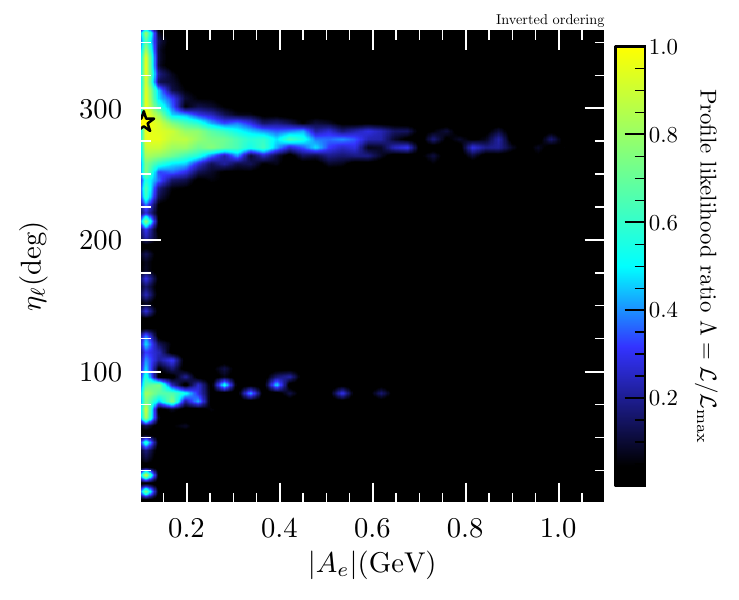}
		\hspace{1mm}\includegraphics[scale=0.7]{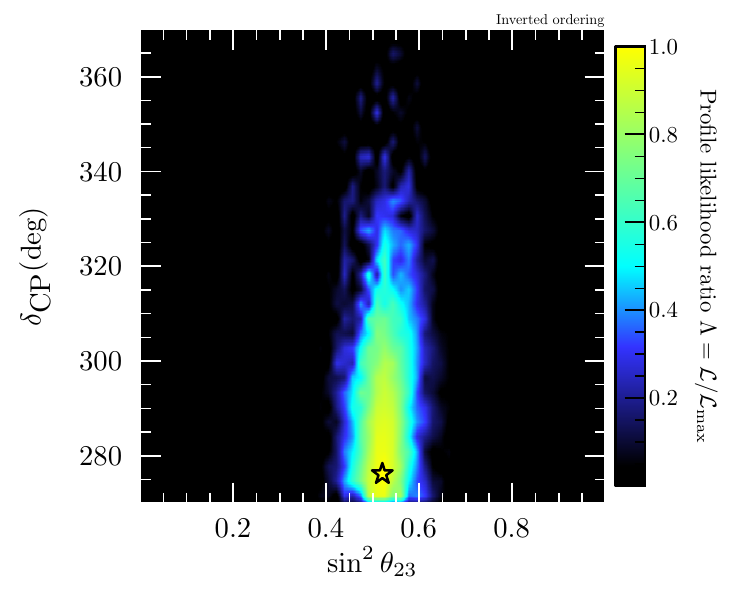} \\
		\includegraphics[scale=0.7]{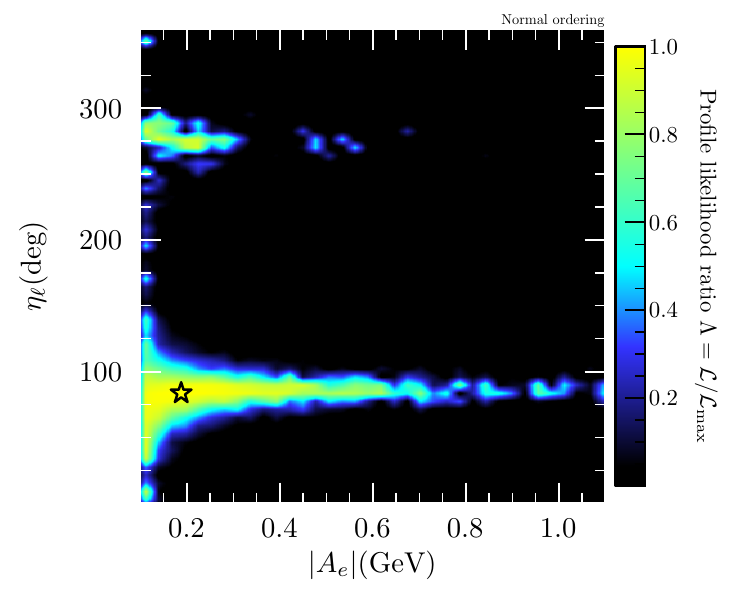}
		\hspace{1mm}\includegraphics[scale=0.7]{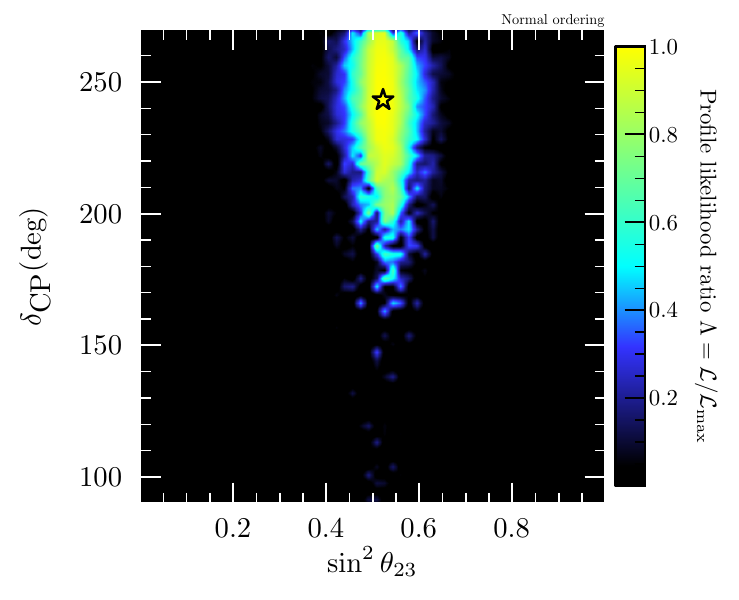}  
		\caption{\label{fig:LFV_obs} 
			Regions of the free parameters where the model can fit accurately the experimental observation regarding the observables in the lepton sector (left panels), the right panels show the values of $\sin^2\theta_{23}$ and $\delta_{CP}$ predicted by the model that are most compatible with current observations. Top (bottom) panel is for Inverted (Normal) Hierarchy. Dark regions are not compatible with observations at all, while the best fit point (BFP) is depicted with a star. }
\end{figure}

\begin{figure}[ht]\centering
	\includegraphics[scale=0.46]{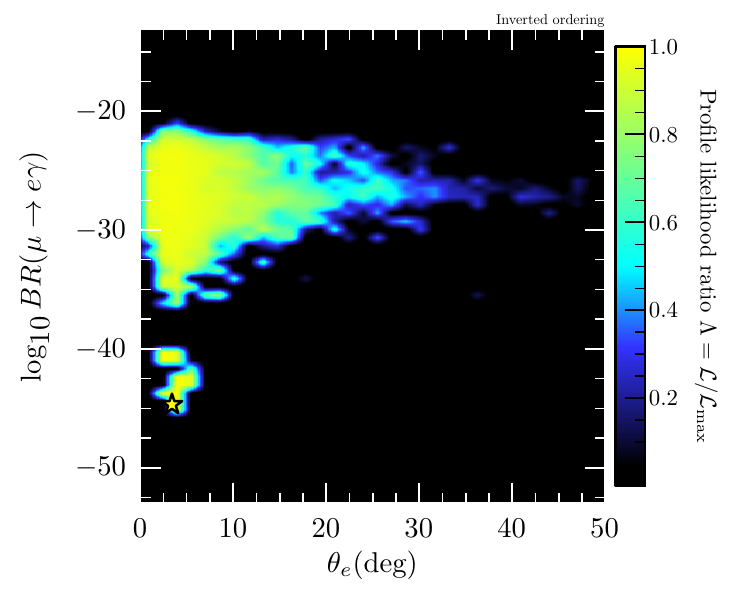}
	\hspace{1mm}\includegraphics[scale=0.46]{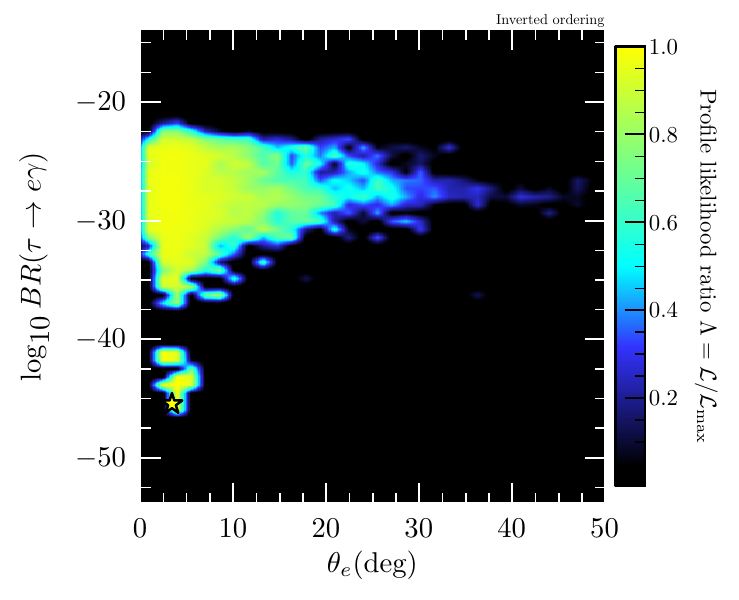}
	\hspace{1mm}\includegraphics[scale=0.46]{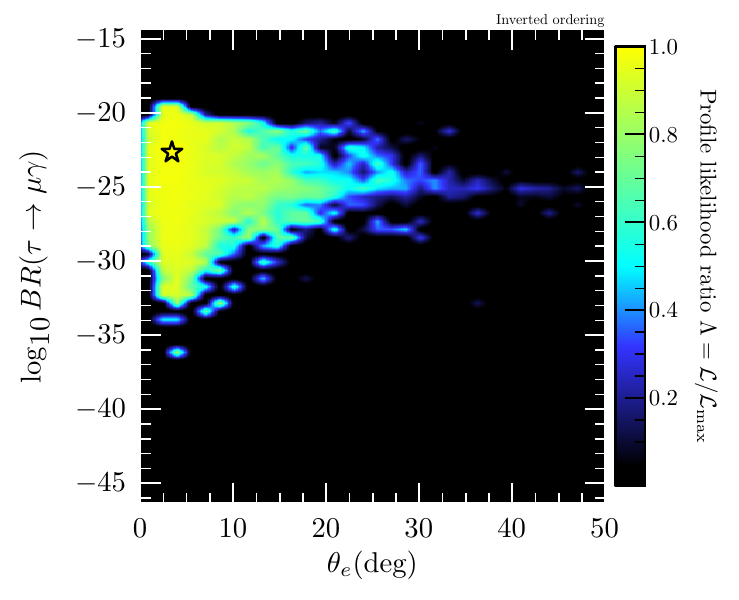} \\
	\includegraphics[scale=0.46]{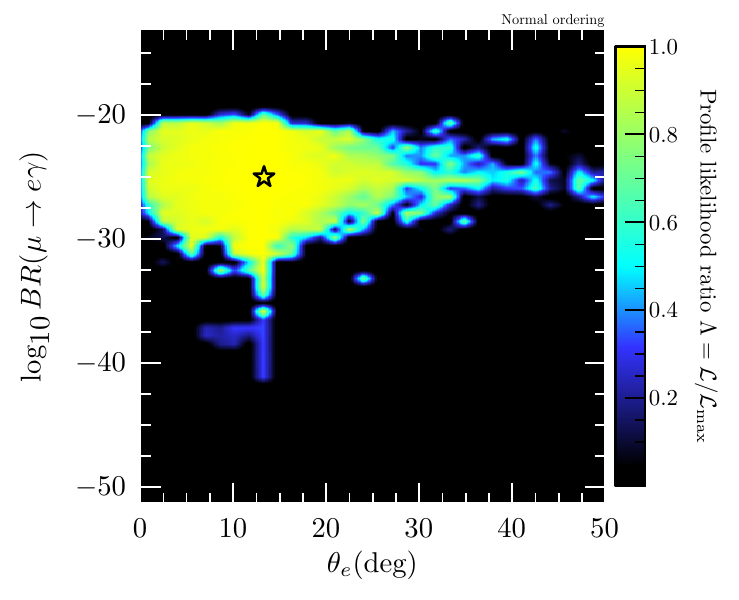}
	\hspace{1mm}\includegraphics[scale=0.46]{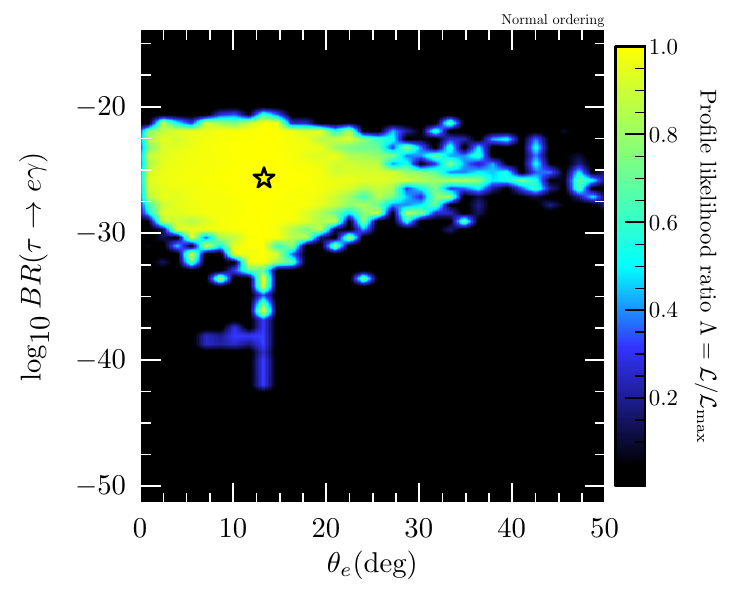}
	\hspace{1mm}\includegraphics[scale=0.46]{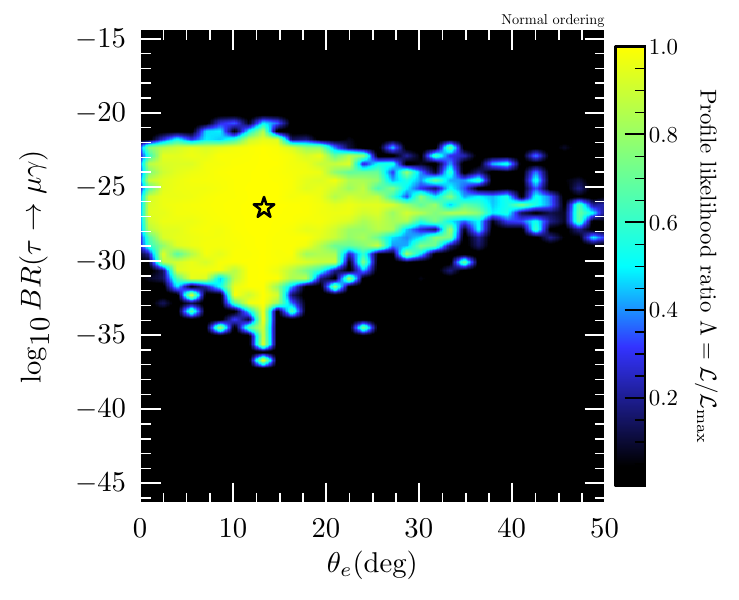}
	\caption{\label{fig:LFV_Bs} 
		Predicted branching fractions as functions of $\theta_e$. Dark regions are not compatible with observations at all, while the best fit point (BFP) is depicted with a star. }
\end{figure}
We present our main results in figures (\ref{fig:LFV_obs}) and (\ref{fig:LFV_Bs}), other plots can be found in appendix (\ref{appFits2}).

\subsection{Neutrinoless double beta decay}
An observable that can be calculated straight away is the $m_{ee}$ mass that takes place in the neutrinoless double beta decay~\cite{Vergados:2002pv,Blennow:2010th, Rodejohann:2011mu,Vergados:2012xy,Pas:2015eia}. The observation of such process would be a direct signal that active neutrinos are Majorana fermions. So far,  there is an upper limit  $\vert m_{ee}\vert< 0.06-0.2~eV$ given by GERDA collaboration~\cite{Agostini:2013mzu}.

In the ISSM framework, the heavy neutrinos contribute to effective neutrino mass~\cite{Blennow:2010th} as follows
\begin{eqnarray}\label{mee}
	\big|m_{ee}\big|&\approx&\bigg| \sum^{3}_{i=1}\left(\mathbf{U}\right)^{2}_{e i} m_{i}+\sum^{6}_{i=1}\left(\mathcal{K}\right)^{2}_{ei} p^{2}\frac{m_{i R}}{p^{2}-m^{2}_{i R}}\bigg|;\nn\\
	&\approx&\bigg |\sum^{3}_{i=1}\left(\mathbf{U}\right)^{2}_{e i} m_{i}-p^{2}\sum^{6}_{i=1}\frac{\left(\mathcal{K}\right)^{2}_{ei} }{m_{i R}}\bigg|,
\end{eqnarray}
where $\mathbf{U}$ and $\mathcal{K}$ have been already defined. Here, $p^{2} \approx -(125~ MeV)^{2}$ stands for the virtual momentum of the neutrino, and we have assumed that $m_{i R}\gg p^{2}$ in the second term.

In the above expression, the first term is the usual contribution due to the lightest neutrinos, explicitly we have
\begin{equation}
	m^{\nu}_{ee}=\cos^{2}{\theta_{13}}\left(\vert m_{1}\vert \cos^{2}{\theta_{12}}+\vert m_{2}\vert\sin^{2}{\theta_{12}}\right)+\vert m_{3}\vert \sin^{2}{\theta_{13}},
\end{equation}
that quantity can be performed by knowing the absolute neutrino masses, the solar and reactor angles. The available information on the neutrino masses are given by the squared mass difference $\Delta m^{2}_{21}= \vert m_{2} \vert^{2}-\vert m_{1} \vert^{2}$ and  $\Delta m^{2}_{31}= \vert m_{3} \vert^{2}-\vert m_{1} \vert^{2}$ ($\Delta m^{2}_{13}= \vert m_{1} \vert^{2}-\vert m_{3} \vert^{2}$) for the normal (inverted)
ordering~\cite{deSalas:2020pgw,Esteban:2020cvm}. Fixing two neutrino masses in terms of the lightest one, we get for the normal and inverted ordering, respectively.
\begin{eqnarray}
	\vert m_{3} \vert^{2}&=& \Delta m^{2}_{31}+\vert m_{1} \vert^{2},\qquad \vert m_{2} \vert^{2}= \Delta m^{2}_{21}+\vert m_{1} \vert^{2};\nn\\
	\vert m_{2} \vert^{2}&=& \Delta m^{2}_{21}+\Delta m^{2}_{13}+\vert m_{3} \vert^{2},\qquad \vert m_{1} \vert^{2}= \Delta m^{2}_{13}+\vert m_{3} \vert^{2}.
\end{eqnarray}

The  second term, in the $\vert m_{ee} \vert$ effective neutrino, has the contribution of the heavy sector. As was shown in section II, $\mathcal{K}=\mathbf{U}^{\dagger}_{\ell}\mathcal{A}\mathbf{V}_{R}$ so that
\begin{equation}
\mathcal{K}\approx \frac{1}{\sqrt{2}}\mathbf{U}^{\dagger}_{\ell}\mathbf{A}
 \begin{pmatrix}
		-\mathbf{V}_{1}	& \mathbf{V}_{2} 
\end{pmatrix},
\end{equation}
where $\mathbf{V}_{(1,2)}$ have been obtained in appendix~\ref{AppA}. In addition to this, $\mathbf{U}_{\ell}$ and $\mathbf{A}$ were obtained in~\ref{LSEC}. 

\begin{figure}[ht]\centering
	\includegraphics[scale=0.7]{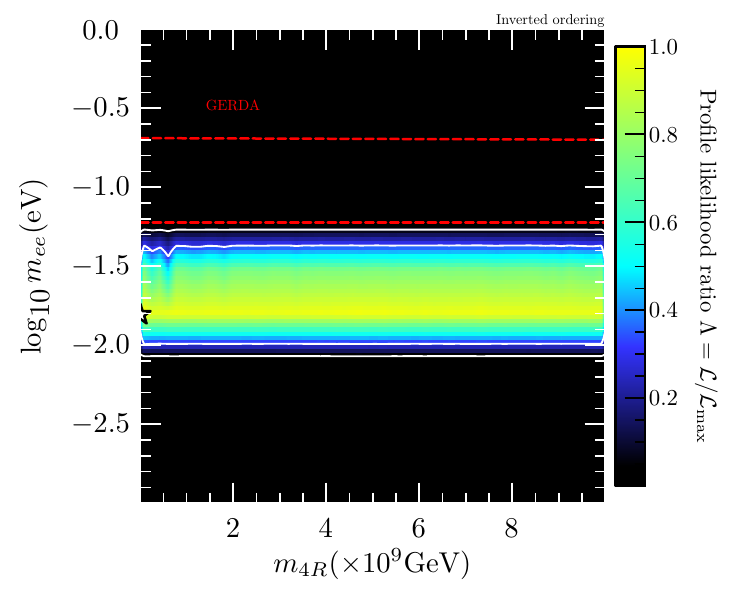}
	\hspace{1mm}\includegraphics[scale=0.7]{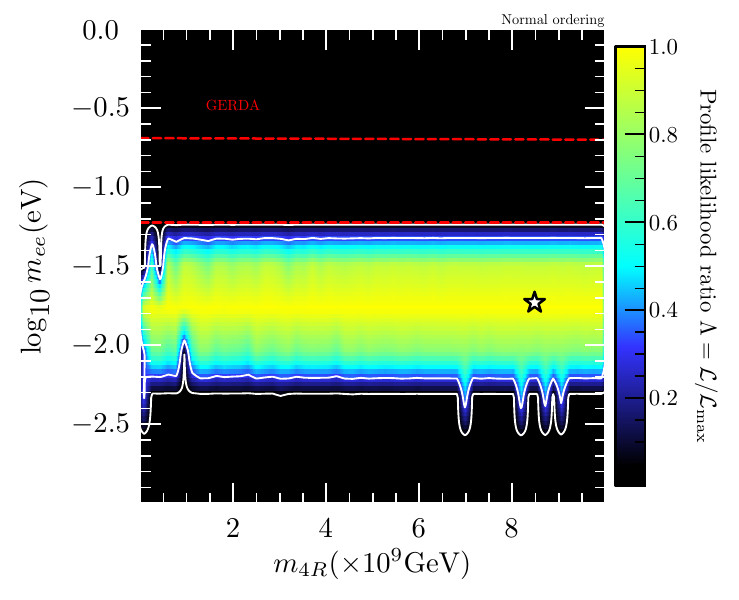} \\
	\caption{\label{fig:DBeta} 
		Region of parameter space where the model is compatible with CLFV observables is depicted as bright. Dark regions are not compatible with observations at all, while the best fit point (BFP) with regards equation (\ref{likeli}) is depicted with a star. The entire compatible region lies below the GERDA interval which is enclosed by the 2 red-dotted lines.}
\end{figure}

We note that in the expression for $\vert m_{ee} \vert$ only the matrix elements $\left(\mathcal{K}\right)_{ei}$ occur, therefore this observable does not depend on the lepton sector free parameters discussed in the previous section. 
For the numerical evaluation of equation (\ref{mee}) we again use the results of appendix \ref{AppA} and calculate the value of this observable for every point of the set of data obtained in the previous subsection. This in order to check that the region of parameter space consistent with the CLFV observables and equation (\ref{likeli}) is also compatible with the GERDA interval. 
In figure \ref{fig:DBeta} we show these regions for both neutrino hierarchies. We can see that the numerical results are fully compatible with the GERDA limit.

\section{Conclusions}
The $\mathbf{S}_{3}\otimes \mathbf{Z}_{2}$ discrete symmetry has been studied in several models beyond the SM to accommodate the fermion mixings. In this work, the cited symmetry was explored in the $B-L$ gauge model which has the peculiarity of explaining the small active neutrino masses by means of the testable ISSM. To attempt to understand the CKM and PMNS patterns, a scalar extension was realized such that three Higgs doublets and singlets were included to the matter content, and under the flavor symmetry, the quark and the scalar sectors are treated in a different manner than the leptons. Along with this, by considering complex vev's on the Higgs doublets and certain assumptions on the Yukawa couplings for the quark and lepton sector, our findings are the following. 

In the quark sector, we considered a benchmark where the mass matrices acquire adequate textures which provide a theoretical CKM matrix with six free parameters. An analytic study allows to obtain the Gatto-Sartori-Toni relations as a limiting case for specific values of the free parameters. Motivated by this, a numerical calculation permits to constrain the region of values for the free parameters such that the quark mixing angles and the Jarlskog invariant are compatible with the latest quark data. 

On the other hand, due to the ISSM the proliferation of free parameters exists in the active light neutrino mass matrix. Despite this, the light neutrino mass matrix ends up having only six effective parameters by assuming real Yukawa couplings and complex vev's in the Higgs doublets. This makes possible to identify the Cobimaximal pattern in the neutrino mass matrix. Whereas, the charged lepton sector contributes to the PMNS in such a way the Cobimaximal pattern controls the mixing angles. At the end of the day, as it was shown, the lepton sector breaks the well known Cobimaximal predictions on the atmospheric angle and Dirac CP-violating phase, and such breaking is managed by two free parameters. Therefore, the PMNS matrix has notable fewer free parameters ($\eta_{\ell}$ and $\theta_{e}$) than the CKM  matrix. Once the lepton mixing is performed, on the phenomenological side, the branching ratios for the charged lepton violation processes as well as the effective neutrino mass from neutrinoless double beta decay get notable contribution due to the heavy neutrino mixings ($\mathcal{K}$) with masses at the $TeV$ scale. The $\mathcal{K}$ mixing has many free parameters, some of which are written in terms of others in such a way that $\mathcal{K}$ depends on a reduced number of free parameters.

From the numerical results, a set of values  for $\eta_{\ell}$ and $\theta_{e}$ (or $\vert A_{e}\vert$) were found for the normal and inverted hierarchy. These values pointed out to a soft breaking on the Cobimaximal predictions, as we expected.   In the charged lepton decays, the constrained values for the free parameters allow to get branching ratios below  the  current experimental bounds, however the best fit indicates  that the normal hierarchy is favored to the inverted one, since we obtained $BR\left(\mu\rightarrow e\gamma\right)=1.155\times  10^{-25}$, $BR\left(\tau\rightarrow e\gamma\right)=2.626\times 10^{-26}$ and $BR\left(\tau\rightarrow \mu\gamma\right)=4.444\times 10^{-27}$ as we can see in \ref{tab1}.  Having fixed the free parameters by constraining the lepton mixing  and branching ratios, the region of parameter space for the $m_{ee}$ effective neutrino mass lies below the GERDA bounds for the normal and inverted hierarchies. Remarkably,  the heavy neutrinos contribution is notable in comparison to the active ones.

Even though some assumptions were considered along the work, the combination of the $\mathbf{S}_{3}\otimes \mathbf{Z}_{2}$ flavor symmetry together with the $B-L$ gauge model, provides an appealing fermion mixing explanation and a rich phenomenology  that might be possible to test in the future.

\section*{Acknowledgements}
This work was partially financed by Instituto Politécnico Nacional project SIP 20242130, by UNAM projects DGAPA PAPIIT IN111224 and IA104223, and by CONAHCYT project CBF2023-2024-548. C.E. acknowledges the support of CONAHCYT C\'atedra no. 341. LEGL acknowledges financial support from CONACyT graduate grants program 786444.

\appendix
\section{Heavy neutrino sector}\label{AppA}
As it was mentioned, the heavy neutrino sector is controlled by the $\mathbf{M}_{R}$ and the block mass matrix, $\mathcal{M}_{R}$, is given by 
\begin{equation}
\mathcal{M}_{R}= \begin{pmatrix}
	0& \mathbf{M}_{R} \\
	\mathbf{M}^{T}_{R} & \mathbf{M}_{2}
\end{pmatrix}.
\end{equation}
In our model, $\mathbf{M}_{R}$ is not symmetric so that there is a clear difference with the work~\cite{Karmakar:2016cvb}. Then, $\mathcal{M}_{R}$ is diagonalized approximately by $\mathbf{V}_{R}$ such that
\begin{equation}
	\mathbf{V}^{\dagger}_{R} \mathcal{M}_{R} \mathbf{V}^{\ast}_{R}=\hat{\mathbf{M}}_{R}\approx\begin{pmatrix}
		-\mathbf{V}^{\dagger}_{1}\overbrace{\left[\frac{1}{2}\left(\mathbf{M}_{R}+\mathbf{M}^{T}_{R}\right)-\frac{1}{2}\mathbf{M}_{2}\right]}^{\mathbf{M}_{1 R}} \mathbf{V}^{\ast}_{1}	&  0  \\
		0	&  \mathbf{V}^{\dagger}_{2}\overbrace{\left[\frac{1}{2}\left(\mathbf{M}_{R}+\mathbf{M}^{T}_{R}\right)+\frac{1}{2}\mathbf{M}_{2}\right]}^{\mathbf{M}_{2 R}} \mathbf{V}^{\ast}_{2}
	\end{pmatrix},
\end{equation}
in good approximation, we have 
\begin{equation}
 	\mathbf{V}_{R}\approx\frac{1}{\sqrt{2}}\begin{pmatrix}
 		\mathbf{1}	& \mathbf{1} \\
 		-\mathbf{1} & \mathbf{1}
 	\end{pmatrix} \begin{pmatrix}
 		\mathbf{V}_{1} & 0 \\
 		0  & \mathbf{V}_{2} 
 	\end{pmatrix},
 \end{equation}
where we have neglected subleading terms. Therefore, $\mathbf{V}_{R}$ is fixed once $\mathbf{V_{1,2}}$ are performed explicitly. To do so, let us diagonalize the $\mathbf{M}_{(1,2)R}$ heavy neutrino mass matrices. Then,
\begin{equation}
\mathbf{M}_{1R}=\begin{pmatrix}
a_{1R}	& B_{R} & B_{R} \\
B_{R} & c_{1R} & 0 \\
B_{R} & 0  & c_{1R}
\end{pmatrix},\qquad \mathbf{M}_{2R}=\begin{pmatrix}
a_{2 R}	& B_{R} & B_{R} \\
B_{R} & c_{2 R} & 0 \\
B_{R} & 0  & c_{2 R}
\end{pmatrix}
\end{equation}
where some parameters have been defined $a_{1 R}=a_{R}-\frac{1}{2}M_{1}$, $c_{1 R}=c_{R}-\frac{1}{2}M_{2}$; $a_{2 R}=a_{R}+\frac{1}{2}M_{1}$,  $c_{2 R}=c_{R}+\frac{1}{2}M_{2}$ and $B_{R}=\frac{1}{2}\left(b_{R}+b^{\prime}_{R}\right)$. Having done this, we have to keep in mind that the Dirac, Pseudo-Dirac and sterile Yukawa couplings were considered real. Consequently, the aforementioned mass matrices are real. As one can verify straightforwardly, $\mathbf{M}_{(1, 2) R}$ is diagonalized by
\begin{equation}\label{A4}
\mathbf{V}_{1,2}=\begin{pmatrix}
\cos{\sigma_{1,2}}	& \sin{\sigma_{1,2}} & 0 \\
-\frac{\sin{\sigma_{1,2}}}{\sqrt{2}}	& \frac{\cos{\sigma_{1,2}}}{\sqrt{2}} & -\frac{1}{\sqrt{2}} \\
-\frac{\sin{\sigma_{1,2}}}{\sqrt{2}}	& \frac{\cos{\sigma_{1,2}}}{\sqrt{2}} & \frac{1}{\sqrt{2}}
\end{pmatrix},\qquad \cos{\sigma_{(1,2)}}=\sqrt{\frac{m_{(3,6)R}-m_{(1,4)R}}{m_{(2,5)R}-m_{(1,4)R}}}
\end{equation}
where
\begin{eqnarray}\label{A5}
a_{(1,2)R}&=&m_{(1,4)R} \cos^{2}{\sigma_{(1,2})}+m_{(2,5) R} \sin^{2}{\sigma_{(1,2)}};\nn\\
B_{R}&=&\frac{m_{(2,5)R}-m_{(1,4)R}}{\sqrt{2}}\sin{\sigma_{(1,2)}}\cos{\sigma_{(1,2)}};\nn\\
c_{(1,2)R}&=&m_{(1,4)R} \sin^{2}{\sigma_{(1,2)}}+m_{(2,5)R} \cos^{2}{\sigma_{(1,2)}};\nn\\
c_{(1,2)R}&=&m_{(3,6)R}.
\end{eqnarray}
Then, $\mathbf{V}^{\dagger}_{1}\mathbf{M}_{1 R}\mathbf{V}^{\ast}_{1}=\textrm{Diag.}\left(m_{1 R},m_{2 R}, m_{3 R} \right)$ and $\mathbf{V}^{\dagger}_{2}\mathbf{M}_{1 R}\mathbf{V}^{\ast}_{2}=\textrm{Diag.}\left(m_{4 R},m_{5 R}, m_{6 R} \right)$. As a result of this, the $\mathcal{K}$ mixing matrix, that takes places in the neutrinoless double beta decay, is obtained explicitly by means of $\mathbf{U}_{\ell}$, $\mathbf{A}$ and $\mathbf{V}_{R}$.
As one notices, there are many free parameters in the $\mathcal{K}$ matrix so we will try of reducing some of them by writing some in terms of other ones.

We note from (\ref{A4}), that the last two equations of (\ref{A5}) are identical, and therefore:
\begin{eqnarray}\label{A6}
		c_{R}&=&m_{3R} + M_2/2;\nn\\
		c_{R}&=&m_{6R} - M_2/2;
\end{eqnarray}
which implies:
	\begin{equation}\label{A7}
		m_{6R}=m_{3R} + M_2.
\end{equation}
In a similar manner, consistency of the rest of equations of (\ref{A5}) requires:
\begin{eqnarray}\label{A8}
		m_{5R}&=& M_1 + M_2 + m_{1R} + m_{2R} - m_{4R};\nn\\
		(m_{1R} - m_{2R}) \sqrt{\frac{(m_{1R} - m_{3R})(m_{3R} - m_{2R})}{(m_{1R} - m_{2R})^2}} 
		&=& (m_{4R} - m_{5R}) \sqrt{\frac{(m_{4R} - m_{6R})(m_{6R} - m_{5R})}{(m_{4R} - m_{5R})^2}}.
\end{eqnarray}
Assuming the mass hierarchies:
\begin{equation}\label{A9}
		m_{1R} > m_{3R} > m_{2R} \quad , \quad m_{4R} > m_{6R} > m_{5R},
\end{equation}
we are led to the condition:
	\begin{equation}
		(m_{1R} - m_{3R})(m_{3R} - m_{2R}) = (m_{4R} - m_{6R})(m_{6R} - m_{5R}).
	\end{equation}
Substituting the previous expressions for $m_{5R}$ and $m_{6R}$, we can ensure the fulfillment of this condition by choosing $m_{3R}$ such that:
\begin{equation}
		m_{3R} = \frac{ (m_{4R} - M_2 )(M_1 + m_{1R} + m_{2R} - m_{4R}) - m_{1R} m_{2R} }{M_1 - M_2}
\end{equation}
In this manner, the elements of the matrix $\mathbf{M}_{R}$ will depend on the free parameters $m_{1R},m_{2R},m_{4R},M_1, M_2$ and $b_{R}$. In addition, $\mathcal{K}$ depends on a reduced number of free parameters namely: $\vert A_{e}\vert$, $\eta_{\ell}$, $m_{iD}$, $b_{D}$, $m_{jR}$, $b_{R}$, $M_1$ and $M_2$ where $i=1-3$ and $j=1,2,4$.

\section{Quark sector likelihood profiles}\label{appFits}
In this appendix we present additional plots featuring characteristics of the regions of the parameter space where the model fits accurately the observations.
	
\begin{figure}[ht]\centering
		\includegraphics[scale=0.7]{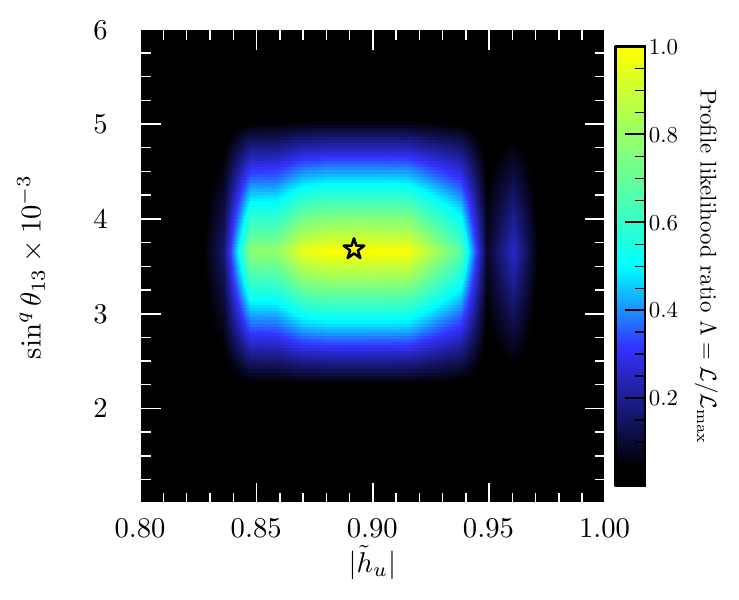}
		\hspace{1mm}\includegraphics[scale=0.7]{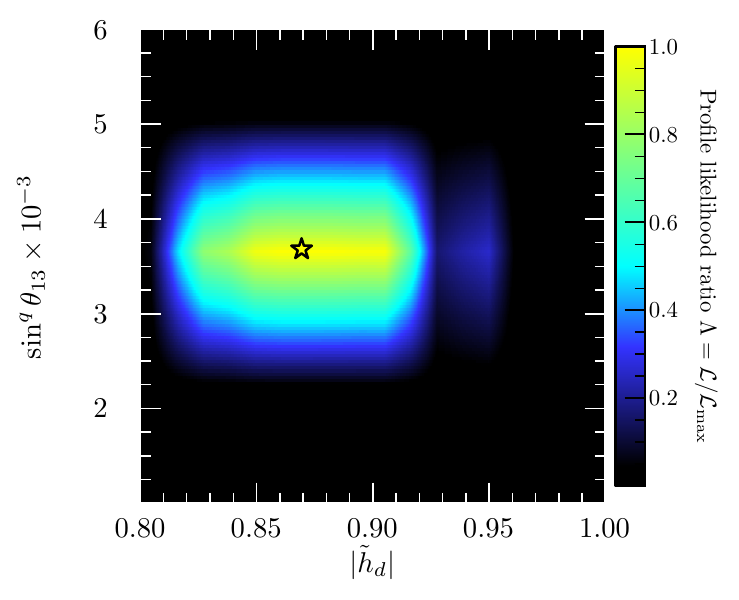} \\
		\includegraphics[scale=0.7]{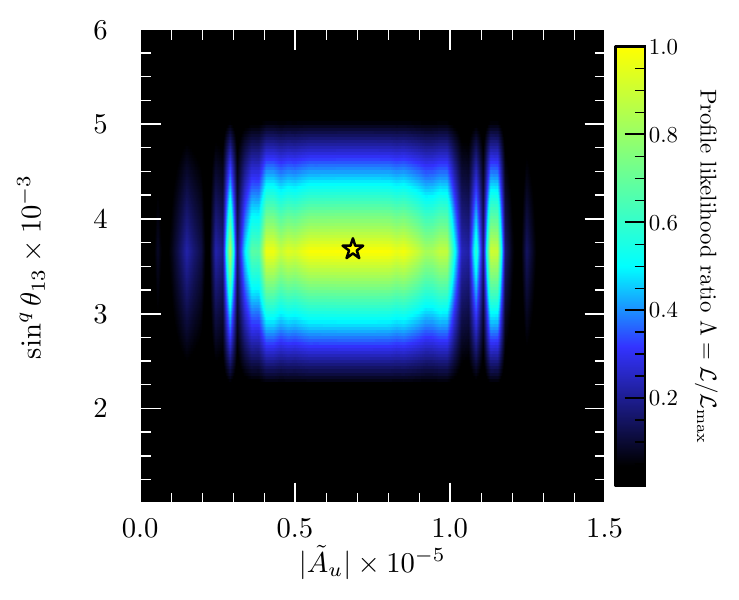}
		\hspace{1mm}\includegraphics[scale=0.7]{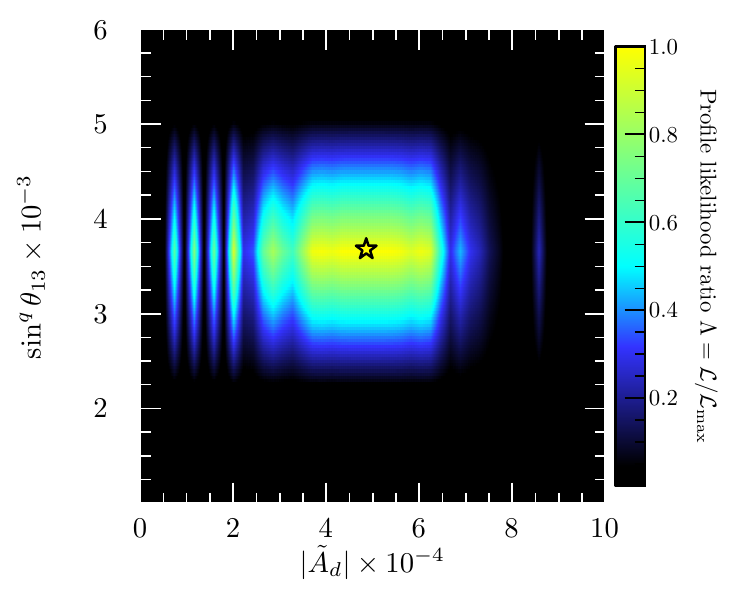} \\
		\includegraphics[scale=0.7]{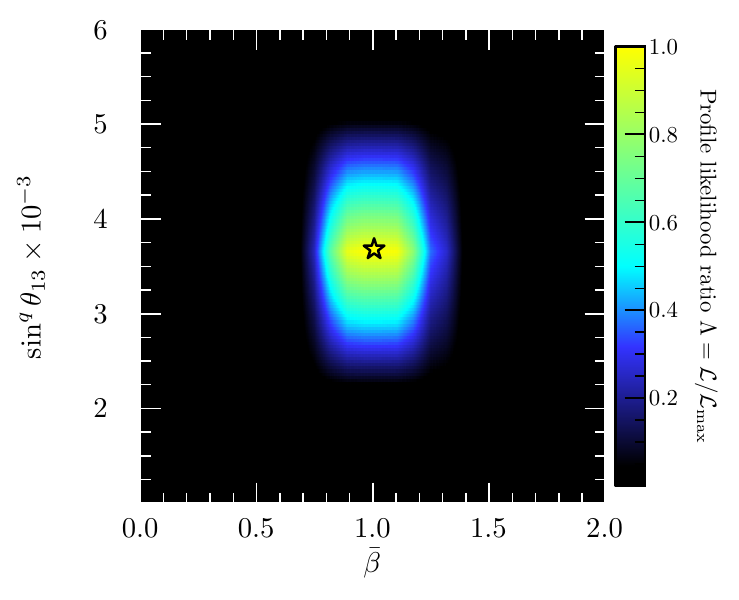}
		\hspace{1mm}\includegraphics[scale=0.7]{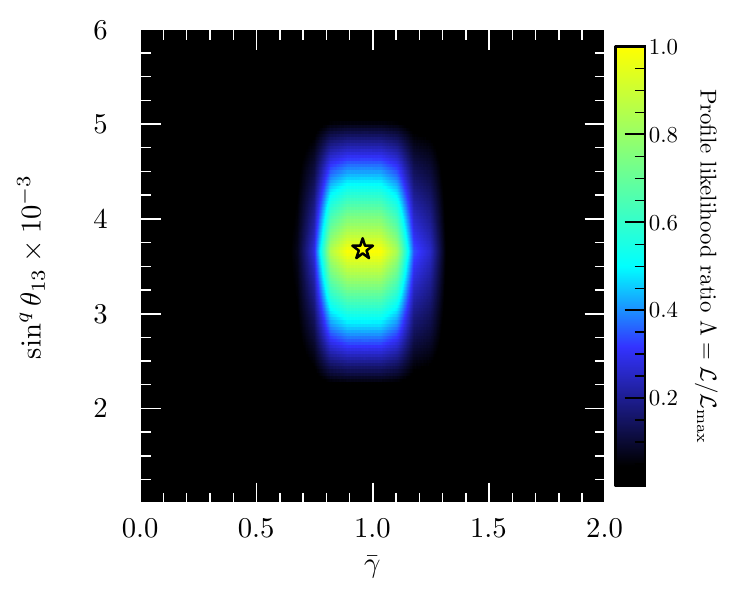} 
		\caption{\label{fig:Qfits_15} 
			Regions of the free parameters where the model can fit accurately the experimental observation regarding the $\sin^q\theta_{13}$ observable. Dark regions are not compatible with observations at all, while the best fit point (BFP) is depicted with a star. }
	\end{figure}

	\begin{figure}[ht]\centering
		\includegraphics[scale=0.7]{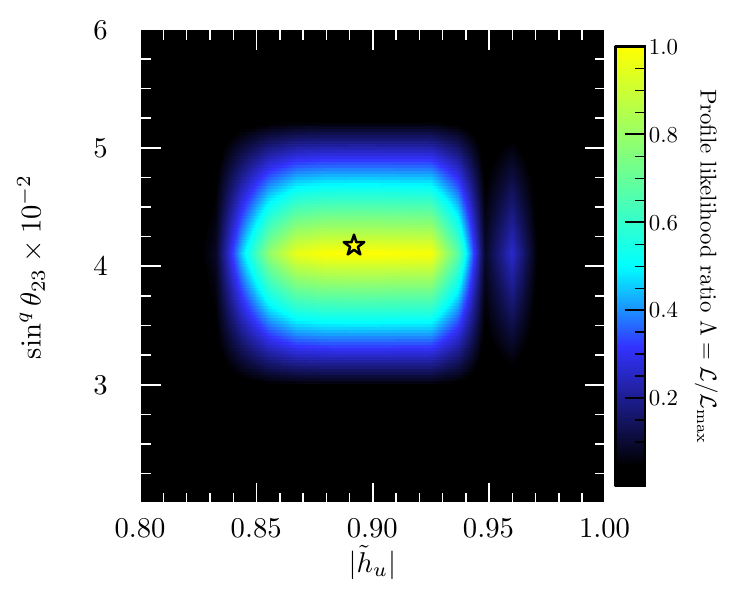}
		\hspace{1mm}\includegraphics[scale=0.7]{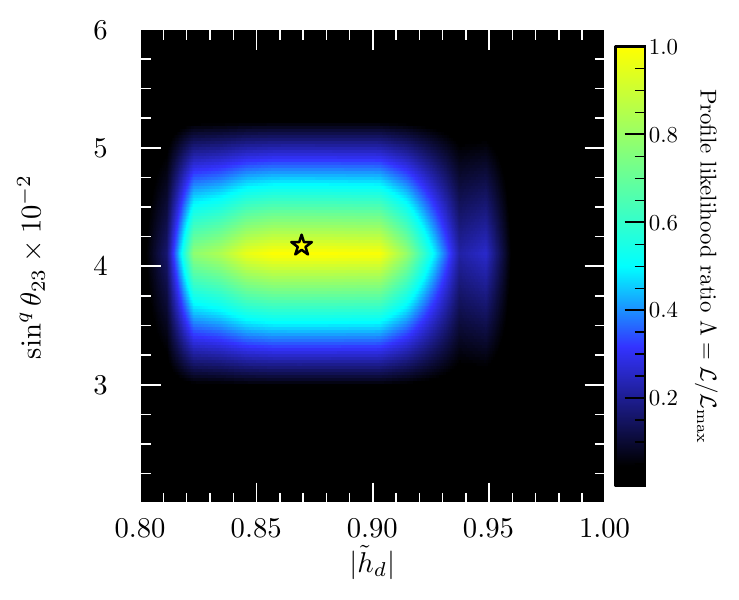} \\
		\includegraphics[scale=0.7]{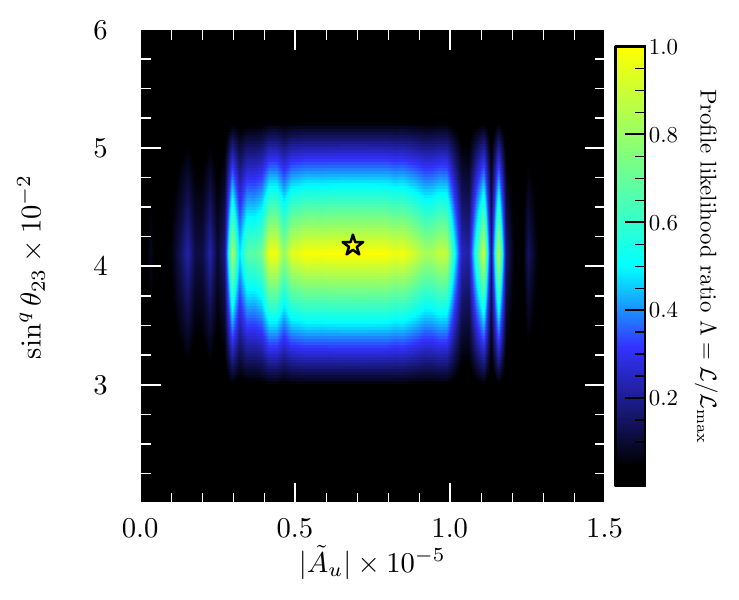}
		\hspace{1mm}\includegraphics[scale=0.7]{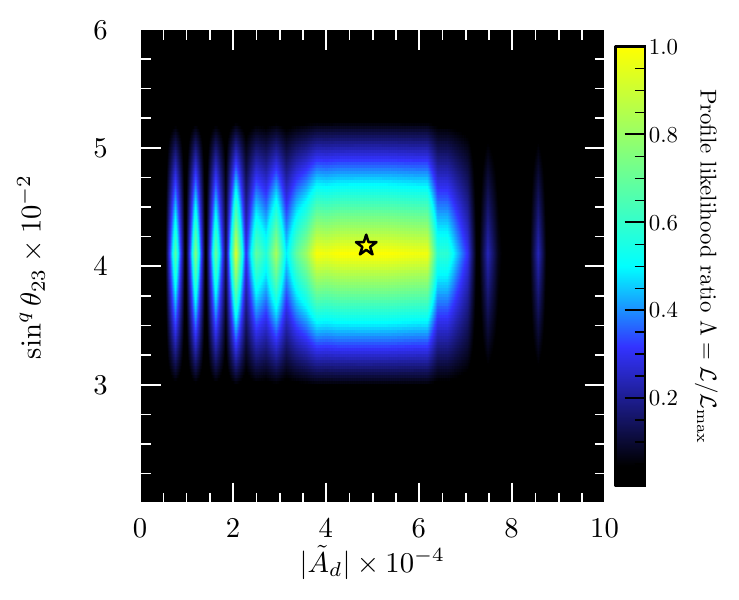} \\
		\includegraphics[scale=0.7]{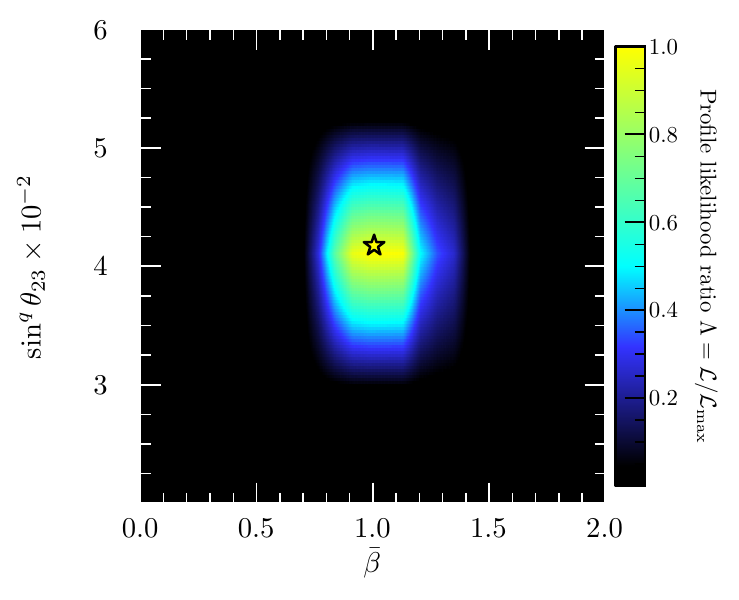}
		\hspace{1mm}\includegraphics[scale=0.7]{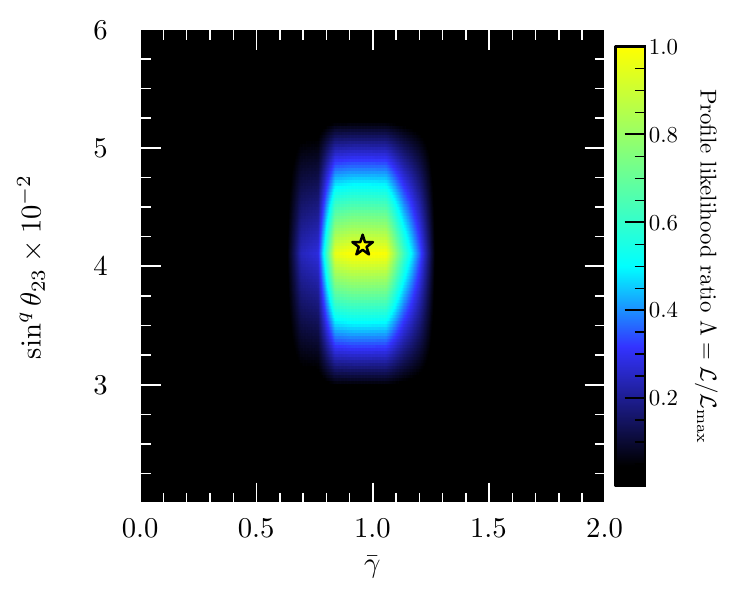} 
		\caption{\label{fig:Qfits_16} 
			Regions of the free parameters where the model can fit accurately the experimental observation regarding the $\sin^q\theta_{23}$ observable. Dark regions are not compatible with observations at all, while the best fit point (BFP) is depicted with a star. }
	\end{figure}

\begin{figure}[ht]\centering
	\includegraphics[scale=0.7]{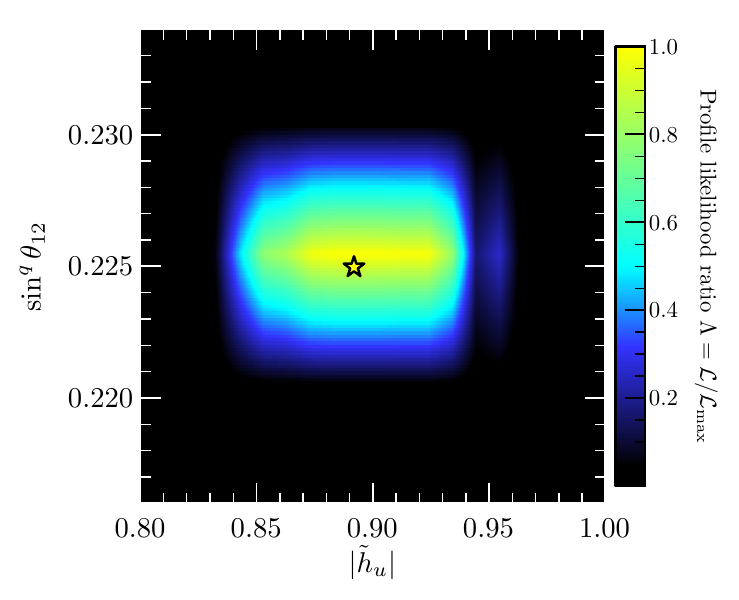}
	\hspace{1mm}\includegraphics[scale=0.7]{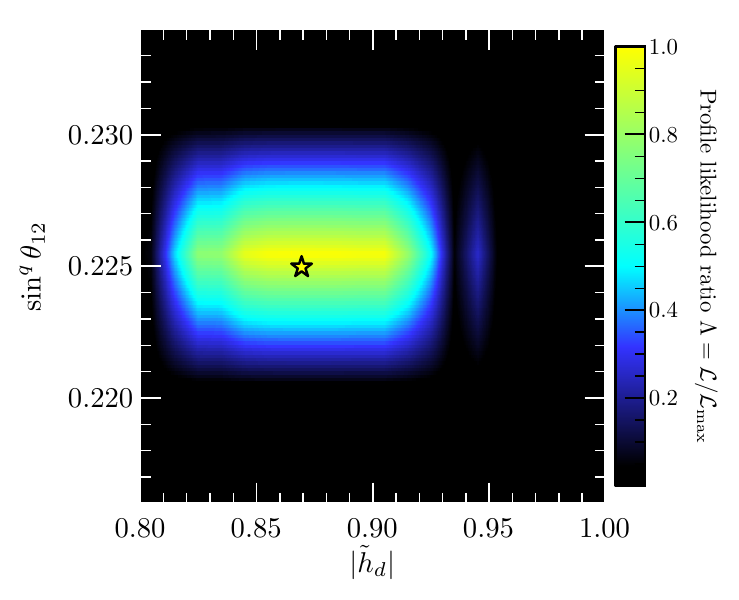} \\
	\includegraphics[scale=0.7]{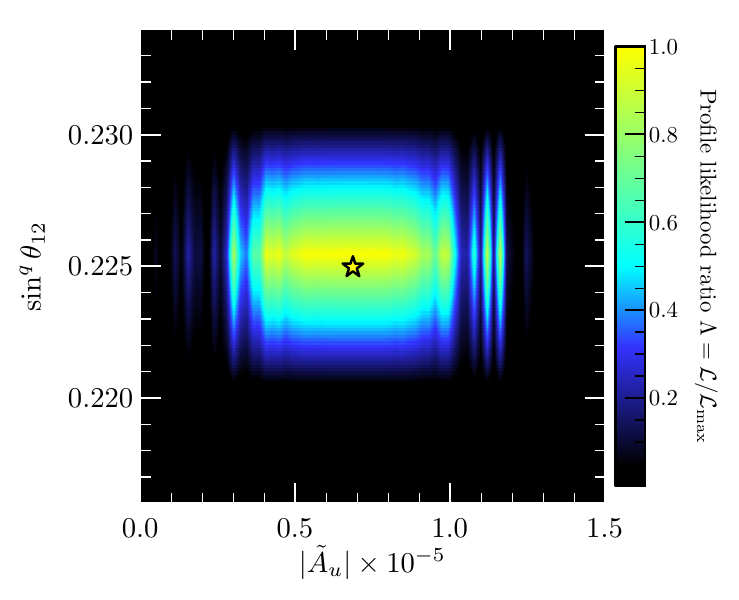}
	\hspace{1mm}\includegraphics[scale=0.7]{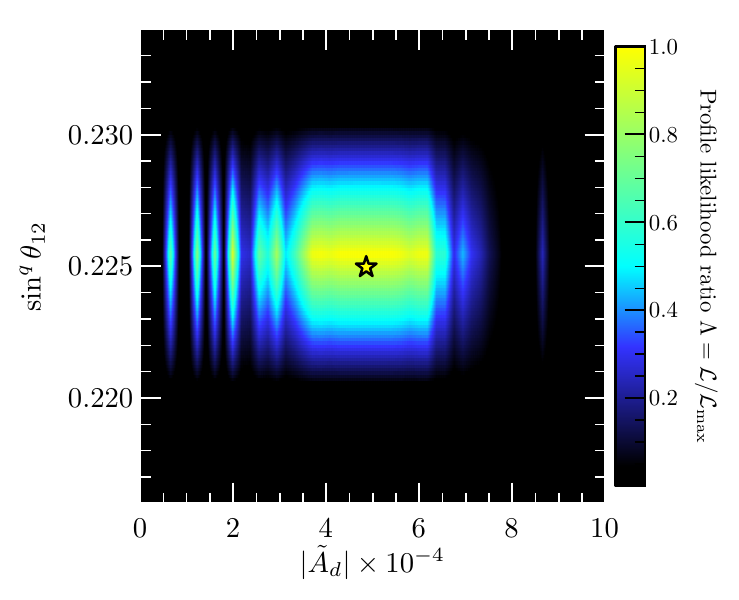} \\
	\includegraphics[scale=0.7]{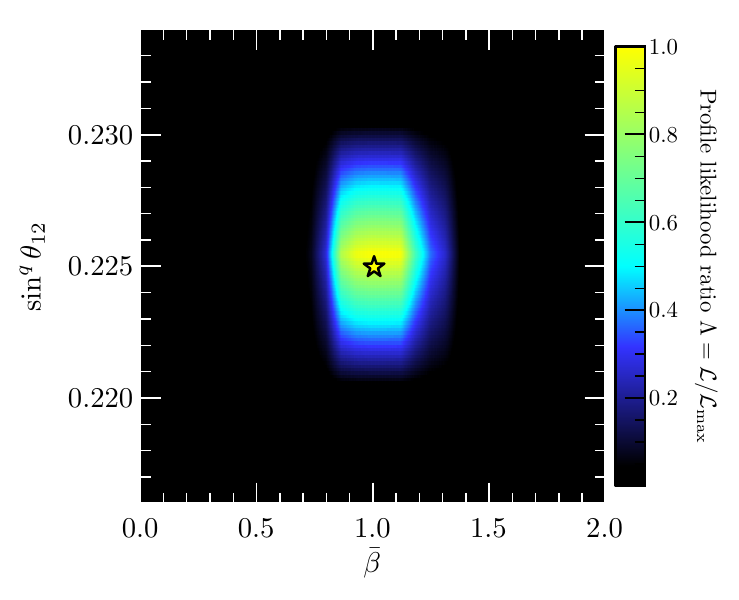}
	\hspace{1mm}\includegraphics[scale=0.7]{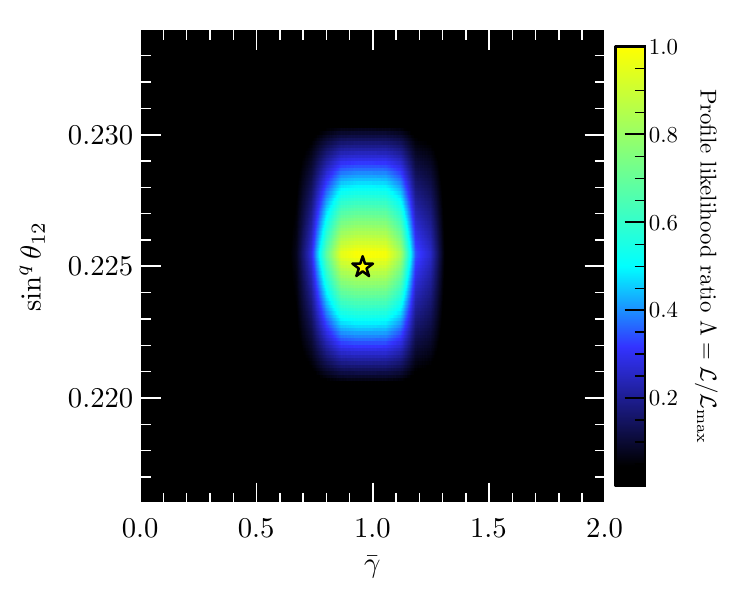} 
	\caption{\label{fig:Qfits_17} 
		Regions of the free parameters where the model can fit accurately the experimental observation regarding the $\sin^q\theta_{12}$ observable. Dark regions are not compatible with observations at all, while the best fit point (BFP) is depicted with a star. }
\end{figure}

\begin{figure}[ht]\centering
	\includegraphics[scale=0.7]{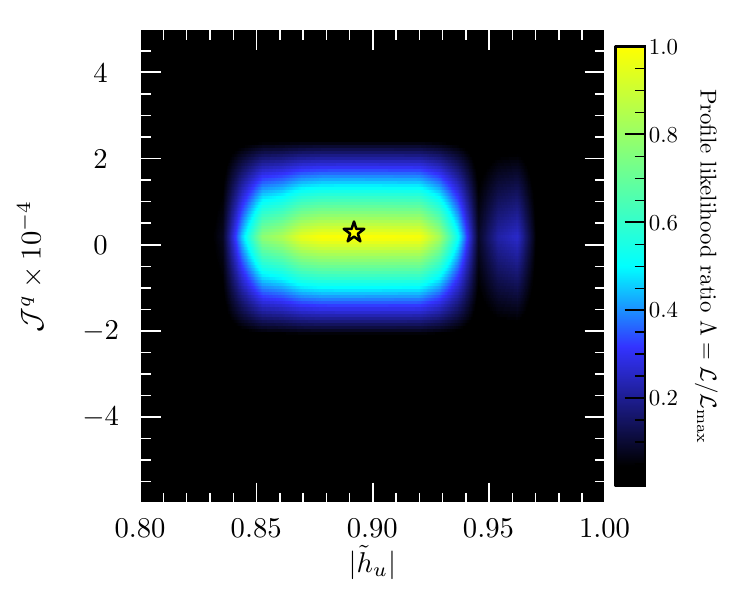}
	\hspace{1mm}\includegraphics[scale=0.7]{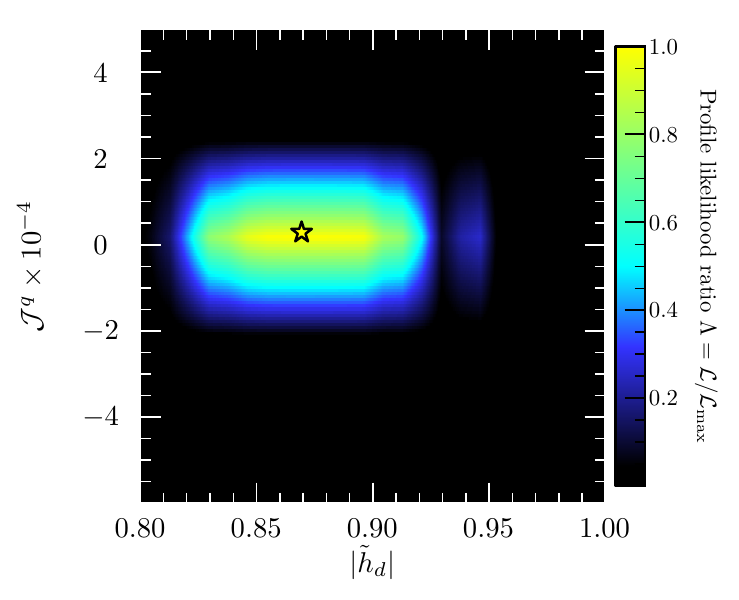} \\
	\includegraphics[scale=0.7]{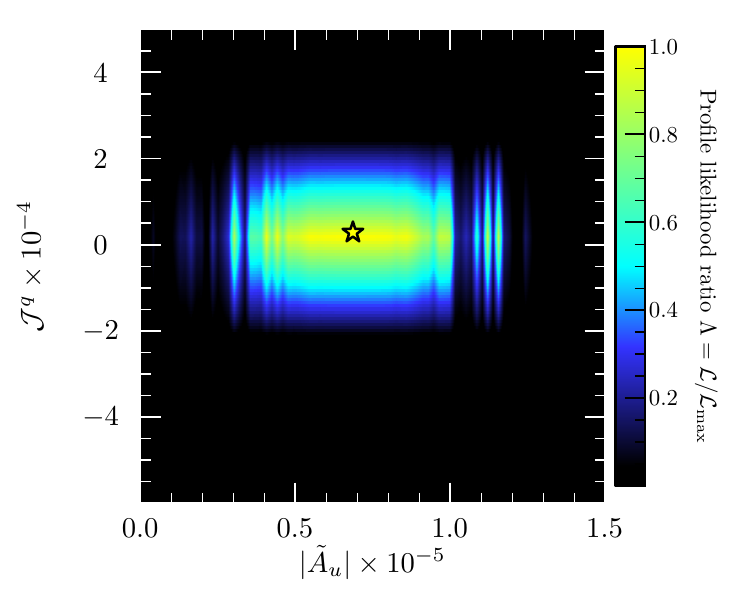}
	\hspace{1mm}\includegraphics[scale=0.7]{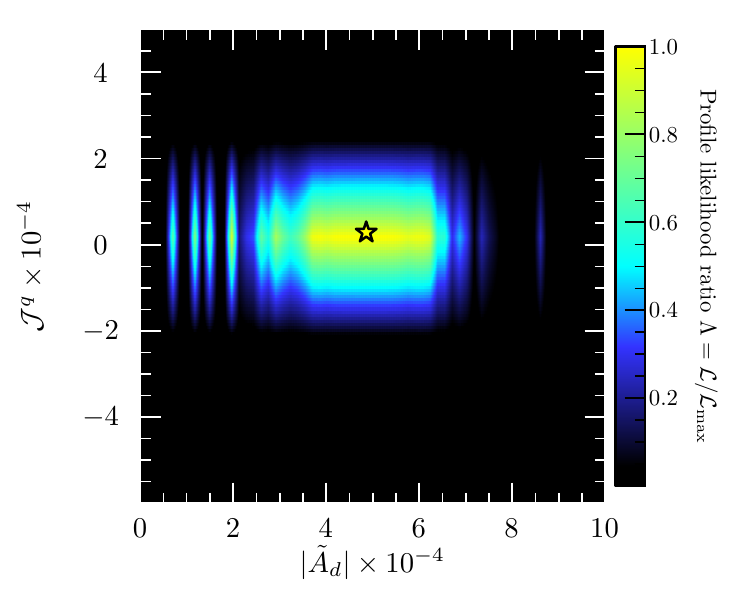} \\
	\includegraphics[scale=0.7]{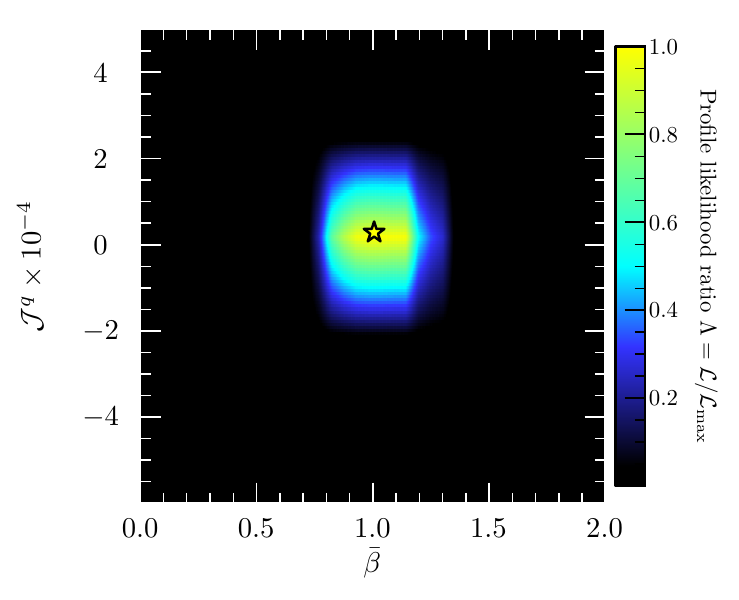}
	\hspace{1mm}\includegraphics[scale=0.7]{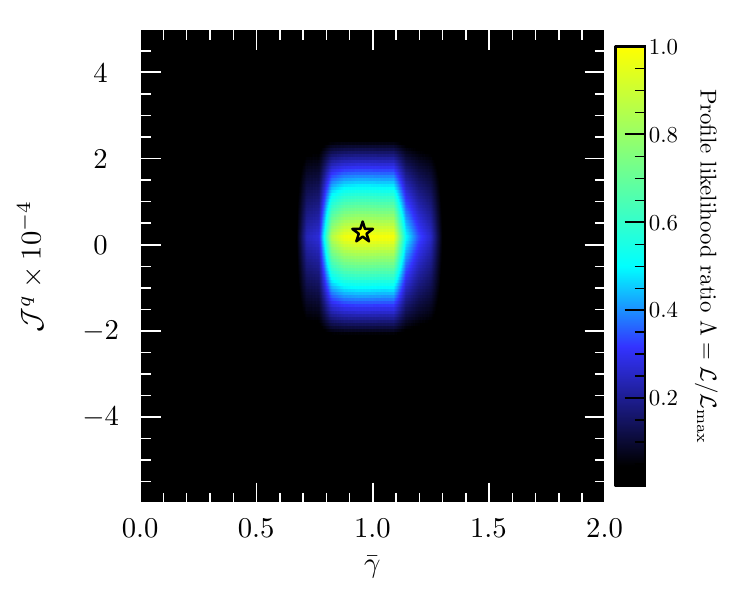} 
	\caption{\label{fig:Qfits_18} 
		Regions of the free parameters where the model can fit accurately the experimental observation regarding the ${\cal J}^q$ observable. Dark regions are not compatible with observations at all, while the best fit point (BFP) is depicted with a star. }
\end{figure}

\section{Lepton sector likelihood profiles}\label{appFits2}
In this appendix we show some other plots of the regions in parameter space of the model where observables can be accurately fitted.
	
\begin{figure}[ht]\centering
		\includegraphics[scale=0.6]{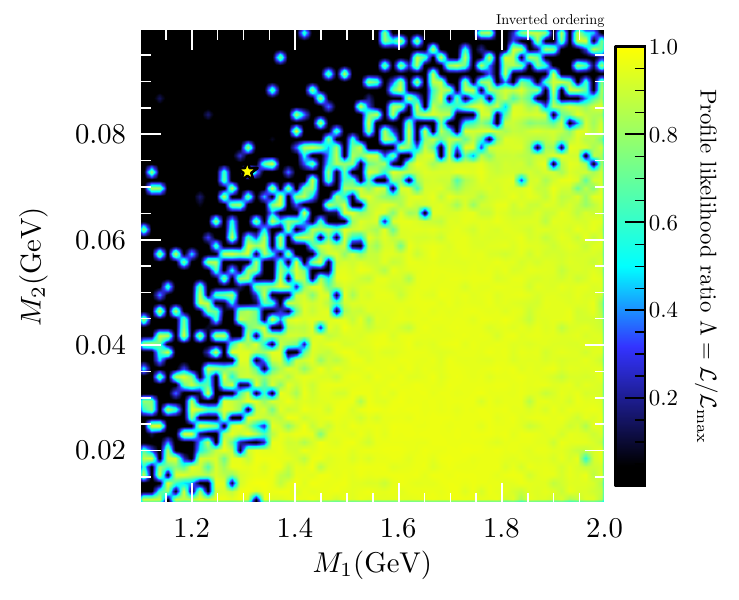}
		\hspace{1mm}\includegraphics[scale=0.6]{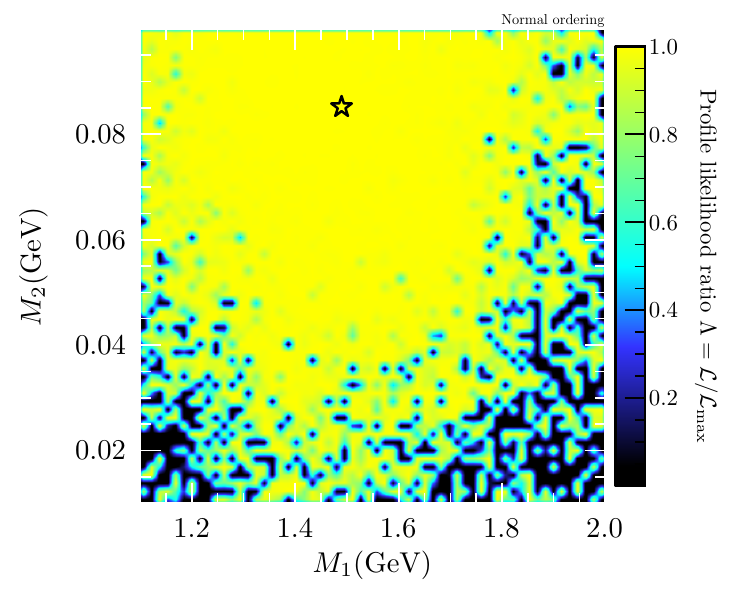} \\
		\includegraphics[scale=0.6]{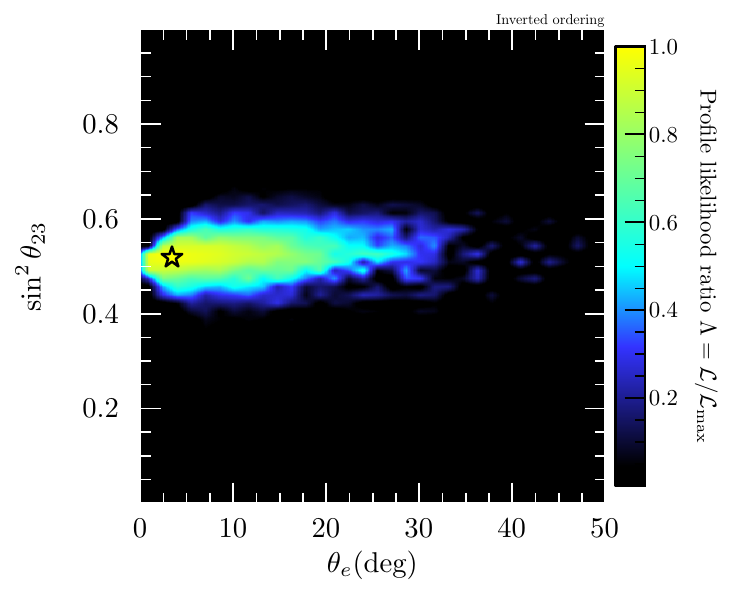}
		\hspace{1mm}\includegraphics[scale=0.6]{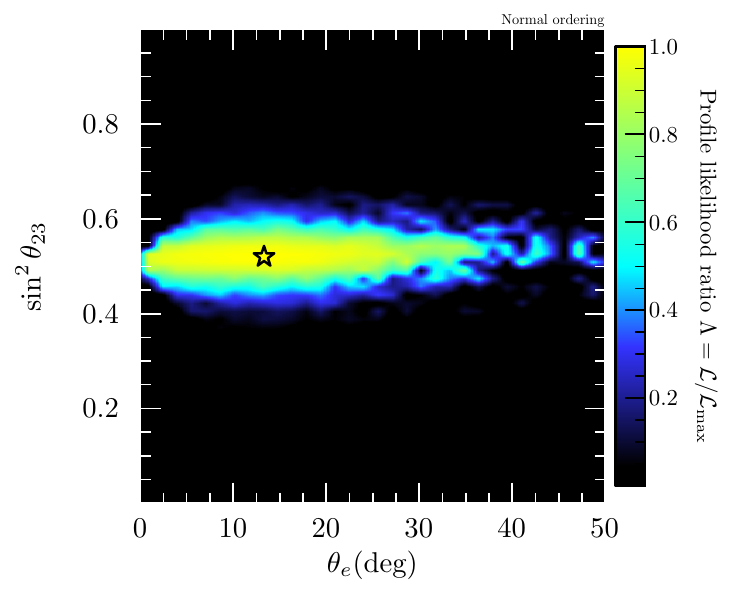} \\
		\includegraphics[scale=0.6]{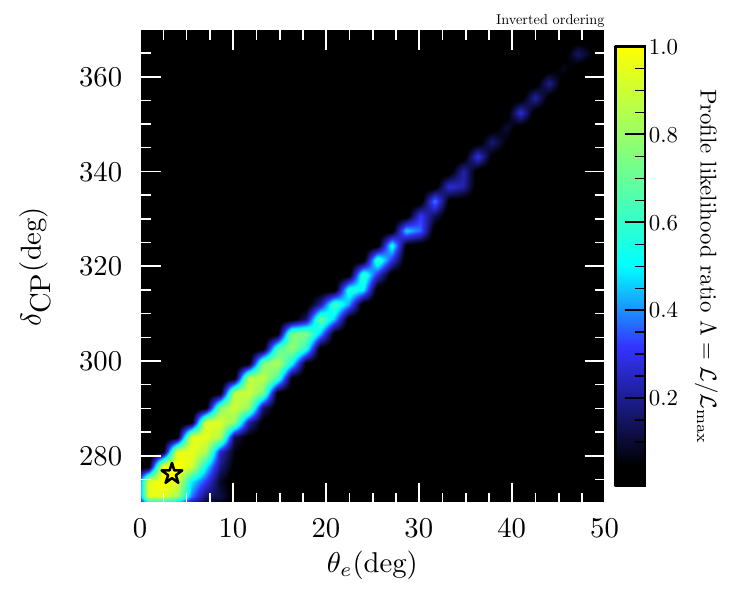}
		\hspace{1mm}\includegraphics[scale=0.6]{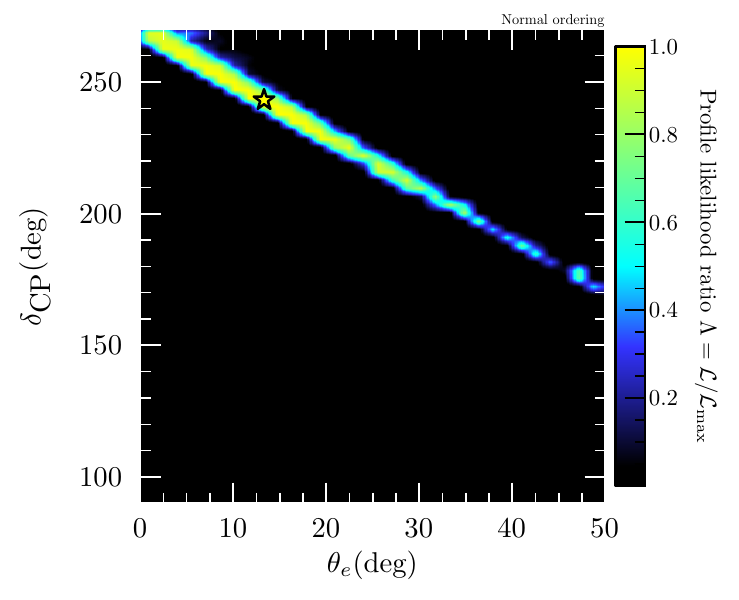} 
		\caption{\label{fig:LFV_C} 
			Regions of the free parameters where the model can fit accurately the experimental observation regarding the lepton sector observables. Dark regions are not compatible with observations at all, while the best fit point (BFP) is depicted with a star. }
\end{figure}
	
\begin{figure}[ht]\centering
		\includegraphics[scale=0.7]{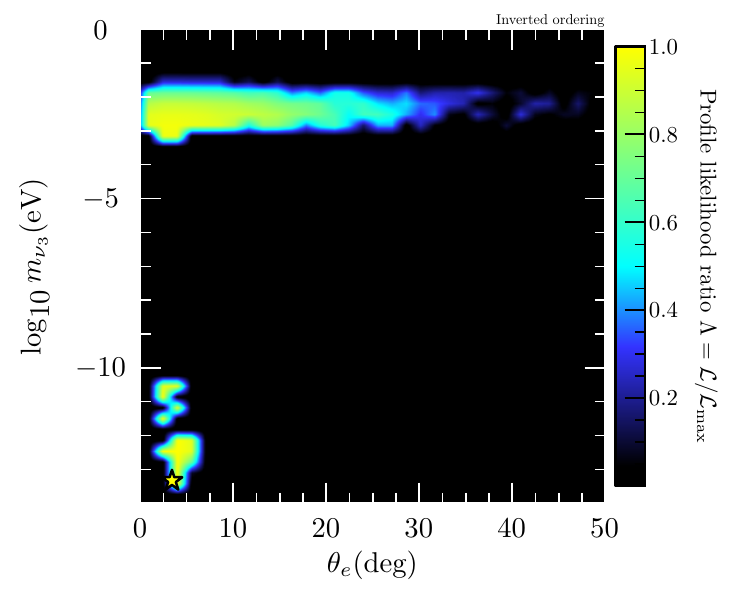}
		\hspace{1mm}\includegraphics[scale=0.7]{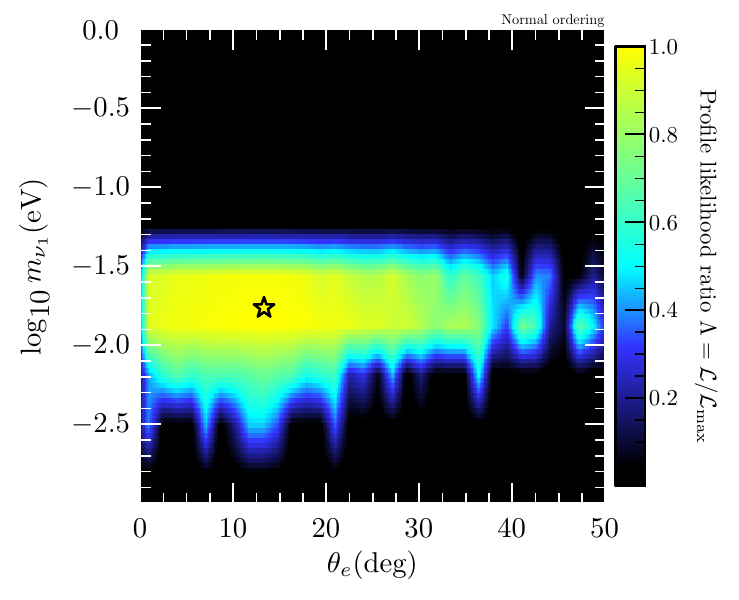} \\
		\includegraphics[scale=0.7]{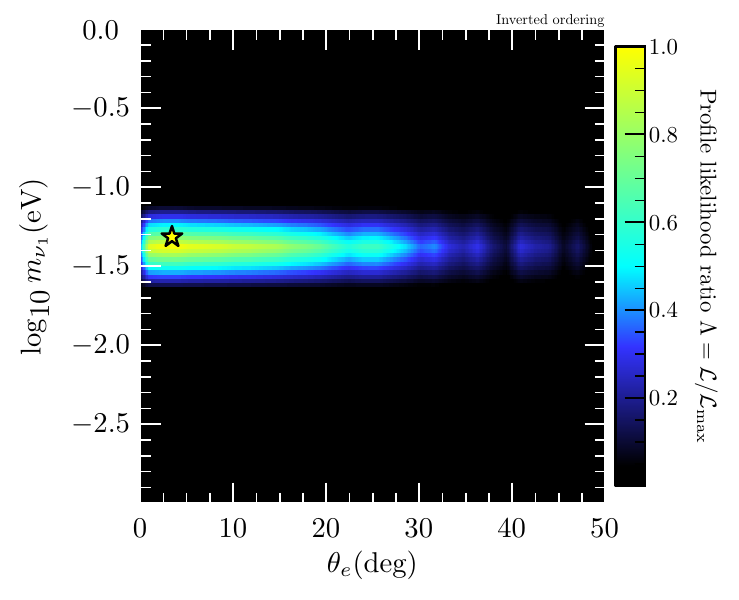}
		\hspace{1mm}\includegraphics[scale=0.7]{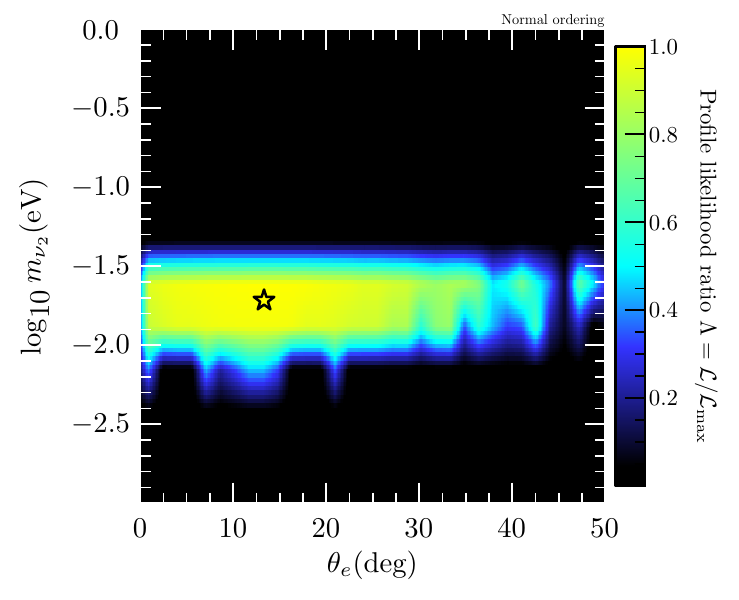} \\
		\includegraphics[scale=0.7]{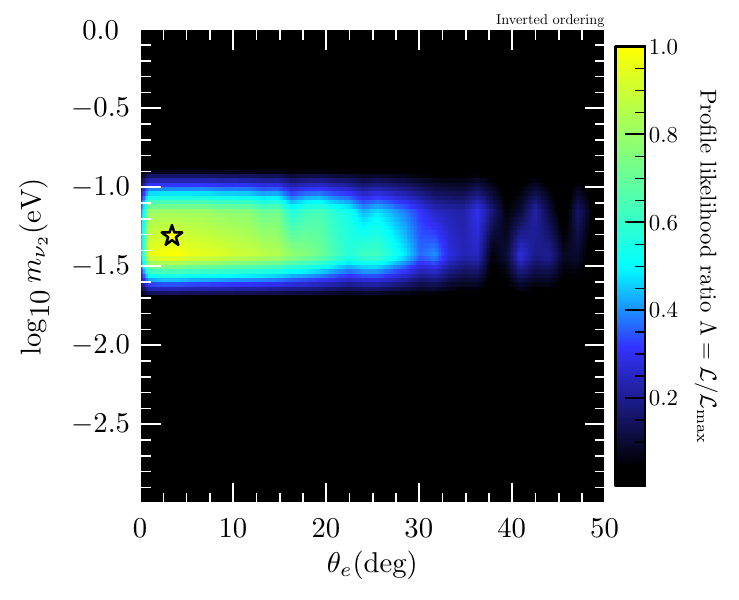}
		\hspace{1mm}\includegraphics[scale=0.7]{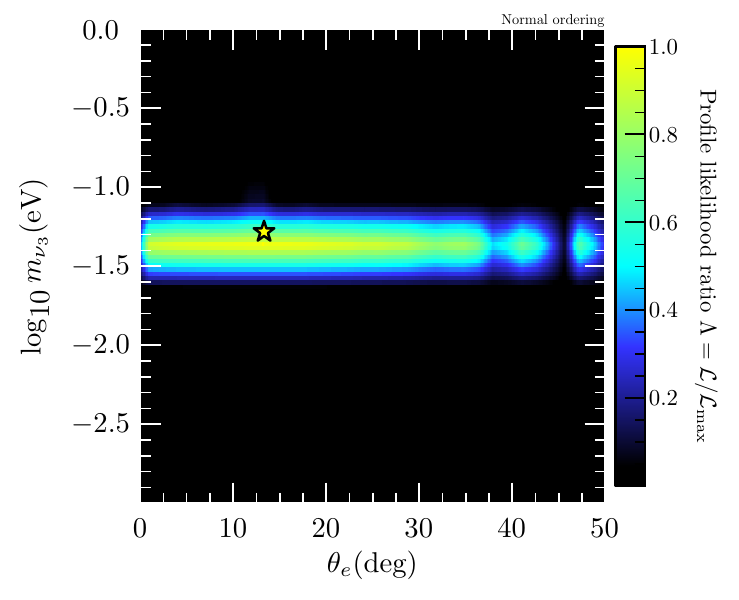} 
		\caption{\label{fig:LFV_D} 
			Predicted neutrino masses as functions of $\theta_e$, arranged from lightest (top) to heaviest (bottom). Dark regions are not compatible with observations at all, while the best fit point (BFP) is depicted with a star. }
	\end{figure}

\newpage

\section{Scalar Potential} \label{potencial}

As the potential with three doublets of Higgs is already a well-studied topic in literature, we will only focus on the potential for the $\phi$ fields.

Keep in mind the general structure of the $V(\phi)$ potential that is depicted in the equation \eqref{potencialescalar}. Then, applying the assignment shown in the table \ref{TAB2} and the multiplication rules of the irreducible representations of the group $\mathbf{S}_{3}$, we get the following potential,
\begin{equation}
\begin{aligned}
V(\phi) & = \mu_{1}^2 (\phi_1^{\dagger} \phi_1 + \phi_2^{\dagger} \phi_2 ) + \mu_{2}^2 (\phi_3^{\dagger} \phi_3 ) \\
& +\lambda_1 (\phi_1^{\dagger} \phi_1 + \phi_2^{\dagger} \phi_2 )^2 +\lambda_2 (\phi_1^{\dagger} \phi_2 - \phi_2^{\dagger} \phi_1 )^2+\lambda_3 \left[ (\phi_1^{\dagger} \phi_2 + \phi_2^{\dagger} \phi_1 )^2 + (\phi_1^{\dagger} \phi_1 - \phi_2^{\dagger} \phi_2 )^2 \right]\\ & + \lambda_4 \left[ (\phi_3^{\dagger} \phi_1)(\phi_1^{\dagger} \phi_2 + \phi_2^{\dagger} \phi_1 )+ (\phi_3^{\dagger} \phi_2)(\phi_1^{\dagger} \phi_1 - \phi_2^{\dagger} \phi_2) + (\phi_1^{\dagger} \phi_2 + \phi_2^{\dagger} \phi_1)(\phi_1^{\dagger} \phi_3) + (\phi_1^{\dagger} \phi_1 - \phi_2^{\dagger} \phi_2) (\phi_2^{\dagger} \phi_3) \right] \\
& + \lambda_5 (\phi_3^{\dagger} \phi_3) (\phi_1^{\dagger} \phi_1 + \phi_2^{\dagger} \phi_2 ) + \lambda_6 \left[ (\phi_1^{\dagger} \phi_3)(\phi_3^{\dagger} \phi_1) + (\phi_2^{\dagger} \phi_3)(\phi_3^{\dagger} \phi_2 )\right] \\
& + \lambda_7 \left[(\phi_1^{\dagger} \phi_3)^2 + (\phi_2^{\dagger} \phi_3)^2 + (\phi_3^{\dagger} \phi_1)^2 + (\phi_3^{\dagger} \phi_2)^2 \right] + \lambda_8 (\phi_3^{\dagger} \phi_3)^2,
\end{aligned}
\end{equation}
where the parameters $\mu_i$, $i, =1,2$, have dimensions of mass squared and the eight real couplings $\lambda_1,...,\lambda_8$ are dimensionless free parameters.

To manage the increase of free parameters in the Yukawa sector, it is necessary to introduce a symmetry $\mathbf{Z}_{2}$. Refer to table \ref{TAB2}. This new discrete symmetry prohibits some couplings in the scalar potential, particularly we get $\lambda_4=0$. So the more general potential, invariant under $\mathbf{S}_{3} \otimes \mathbf{Z}_{2}$ symmetry is as follows,
\begin{equation}
\begin{aligned}
V(\phi) & = \mu_{1}^2 (\phi_1^{\dagger} \phi_1 + \phi_2^{\dagger} \phi_2 ) + \mu_{2}^2 (\phi_3^{\dagger} \phi_3 ) \\
&+\lambda_1 (\phi_1^{\dagger} \phi_1 + \phi_2^{\dagger} \phi_2 )^2 +\lambda_2 (\phi_1^{\dagger} \phi_2 - \phi_2^{\dagger} \phi_1 )^2  +\lambda_3 \left[ (\phi_1^{\dagger} \phi_2 + \phi_2^{\dagger} \phi_1 )^2 + (\phi_1^{\dagger} \phi_1 - \phi_2^{\dagger} \phi_2 )^2 \right]\\
& + \lambda_5 (\phi_3^{\dagger} \phi_3) (\phi_1^{\dagger} \phi_1 + \phi_2^{\dagger} \phi_2 ) + \lambda_6 \left[ (\phi_1^{\dagger} \phi_3)(\phi_3^{\dagger} \phi_1) + (\phi_2^{\dagger} \phi_3)(\phi_3^{\dagger} \phi_2 )\right] \\
& +\lambda_7 \left[(\phi_1^{\dagger} \phi_3)^2 + (\phi_2^{\dagger} \phi_3)^2 + (\phi_3^{\dagger} \phi_1)^2 + (\phi_3^{\dagger} \phi_2)^2 \right] + \lambda_8 (\phi_3^{\dagger} \phi_3)^2 . \\
	\end{aligned}
\end{equation} 
If we minimize this potential, applying the parameterization for singlet fields shown in equation \eqref{componentes} and the vacuum expectation values as shown in equation \eqref{vevs}, we get the alignment $w_1 = w_2$.
Nevertheless, when examining the mass spectrum generated by $V(\phi)$, two scalar bosons without mass appear.
To avoid this problem, we introduce the following explicit breaking term of $\mathbf{S}_{3}$ symmetry, 
\begin{equation}
 \lambda_9 (\phi^{\dagger}_3 \phi_1)(\phi^{\dagger}_3 \phi_2) .
\end{equation}
This term is invariant under the gauge symmetry of the Standard Model and corresponds to a part of the anti-symmetric singlet, obtained when operating the product $(\mathbf{1}_S \otimes \mathbf{2})^2$ of irreducible representations of group $\mathbf{S}_{3}$, according to the multiplication rules of the group $\mathbf{S}_{3}$ and the assignment shown in the table \ref{TAB2},
\begin{eqnarray}
\left[ \phi^{\dagger}_3 \otimes 
\begin{pmatrix}
\phi_1 \\
\phi_2
\end{pmatrix} \right]^2 &=& 
\begin{pmatrix}
\phi^{\dagger}_3 \phi_1 \\
\phi^{\dagger}_3 \phi_2
\end{pmatrix}  \otimes 
\begin{pmatrix}
\phi^{\dagger}_3 \phi_1 \\
\phi^{\dagger}_3 \phi_2
\end{pmatrix}\nn\\
&=& [(\phi_3^{\dagger} \phi_1)^2 + (\phi_3^{\dagger} \phi_2)^2]_{S} \oplus [ \underline{ (\phi_3^{\dagger} \phi_1) (\phi_3^{\dagger} \phi_2) } - (\phi_3^{\dagger} \phi_2)  (\phi_3^{\dagger} \phi_1)]_{A} \oplus 
\begin{pmatrix}
\phi^{\dagger}_3 \phi_1(\phi^{\dagger}_3 \phi_2) + \phi^{\dagger}_3 \phi_2(\phi^{\dagger}_3 \phi_1) \\
\phi^{\dagger}_3 \phi_1(\phi^{\dagger}_3 \phi_1) - \phi^{\dagger}_3 \phi_2(\phi^{\dagger}_3 \phi_2) 
\end{pmatrix}_2.
\end{eqnarray} 
This term is also invariant under the $\mathbf{Z}_{2}$ symmetry,
\begin{equation}
    \mathbf{Z}_{2} [(\phi^{\dagger}_3 \phi_1)(\phi^{\dagger}_3  \phi_2)]  \rightarrow \exp{(-2i\pi)} \exp{(i\pi)} \exp{(-2i\pi)} \exp{(i\pi)} = \exp{(-2i\pi)} = 1.
\end{equation}
Therefore, we will use for our model, the following expression for potential, 
\begin{equation}
\begin{aligned}
V'(\phi) & =  \mu_{1}^2 (\phi_1^{\dagger} \phi_1 + \phi_2^{\dagger} \phi_2 ) + \mu_{2}^2 (\phi_3^{\dagger} \phi_3 ) \\
&+\lambda_1 (\phi_1^{\dagger} \phi_1 + \phi_2^{\dagger} \phi_2 )^2 +\lambda_2 (\phi_1^{\dagger} \phi_2 - \phi_2^{\dagger} \phi_1 )^2  +\lambda_3 \left[ (\phi_1^{\dagger} \phi_2 + \phi_2^{\dagger} \phi_1 )^2 + (\phi_1^{\dagger} \phi_1 - \phi_2^{\dagger} \phi_2 )^2 \right]\\
& + \lambda_5 (\phi_3^{\dagger} \phi_3) (\phi_1^{\dagger} \phi_1 + \phi_2^{\dagger} \phi_2 ) + \lambda_6 \left[ (\phi_1^{\dagger} \phi_3)(\phi_3^{\dagger} \phi_1) + (\phi_2^{\dagger} \phi_3)(\phi_3^{\dagger} \phi_2 )\right] \\
& +\lambda_7 \left[(\phi_1^{\dagger} \phi_3)^2 + (\phi_2^{\dagger} \phi_3)^2 + (\phi_3^{\dagger} \phi_1)^2 + (\phi_3^{\dagger} \phi_2)^2 \right] + \lambda_8 (\phi_3^{\dagger} \phi_3)^2 \\
& + \lambda_9 (\phi_3^{\dagger} \phi_1) (\phi_3^{\dagger} \phi_2).
\end{aligned}
\end{equation}
Rewriting the potential in terms of scalar fields \eqref{componentes} and minimizing it, we obtain the following three equations, 
\begin{equation}
\begin{aligned}
& \frac{1}{2} w_1 \left(2 \mu_1^2 +2(\lambda_1 +\lambda_3) w_2^2 +(\lambda_5 +\lambda_6 +2 \lambda_7) w_3^2\right) + (\lambda_1 +\lambda_3) w_1^3 +\frac{1}{4} \lambda_9 w_2 w_3^2 = 0 \\
& (\lambda_1+\lambda_3) w_2^3 +(\lambda_1 +\lambda_3) w_1^2 w_2 +\frac{1}{2} \left(\lambda_5 +\lambda_6 +2\lambda_7 \right) w_3^2 w_2 +\frac{1}{4} \lambda_9 w_1 w_3^2 +\mu_1^2 w_2 =0 \\
& \frac{1}{2} w_3 \left(2 (\mu_2^2 +\lambda_8 w_3^2) + (\lambda_5 +\lambda_6 +2 \lambda_7) (w_1^2+w_2^2) +\lambda_9 w_1 w_2\right) = 0 .
\end{aligned}
\end{equation}
By the self-consistency of the equations for $\mu_1^2$ and the requirement that $\lambda_9 \neq 0$, $w_i \neq 0$, $i=1,2,3$, we are able to achieve the following,
\begin{equation}
\lambda_9 \left(w_2^2-w_1^2\right) w_3^2 = 0 \hspace{5mm} \Rightarrow \hspace{5mm} w_1 = \pm w_2. 
\end{equation}
We will consider only in detail the case with $w_1 = w_2$. The negative solution leads exactly to the same results for the masses and couplings as the positive solution.
The mass parameters in this case are as follows,
\begin{equation}
\begin{aligned}
& \mu^2_1 = \frac{1}{4} \left( -8 \left(\lambda_1+\lambda_3\right) w_2^2-\left(2 \lambda_5+2 \lambda_6+4 \lambda_7+\lambda_9\right) w_3^2 \right) \\
& \mu^2_2 = -\frac{1}{2} \left(2 \lambda_5+2 \lambda_6+4 \lambda_7+\lambda_9\right) w_2^2-\lambda_8 w_3^2 .
\end{aligned}
\end{equation}

\subsection{Scalar masses}
We analyze the mass spectrum of the real components of the $\phi_i$ fields, for $V'(\phi)$ and we find the mass-squared matrices in the scalar ($m_s$) and pseudoscalar ($m_p$) sectors,
\begin{equation*}
\begin{aligned}
 &   m_s = \left(
\begin{array}{ccc}
 2 (\lambda_1 +\lambda_3) w_2^2 -\frac{1}{4} \lambda_9 w_3^2 & 2 (\lambda_1 +\lambda_3) w_2^2 +\frac{1}{4} \lambda_9 w_3^2 & \frac{1}{2} (2 \lambda_5 +2\lambda_6 +4\lambda_7+\lambda_9) w_2 w_3 \\
 2 (\lambda_1 +\lambda_3) w_2^2 +\frac{1}{4} \lambda_9 w_3^2 & 2 (\lambda_1 +\lambda_3) w_2^2-\frac{1}{4} \lambda_9 w_3^2 & \frac{1}{2} (2 \lambda_5 +2\lambda_6 +4\lambda_7 +\lambda_9) w_2 w_3 \\
 \frac{1}{2} (2\lambda_5 +2\lambda_6 +4\lambda_7 +\lambda_9) w_2 w_3 & \frac{1}{2} (2\lambda_5 +2\lambda_6 +4\lambda_7 +\lambda_9) w_2 w_3 & 2 \lambda_8 w_3^2 \\
\end{array}
\right)  \\
&  m_p = \left(
\begin{array}{ccc}
 -2 (\lambda_2 +\lambda_3) w_2^2 -\frac{1}{4} (8\lambda_7 +\lambda_9) w_3^2 & 2 (\lambda_2 +\lambda_3) w_2^2 -\frac{1}{4} \lambda_9 w_3^2 & \frac{1}{2} (4\lambda_7 +\lambda_9) w_2 w_3 \\
 2 (\lambda_2+\lambda_3) w_2^2-\frac{1}{4} \lambda_9 w_3^2 & -2(\lambda_2 +\lambda_3) w_2^2 -\frac{1}{4} (8\lambda_7 +\lambda_9) w_3^2 & \frac{1}{2} (4 \lambda_7 +\lambda_9) w_2 w_3 \\
 \frac{1}{2} (4\lambda_7 +\lambda_9) w_2 w_3 & \frac{1}{2} (4\lambda_7 +\lambda_9) w_2 w_3 & -w_2^2 (4\lambda_7 +\lambda_9)  \\
\end{array} 
\right).
\end{aligned}
\end{equation*}
At this point it is convenient to introduce the following geometric parameterization in spherical coordinates,
\begin{equation} 
\begin{aligned}
w_1 & = w \sin{\beta} \cos{\gamma} \\
w_2 & = w \sin{\beta} \sin{\gamma} \\
w_3 & = w \cos{\beta} .
\end{aligned}  
\end{equation}
The inverse transformations are described below,
\begin{equation} 
 \begin{aligned} 
& w = \sqrt{w_1^2 + w_2^2 + w_3^2} \\ 
& \tan{\gamma} = \frac{w_2}{w_1}  \\ 
& \tan{\beta} = \sqrt{\dfrac{w_1^2 + w_2^2}{w_3^2}} \hspace{5mm} \text{with} \hspace{5mm} w_1 =w_2 .
\end{aligned}
\end{equation}
Defining matrices that allow us to diagonalize mass matrices is made much easier with these coordinates. For example, $m_s$ and $m_p$ are simultaneously block diagonalizable by the following matrix, 
\begin{equation}
r_1 = \left(
\begin{array}{ccc}
 \cos{\gamma} & \sin{\gamma} & 0 \\
-\sin{\gamma} & \cos{\gamma} & 0 \\
 0 & 0 & 1 \\
\end{array}  
\right) = \left(
\begin{array}{ccc}
 \frac{1}{\sqrt{2}} & \frac{1}{\sqrt{2}} & 0 \\
 -\frac{1}{\sqrt{2}} & \frac{1}{\sqrt{2}} & 0 \\
 0 & 0 & 1 \\
\end{array}
\right) .
\end{equation}
In the case of the pseudo-scalar matrix, we get the following,
\begin{equation}
  {r_1}^T m_p r_1 = \left(
\begin{array}{ccc}
 -2 \left(2 (\lambda_2 +\lambda_3) w_2^2 +\lambda_7 w_3^2\right) & 0 & 0 \\
 0 & -\frac{1}{2} \left(4\lambda_7 +\lambda_9\right) w_3^2 & \frac{1}{\sqrt{2}} ( 4\lambda_7 +\lambda_9) w_2 w_3 \\
 0 & \frac{1}{\sqrt{2}} (4\lambda_7 +\lambda_9) w_2 w_3 & -(4\lambda_7 +\lambda_9) w_2^2 \\
\end{array}
\right) .
\end{equation}
The remaining block $2 \times 2$ is diagonalized by the matrix $r_2$,
\begin{equation}
   r_2 =  \left(
\begin{array}{ccc}
 1 & 0 & 0 \\
 0 & \cos{\beta} & \sin{\beta} \\
 0 & -\sin{\beta} & \cos{\beta} \\
\end{array}
\right) .
\end{equation}
So we have the following states of mass,
\begin{equation} \label{masap}
M_P \equiv {r_2}^T({r_1}^T m_p r_1) r_2 = \left(
\begin{array}{ccc}
 -2 \left( 2(\lambda_2 +\lambda_3) w_2^2 +\lambda_7 w_3^2\right) & 0 & 0 \\
 0 & -\frac{1}{2} (4\lambda_7 +\lambda_9) (2 w_2^2+w_3^2) & 0 \\
 0 & 0 & 0 \\
\end{array}
\right) . 
\end{equation}
Finally, the rotation to obtain the mass matrix directly from the interaction basis is given as
\begin{equation} \label{rot1}
    R_p \equiv r_1 r_2 = \left(
\begin{array}{ccc}
 \cos{\gamma} & \cos{\beta} \sin{\gamma} & \sin{\beta} \sin{\gamma} \\
 -\sin{\gamma} & \cos{\beta} \cos{\gamma} & \sin{\beta} \cos{\gamma} \\
 0 & -\sin{\beta} & \cos{\beta} \\
\end{array}
\right) 
\end{equation}
We have the following results for the scalar matrix, 
\begin{equation}
{r_1}^T m_s r_1 = \left(
\begin{array}{ccc}
 -\frac{1}{2} \lambda_9 w_3^2 & 0 & 0 \\
 0 & 4(\lambda_1 +\lambda_3) w_2^2 & \frac{1}{\sqrt{2}} \left(2 \lambda_5 +2\lambda_6 +4\lambda_7 +\lambda_9\right) w_2 w_3 \\
 0 & \frac{1}{\sqrt{2}} \left(2 \lambda_5+2 \lambda_6+4 \lambda_7+\lambda_9\right) w_2 w_3 & 2 \lambda_8 w_3^2 \\
\end{array}
\right)
\end{equation}
Rewriting the diagonal matrix in blocks, is a convenient way to diagonalize the remaining block of $2 \times 2$.
\begin{equation}
{M}_s = \begin{pmatrix}
m_{s1} & 0  & 0 \\
0      & a  & b \\
0      & b  & c 
\end{pmatrix},
\end{equation}
where,
\begin{equation}
\begin{aligned}
& m_{s1} = -\frac{1}{2} \lambda_9 w_3^2 \\
& a =  4(\lambda_1 +\lambda_3) w_2^2 \\
& b = \frac{1}{\sqrt{2}} \left(2 \lambda_5+2 \lambda_6+4 \lambda_7+\lambda_9\right) w_2 w_3  \\
& c = 2 \lambda_8 w_3^2 .
\end{aligned}
\end{equation}
Diagonalization of this matrix is achieved through the transformation $r_3$,
\begin{equation}
r_3 = 
\begin{pmatrix}
 0 & 0 & 1 \\
 -\sin{\alpha} & \cos{\alpha} & 0 \\
 \cos{\alpha} & \sin{\alpha} & 0
\end{pmatrix}, \hspace{5mm} \text{with} \hspace{5mm} \tan{(2\alpha)} =\dfrac{2b}{a-c}.
\end{equation}
The mass eigenstates are depicted below,
\begin{equation} \label{masas}
M_S \equiv {r_3}^T ({r_1}^T m_s r_1) r_3 \begin{pmatrix}
 \frac{1}{2} \left(a+c-\sqrt{(a-c)^2+4 b^2}\right) & 0 & 0 \\
 0 & \frac{1}{2} \left(a+c + \sqrt{(a-c)^2+4 b^2} \right)  & 0 \\
 0 & 0 & m_{s1}  \\
\end{pmatrix} .
\end{equation}
For the diagonalization of the $m_s$ mass matrix, we will have the following rotation matrix:
\begin{equation} \label{rot2}
   R_s \equiv r_1 r_3 = \left(
\begin{array}{ccc}
 -\sin{\alpha} \sin{\gamma} & \cos{\alpha} \sin{\gamma} & \cos{\gamma} \\
- \sin{\alpha} \cos{\gamma} & \cos{\alpha} \cos{\gamma} & -\sin{\gamma} \\
 \cos{\alpha} & \sin{\alpha} & 0 \\
\end{array}
\right) .
\end{equation}
Define the vectors $\vec{S}$ and $\vec{P}$, as the vectors containing fields at the mass base. Similarly, consider the vectors $\vec{x}$, $\vec{y}$ as the vectors containing fields at the base of the interaction,
\begin{equation}
 \vec{S} =  
\begin{pmatrix}
s_1 \\
s_2 \\
s_3
\end{pmatrix} \hspace{10mm}
 \vec{P} =  
\begin{pmatrix}
p_1 \\
p_2 \\
p_3
\end{pmatrix}   \hspace{10mm}
 \vec{x} =  
\begin{pmatrix}
x_1 \\
x_2 \\
x_3
\end{pmatrix} \hspace{10mm}
 \vec{y} =  
\begin{pmatrix}
y_1 \\
y_2 \\
y_3
\end{pmatrix} .
\end{equation}
The mass states in terms of the fields at the base of the interaction are related through the matrices \eqref{rot1},  \eqref{rot2}, by the equations $\vec{S} = R_s \vec{x}$, $\vec{P} = R_p \vec{y}$. Therefore, 
\begin{equation}
\begin{aligned}
& s_1 = - x_1 \sin{\alpha} \sin{\gamma} +x_2 \cos{\alpha} \sin{\gamma} +x_3 \cos{\gamma} \\
& s_2 = x_2 \cos{\alpha} \cos{\gamma} -x_1 \sin{\alpha} \cos{\gamma} -x_3 \sin{\gamma} \\
& s_3 = x_2 \sin{\alpha} +x_1 \cos{\alpha} \\
& p_1 =  y_3 \sin{\beta} \sin{\gamma} +y_2 \cos{\beta} \sin{\gamma} +y_1 \cos{\gamma} \\
& p_2 =  y_2 \cos{\beta} \cos{\gamma} +y_3 \sin{\beta} \cos{\gamma} -y_1 \sin{\gamma}  \\
& p_3 = y_3 \cos{\beta} -y_2 \sin{\beta} .
\end{aligned}
\end{equation}
Finally, we can observe in the matrices \eqref{masap} and \eqref{masas} the appearance of five massive scalar particles and one Goldstone boson. The scalar particle without mass is associated to one of the imaginary components of the $\phi_i$ fields and results from the breaking of the gauge symmetry $U(1)_{B-L}$.

\bibliographystyle{bib_style_T1}
\bibliography{references.bib}

\begin{thebibliography}{100}
\providecommand{\url}[1]{\texttt{#1}}
\providecommand{\urlprefix}{URL }
\providecommand{\eprint}[2][]{\url{#2}}

\bibitem{Minkowski:1977sc}
P.~Minkowski, \emph{{mu $\to$ e gamma at a Rate of One Out of 1-Billion Muon
  Decays?}},
  \MYhref[journalLinks]{http://dx.doi.org/10.1016/0370-2693(77)90435-X}{Phys.
  Lett.
  }\MYhref[journalLinks]{http://dx.doi.org/10.1016/0370-2693(77)90435-X}{\textbf{B67}
  (1977) 421}.

\bibitem{GellMann:1980vs}
M.~Gell-Mann, P.~Ramond and R.~Slansky, \emph{{Complex Spinors and Unified
  Theories}}, Conf.Proc. \textbf{C790927} (1979) 315--321,
  \MYhref[eprintLinks]{http://arxiv.org/abs/1306.4669}{{\ttfamily
  arXiv:1306.4669 [hep-th]}}.

\bibitem{Yanagida:1979as}
T.~Yanagida, \emph{{Horizontal gauge symmetry and masses of neutrinos}} In
  Proceedings of the Workshop on the Baryon Number of the Universe and Unified
  Theories, Tsukuba, Japan, 13-14 Feb 1979.

\bibitem{Mohapatra:1980yp}
R.~N. Mohapatra and G.~Senjanovic, \emph{{Neutrino Masses and Mixings in Gauge
  Models with Spontaneous Parity Violation}},
  \MYhref[journalLinks]{http://dx.doi.org/10.1103/PhysRevD.23.165}{Phys. Rev.
  }\MYhref[journalLinks]{http://dx.doi.org/10.1103/PhysRevD.23.165}{\textbf{D23}
  (1981) 165}.

\bibitem{Mohapatra:1979ia}
R.~N. Mohapatra and G.~Senjanovic, \emph{{Neutrino Mass and Spontaneous Parity
  Violation}},
  \MYhref[journalLinks]{http://dx.doi.org/10.1103/PhysRevLett.44.912}{Phys.Rev.Lett.
  }\MYhref[journalLinks]{http://dx.doi.org/10.1103/PhysRevLett.44.912}{\textbf{44}
  (1980) 912}.

\bibitem{Schechter:1980gr}
J.~Schechter and J.~W.~F. Valle, \emph{{Neutrino Masses in SU(2) x U(1)
  Theories}},
  \MYhref[journalLinks]{http://dx.doi.org/10.1103/PhysRevD.22.2227}{Phys. Rev.
  }\MYhref[journalLinks]{http://dx.doi.org/10.1103/PhysRevD.22.2227}{\textbf{D22}
  (1980) 2227}.

\bibitem{Schechter:1981cv}
J.~Schechter and J.~W.~F. Valle, \emph{{Neutrino Decay and Spontaneous
  Violation of Lepton Number}},
  \MYhref[journalLinks]{http://dx.doi.org/10.1103/PhysRevD.25.774}{Phys. Rev.
  }\MYhref[journalLinks]{http://dx.doi.org/10.1103/PhysRevD.25.774}{\textbf{D25}
  (1982) 774}.

\bibitem{Feruglio:2015jfa}
F.~Feruglio, \emph{{Pieces of the Flavour Puzzle}},
  \MYhref[journalLinks]{http://dx.doi.org/10.1140/epjc/s10052-015-3576-5}{Eur.
  Phys. J.
  }\MYhref[journalLinks]{http://dx.doi.org/10.1140/epjc/s10052-015-3576-5}{\textbf{C75}
  (2015) 8 373},
  \MYhref[eprintLinks]{http://arxiv.org/abs/1503.04071}{{\ttfamily
  arXiv:1503.04071 [hep-ph]}}.

\bibitem{Abbas:2023ivi}
G.~Abbas, R.~Adhikari, E.~J. Chun and N.~Singh, \emph{{The problem of flavour}}
  (2023)  \MYhref[eprintLinks]{http://arxiv.org/abs/2308.14811}{{\ttfamily
  arXiv:2308.14811 [hep-ph]}}.

\bibitem{Nilles:2023shk}
H.~P. Nilles and S.~Ramos-Sanchez, \emph{{The Flavor Puzzle: Textures and
  Symmetries}} (2023)
  \MYhref[eprintLinks]{http://arxiv.org/abs/2308.14810}{{\ttfamily
  arXiv:2308.14810 [hep-ph]}}.

\bibitem{Wyler:1982dd}
D.~Wyler and L.~Wolfenstein, \emph{{Massless Neutrinos in Left-Right Symmetric
  Models}},
  \MYhref[journalLinks]{http://dx.doi.org/10.1016/0550-3213(83)90482-0}{Nucl.
  Phys. B
  }\MYhref[journalLinks]{http://dx.doi.org/10.1016/0550-3213(83)90482-0}{\textbf{218}
  (1983) 205--214}.

\bibitem{Akhmedov:1995vm}
E.~K. Akhmedov, M.~Lindner, E.~Schnapka and J.~W.~F. Valle, \emph{{Dynamical
  left-right symmetry breaking}},
  \MYhref[journalLinks]{http://dx.doi.org/10.1103/PhysRevD.53.2752}{Phys. Rev.
  D
  }\MYhref[journalLinks]{http://dx.doi.org/10.1103/PhysRevD.53.2752}{\textbf{53}
  (1996) 2752--2780},
  \MYhref[eprintLinks]{http://arxiv.org/abs/hep-ph/9509255}{{\ttfamily
  arXiv:hep-ph/9509255}}.

\bibitem{Barr:2005ss}
S.~M. Barr and I.~Dorsner, \emph{{A Prediction from the type III see-saw
  mechanism}},
  \MYhref[journalLinks]{http://dx.doi.org/10.1016/j.physletb.2005.10.080}{Phys.
  Lett. B
  }\MYhref[journalLinks]{http://dx.doi.org/10.1016/j.physletb.2005.10.080}{\textbf{632}
  (2006) 527--531},
  \MYhref[eprintLinks]{http://arxiv.org/abs/hep-ph/0507067}{{\ttfamily
  arXiv:hep-ph/0507067}}.

\bibitem{Romao:2007ny}
J.~C. Romao, \emph{{Supersymmetric Models for Neutrino Mass}}, in \emph{{6th
  International Workshop on New Worlds in Astroparticle Physics}} (2007)
  \MYhref[eprintLinks]{http://arxiv.org/abs/0710.5730}{{\ttfamily
  arXiv:0710.5730 [hep-ph]}}.

\bibitem{Xing:2009in}
Z.-z. Xing, \emph{{Naturalness and Testability of TeV Seesaw Mechanisms}},
  \MYhref[journalLinks]{http://dx.doi.org/10.1143/PTPS.180.112}{Prog. Theor.
  Phys. Suppl.
  }\MYhref[journalLinks]{http://dx.doi.org/10.1143/PTPS.180.112}{\textbf{180}
  (2009) 112--127},
  \MYhref[eprintLinks]{http://arxiv.org/abs/0905.3903}{{\ttfamily
  arXiv:0905.3903 [hep-ph]}}.

\bibitem{Mohapatra:1986aw}
R.~N. Mohapatra, \emph{{Mechanism for Understanding Small Neutrino Mass in
  Superstring Theories}},
  \MYhref[journalLinks]{http://dx.doi.org/10.1103/PhysRevLett.56.561}{Phys.
  Rev. Lett.
  }\MYhref[journalLinks]{http://dx.doi.org/10.1103/PhysRevLett.56.561}{\textbf{56}
  (1986) 561--563}.

\bibitem{Mohapatra:1986bd}
R.~N. Mohapatra and J.~W.~F. Valle, \emph{{Neutrino Mass and Baryon Number
  Nonconservation in Superstring Models}},
  \MYhref[journalLinks]{http://dx.doi.org/10.1103/PhysRevD.34.1642}{Phys. Rev.
  D
  }\MYhref[journalLinks]{http://dx.doi.org/10.1103/PhysRevD.34.1642}{\textbf{34}
  (1986) 1642}.

\bibitem{Bernabeu:1987gr}
J.~Bernabeu et~al., \emph{{Lepton Flavor Nonconservation at High-Energies in a
  Superstring Inspired Standard Model}},
  \MYhref[journalLinks]{http://dx.doi.org/10.1016/0370-2693(87)91100-2}{Phys.
  Lett. B
  }\MYhref[journalLinks]{http://dx.doi.org/10.1016/0370-2693(87)91100-2}{\textbf{187}
  (1987) 303--308}.

\bibitem{Xing:2009hx}
Z.-z. Xing and S.~Zhou, \emph{{Multiple seesaw mechanisms of neutrino masses at
  the TeV scale}},
  \MYhref[journalLinks]{http://dx.doi.org/10.1016/j.physletb.2009.07.051}{Phys.
  Lett. B
  }\MYhref[journalLinks]{http://dx.doi.org/10.1016/j.physletb.2009.07.051}{\textbf{679}
  (2009) 249--254},
  \MYhref[eprintLinks]{http://arxiv.org/abs/0906.1757}{{\ttfamily
  arXiv:0906.1757 [hep-ph]}}.

\bibitem{Malinsky:2009df}
M.~Malinsky, T.~Ohlsson, Z.-z. Xing and H.~Zhang, \emph{{Non-unitary neutrino
  mixing and CP violation in the minimal inverse seesaw model}},
  \MYhref[journalLinks]{http://dx.doi.org/10.1016/j.physletb.2009.07.038}{Phys.
  Lett. B
  }\MYhref[journalLinks]{http://dx.doi.org/10.1016/j.physletb.2009.07.038}{\textbf{679}
  (2009) 242--248},
  \MYhref[eprintLinks]{http://arxiv.org/abs/0905.2889}{{\ttfamily
  arXiv:0905.2889 [hep-ph]}}.

\bibitem{Abada:2014vea}
A.~Abada and M.~Lucente, \emph{{Looking for the minimal inverse seesaw
  realisation}},
  \MYhref[journalLinks]{http://dx.doi.org/10.1016/j.nuclphysb.2014.06.003}{Nucl.
  Phys. B
  }\MYhref[journalLinks]{http://dx.doi.org/10.1016/j.nuclphysb.2014.06.003}{\textbf{885}
  (2014) 651--678},
  \MYhref[eprintLinks]{http://arxiv.org/abs/1401.1507}{{\ttfamily
  arXiv:1401.1507 [hep-ph]}}.

\bibitem{Abada:2014zra}
A.~Abada, G.~Arcadi and M.~Lucente, \emph{{Dark Matter in the minimal Inverse
  Seesaw mechanism}},
  \MYhref[journalLinks]{http://dx.doi.org/10.1088/1475-7516/2014/10/001}{JCAP
  }\MYhref[journalLinks]{http://dx.doi.org/10.1088/1475-7516/2014/10/001}{\textbf{10}
  (2014) 001}, \MYhref[eprintLinks]{http://arxiv.org/abs/1406.6556}{{\ttfamily
  arXiv:1406.6556 [hep-ph]}}.

\bibitem{Boucenna:2014zba}
S.~M. Boucenna, S.~Morisi and J.~W.~F. Valle, \emph{{The low-scale approach to
  neutrino masses}},
  \MYhref[journalLinks]{http://dx.doi.org/10.1155/2014/831598}{Adv. High Energy
  Phys.
  }\MYhref[journalLinks]{http://dx.doi.org/10.1155/2014/831598}{\textbf{2014}
  (2014) 831598},
  \MYhref[eprintLinks]{http://arxiv.org/abs/1404.3751}{{\ttfamily
  arXiv:1404.3751 [hep-ph]}}.

\bibitem{Abada:2014nwa}
A.~Abada, V.~De~Romeri and A.~M. Teixeira, \emph{{Effect of steriles states on
  lepton magnetic moments and neutrinoless double beta decay}},
  \MYhref[journalLinks]{http://dx.doi.org/10.1007/JHEP09(2014)074}{JHEP
  }\MYhref[journalLinks]{http://dx.doi.org/10.1007/JHEP09(2014)074}{\textbf{09}
  (2014) 074}, \MYhref[eprintLinks]{http://arxiv.org/abs/1406.6978}{{\ttfamily
  arXiv:1406.6978 [hep-ph]}}.

\bibitem{Abada:2016awd}
A.~Abada and T.~Toma, \emph{{Electron electric dipole moment in Inverse Seesaw
  models}},
  \MYhref[journalLinks]{http://dx.doi.org/10.1007/JHEP08(2016)079}{JHEP
  }\MYhref[journalLinks]{http://dx.doi.org/10.1007/JHEP08(2016)079}{\textbf{08}
  (2016) 079}, \MYhref[eprintLinks]{http://arxiv.org/abs/1605.07643}{{\ttfamily
  arXiv:1605.07643 [hep-ph]}}.

\bibitem{Khalil:2010iu}
S.~Khalil, \emph{{TeV-scale gauged B-L symmetry with inverse seesaw
  mechanism}},
  \MYhref[journalLinks]{http://dx.doi.org/10.1103/PhysRevD.82.077702}{Phys.
  Rev. D
  }\MYhref[journalLinks]{http://dx.doi.org/10.1103/PhysRevD.82.077702}{\textbf{82}
  (2010) 077702},
  \MYhref[eprintLinks]{http://arxiv.org/abs/1004.0013}{{\ttfamily
  arXiv:1004.0013 [hep-ph]}}.

\bibitem{Abdallah:2011ew}
W.~Abdallah, A.~Awad, S.~Khalil and H.~Okada, \emph{{Muon Anomalous Magnetic
  Moment and mu -\ensuremath{>} e gamma in B-L Model with Inverse Seesaw}},
  \MYhref[journalLinks]{http://dx.doi.org/10.1140/epjc/s10052-012-2108-9}{Eur.
  Phys. J. C
  }\MYhref[journalLinks]{http://dx.doi.org/10.1140/epjc/s10052-012-2108-9}{\textbf{72}
  (2012) 2108}, \MYhref[eprintLinks]{http://arxiv.org/abs/1105.1047}{{\ttfamily
  arXiv:1105.1047 [hep-ph]}}.

\bibitem{Abdallah:2019svm}
W.~Abdallah, S.~Choubey and S.~Khan, \emph{{FIMP dark matter candidate(s) in a
  $B ? L$ model with inverse seesaw mechanism}},
  \MYhref[journalLinks]{http://dx.doi.org/10.1007/JHEP06(2019)095}{JHEP
  }\MYhref[journalLinks]{http://dx.doi.org/10.1007/JHEP06(2019)095}{\textbf{06}
  (2019) 095}, \MYhref[eprintLinks]{http://arxiv.org/abs/1904.10015}{{\ttfamily
  arXiv:1904.10015 [hep-ph]}}.

\bibitem{Ibarra:2010xw}
A.~Ibarra, E.~Molinaro and S.~T. Petcov, \emph{{TeV Scale See-Saw Mechanisms of
  Neutrino Mass Generation, the Majorana Nature of the Heavy Singlet Neutrinos
  and $(\beta\beta)_{0\nu}$-Decay}},
  \MYhref[journalLinks]{http://dx.doi.org/10.1007/JHEP09(2010)108}{JHEP
  }\MYhref[journalLinks]{http://dx.doi.org/10.1007/JHEP09(2010)108}{\textbf{09}
  (2010) 108}, \MYhref[eprintLinks]{http://arxiv.org/abs/1007.2378}{{\ttfamily
  arXiv:1007.2378 [hep-ph]}}.

\bibitem{Ibarra:2011xn}
A.~Ibarra, E.~Molinaro and S.~T. Petcov, \emph{{Low Energy Signatures of the
  TeV Scale See-Saw Mechanism}},
  \MYhref[journalLinks]{http://dx.doi.org/10.1103/PhysRevD.84.013005}{Phys.
  Rev. D
  }\MYhref[journalLinks]{http://dx.doi.org/10.1103/PhysRevD.84.013005}{\textbf{84}
  (2011) 013005},
  \MYhref[eprintLinks]{http://arxiv.org/abs/1103.6217}{{\ttfamily
  arXiv:1103.6217 [hep-ph]}}.

\bibitem{Pinheiro:2021mps}
J.~a.~P. Pinheiro, C.~A. de~S.~Pires, F.~S. Queiroz and Y.~S. Villamizar,
  \emph{{Confronting the inverse seesaw mechanism with the recent muon g-2
  result}},
  \MYhref[journalLinks]{http://dx.doi.org/10.1016/j.physletb.2021.136764}{Phys.
  Lett. B
  }\MYhref[journalLinks]{http://dx.doi.org/10.1016/j.physletb.2021.136764}{\textbf{823}
  (2021) 136764},
  \MYhref[eprintLinks]{http://arxiv.org/abs/2107.01315}{{\ttfamily
  arXiv:2107.01315 [hep-ph]}}.

\bibitem{Barroso:2005da}
A.~Barroso, P.~M. Ferreira and R.~Santos, \emph{{Tree-level vacuum stability in
  multi Higgs models}},
  \MYhref[journalLinks]{http://dx.doi.org/10.22323/1.021.0337}{PoS
  }\MYhref[journalLinks]{http://dx.doi.org/10.22323/1.021.0337}{\textbf{HEP2005}
  (2006) 337},
  \MYhref[eprintLinks]{http://arxiv.org/abs/hep-ph/0512037}{{\ttfamily
  arXiv:hep-ph/0512037}}.

\bibitem{Mantry:2007ar}
S.~Mantry, M.~Trott and M.~B. Wise, \emph{{The Higgs decay width in
  multi-scalar doublet models}},
  \MYhref[journalLinks]{http://dx.doi.org/10.1103/PhysRevD.77.013006}{Phys.
  Rev. D
  }\MYhref[journalLinks]{http://dx.doi.org/10.1103/PhysRevD.77.013006}{\textbf{77}
  (2008) 013006},
  \MYhref[eprintLinks]{http://arxiv.org/abs/0709.1505}{{\ttfamily
  arXiv:0709.1505 [hep-ph]}}.

\bibitem{Ferreira:2008zy}
P.~M. Ferreira and J.~P. Silva, \emph{{Discrete and continuous symmetries in
  multi-Higgs-doublet models}},
  \MYhref[journalLinks]{http://dx.doi.org/10.1103/PhysRevD.78.116007}{Phys.
  Rev. D
  }\MYhref[journalLinks]{http://dx.doi.org/10.1103/PhysRevD.78.116007}{\textbf{78}
  (2008) 116007},
  \MYhref[eprintLinks]{http://arxiv.org/abs/0809.2788}{{\ttfamily
  arXiv:0809.2788 [hep-ph]}}.

\bibitem{Botella:2009pq}
F.~J. Botella, G.~C. Branco and M.~N. Rebelo, \emph{{Minimal Flavour Violation
  and Multi-Higgs Models}},
  \MYhref[journalLinks]{http://dx.doi.org/10.1016/j.physletb.2010.03.014}{Phys.
  Lett. B
  }\MYhref[journalLinks]{http://dx.doi.org/10.1016/j.physletb.2010.03.014}{\textbf{687}
  (2010) 194--200},
  \MYhref[eprintLinks]{http://arxiv.org/abs/0911.1753}{{\ttfamily
  arXiv:0911.1753 [hep-ph]}}.

\bibitem{Ferreira:2010xe}
P.~M. Ferreira, L.~Lavoura and J.~P. Silva, \emph{{Renormalization-group
  constraints on Yukawa alignment in multi-Higgs-doublet models}},
  \MYhref[journalLinks]{http://dx.doi.org/10.1016/j.physletb.2010.04.033}{Phys.
  Lett. B
  }\MYhref[journalLinks]{http://dx.doi.org/10.1016/j.physletb.2010.04.033}{\textbf{688}
  (2010) 341--344},
  \MYhref[eprintLinks]{http://arxiv.org/abs/1001.2561}{{\ttfamily
  arXiv:1001.2561 [hep-ph]}}.

\bibitem{Blechman:2010cs}
A.~E. Blechman, A.~A. Petrov and G.~Yeghiyan, \emph{{The Flavor puzzle in
  multi-Higgs models}},
  \MYhref[journalLinks]{http://dx.doi.org/10.1007/JHEP11(2010)075}{JHEP
  }\MYhref[journalLinks]{http://dx.doi.org/10.1007/JHEP11(2010)075}{\textbf{11}
  (2010) 075}, \MYhref[eprintLinks]{http://arxiv.org/abs/1009.1612}{{\ttfamily
  arXiv:1009.1612 [hep-ph]}}.

\bibitem{Diaz-Cruz:2014pla}
J.~L. Diaz-Cruz and U.~J. Salda\~na Salazar, \emph{{Higgs couplings and new
  signals from Flavon\textendash{}Higgs mixing effects within multi-scalar
  models}},
  \MYhref[journalLinks]{http://dx.doi.org/10.1016/j.nuclphysb.2016.10.018}{Nucl.
  Phys. B
  }\MYhref[journalLinks]{http://dx.doi.org/10.1016/j.nuclphysb.2016.10.018}{\textbf{913}
  (2016) 942--963},
  \MYhref[eprintLinks]{http://arxiv.org/abs/1405.0990}{{\ttfamily
  arXiv:1405.0990 [hep-ph]}}.

\bibitem{Ivanov:2017dad}
I.~P. Ivanov, \emph{{Building and testing models with extended Higgs sectors}},
  \MYhref[journalLinks]{http://dx.doi.org/10.1016/j.ppnp.2017.03.001}{Prog.
  Part. Nucl. Phys.
  }\MYhref[journalLinks]{http://dx.doi.org/10.1016/j.ppnp.2017.03.001}{\textbf{95}
  (2017) 160--208},
  \MYhref[eprintLinks]{http://arxiv.org/abs/1702.03776}{{\ttfamily
  arXiv:1702.03776 [hep-ph]}}.

\bibitem{Ishimori:2010au}
H.~Ishimori et~al., \emph{{Non-Abelian Discrete Symmetries in Particle
  Physics}}, \MYhref[journalLinks]{http://dx.doi.org/10.1143/PTPS.183.1}{Prog.
  Theor. Phys. Suppl.
  }\MYhref[journalLinks]{http://dx.doi.org/10.1143/PTPS.183.1}{\textbf{183}
  (2010) 1--163},
  \MYhref[eprintLinks]{http://arxiv.org/abs/1003.3552}{{\ttfamily
  arXiv:1003.3552 [hep-th]}}.

\bibitem{Grimus:2011fk}
W.~Grimus and P.~O. Ludl, \emph{{Finite flavour groups of fermions}},
  \MYhref[journalLinks]{http://dx.doi.org/10.1088/1751-8113/45/23/233001}{J.
  Phys.
  }\MYhref[journalLinks]{http://dx.doi.org/10.1088/1751-8113/45/23/233001}{\textbf{A45}
  (2012) 233001},
  \MYhref[eprintLinks]{http://arxiv.org/abs/1110.6376}{{\ttfamily
  arXiv:1110.6376 [hep-ph]}}.

\bibitem{Altarelli:2012bn}
G.~Altarelli, F.~Feruglio, L.~Merlo and E.~Stamou, \emph{{Discrete Flavour
  Groups, $theta_{13}$ and Lepton Flavour Violation}},
  \MYhref[journalLinks]{http://dx.doi.org/10.1007/JHEP08(2012)021}{JHEP
  }\MYhref[journalLinks]{http://dx.doi.org/10.1007/JHEP08(2012)021}{\textbf{08}
  (2012) 021}, \MYhref[eprintLinks]{http://arxiv.org/abs/1205.4670}{{\ttfamily
  arXiv:1205.4670 [hep-ph]}}.

\bibitem{Altarelli:2012ss}
G.~Altarelli, F.~Feruglio and L.~Merlo, \emph{{Tri-Bimaximal Neutrino Mixing
  and Discrete Flavour Symmetries}},
  \MYhref[journalLinks]{http://dx.doi.org/10.1002/prop.201200117}{Fortsch.
  Phys.
  }\MYhref[journalLinks]{http://dx.doi.org/10.1002/prop.201200117}{\textbf{61}
  (2013) 507--534},
  \MYhref[eprintLinks]{http://arxiv.org/abs/1205.5133}{{\ttfamily
  arXiv:1205.5133 [hep-ph]}}.

\bibitem{King:2013eh}
S.~F. King and C.~Luhn, \emph{{Neutrino Mass and Mixing with Discrete
  Symmetry}},
  \MYhref[journalLinks]{http://dx.doi.org/10.1088/0034-4885/76/5/056201}{Rept.
  Prog. Phys.
  }\MYhref[journalLinks]{http://dx.doi.org/10.1088/0034-4885/76/5/056201}{\textbf{76}
  (2013) 056201},
  \MYhref[eprintLinks]{http://arxiv.org/abs/1301.1340}{{\ttfamily
  arXiv:1301.1340 [hep-ph]}}.

\bibitem{King:2015aea}
S.~F. King, \emph{{Models of Neutrino Mass, Mixing and CP Violation}},
  \MYhref[journalLinks]{http://dx.doi.org/10.1088/0954-3899/42/12/123001}{J.
  Phys.
  }\MYhref[journalLinks]{http://dx.doi.org/10.1088/0954-3899/42/12/123001}{\textbf{G42}
  (2015) 12 123001},
  \MYhref[eprintLinks]{http://arxiv.org/abs/1510.02091}{{\ttfamily
  arXiv:1510.02091 [hep-ph]}}.

\bibitem{Fonseca:2014lfa}
R.~M. Fonseca and W.~Grimus, \emph{{Classification of lepton mixing patterns
  from finite flavour symmetries}},
  \MYhref[journalLinks]{http://dx.doi.org/10.1016/j.nuclphysbps.2015.10.008}{Nucl.
  Part. Phys. Proc.
  }\MYhref[journalLinks]{http://dx.doi.org/10.1016/j.nuclphysbps.2015.10.008}{\textbf{273-275}
  (2016) 2618--2620},
  \MYhref[eprintLinks]{http://arxiv.org/abs/1410.4133}{{\ttfamily
  arXiv:1410.4133 [hep-ph]}}.

\bibitem{Chauhan:2022gkz}
G.~Chauhan et~al., \emph{{Discrete Flavor Symmetries and Lepton Masses and
  Mixings}}, in \emph{{2022 Snowmass Summer Study}} (2022)
  \MYhref[eprintLinks]{http://arxiv.org/abs/2203.08105}{{\ttfamily
  arXiv:2203.08105 [hep-ph]}}.

\bibitem{Kubo:2003iw}
J.~Kubo, A.~Mondragon, M.~Mondragon and E.~Rodriguez-Jauregui, \emph{{The
  Flavor symmetry}},
  \MYhref[journalLinks]{http://dx.doi.org/10.1143/PTP.109.795}{Prog. Theor.
  Phys.
  }\MYhref[journalLinks]{http://dx.doi.org/10.1143/PTP.109.795}{\textbf{109}
  (2003) 795--807}, [Erratum: Prog. Theor. Phys.114,287(2005)],
  \MYhref[eprintLinks]{http://arxiv.org/abs/hep-ph/0302196}{{\ttfamily
  arXiv:hep-ph/0302196 [hep-ph]}}.

\bibitem{Chen:2004rr}
S.-L. Chen, M.~Frigerio and E.~Ma, \emph{{Large neutrino mixing and normal mass
  hierarchy: A Discrete understanding}},
  \MYhref[journalLinks]{http://dx.doi.org/10.1103/PhysRevD.70.079905,
  10.1103/PhysRevD.70.073008}{Phys.Rev.
  }\MYhref[journalLinks]{http://dx.doi.org/10.1103/PhysRevD.70.079905,
  10.1103/PhysRevD.70.073008}{\textbf{D70} (2004) 073008},
  \MYhref[eprintLinks]{http://arxiv.org/abs/hep-ph/0404084}{{\ttfamily
  arXiv:hep-ph/0404084 [hep-ph]}}.

\bibitem{Mondragon:2007af}
A.~Mondragon, M.~Mondragon and E.~Peinado, \emph{{Lepton masses, mixings and
  FCNC in a minimal $S_3$-invariant extension of the Standard Model}},
  \MYhref[journalLinks]{http://dx.doi.org/10.1103/PhysRevD.76.076003}{Phys.
  Rev.
  }\MYhref[journalLinks]{http://dx.doi.org/10.1103/PhysRevD.76.076003}{\textbf{D76}
  (2007) 076003},
  \MYhref[eprintLinks]{http://arxiv.org/abs/0706.0354}{{\ttfamily
  arXiv:0706.0354 [hep-ph]}}.

\bibitem{Mondragon:2007nk}
A.~Mondragon, M.~Mondragon and E.~Peinado, \emph{{S(3)-flavour symmetry as
  realized in lepton flavour violating processes}},
  \MYhref[journalLinks]{http://dx.doi.org/10.1088/1751-8113/41/30/304035}{J.
  Phys.
  }\MYhref[journalLinks]{http://dx.doi.org/10.1088/1751-8113/41/30/304035}{\textbf{A41}
  (2008) 304035},
  \MYhref[eprintLinks]{http://arxiv.org/abs/0712.1799}{{\ttfamily
  arXiv:0712.1799 [hep-ph]}}.

\bibitem{Meloni:2010aw}
D.~Meloni, S.~Morisi and E.~Peinado, \emph{{Fritzsch neutrino mass matrix from
  $S_3$ symmetry}},
  \MYhref[journalLinks]{http://dx.doi.org/10.1088/0954-3899/38/1/015003}{J.
  Phys.
  }\MYhref[journalLinks]{http://dx.doi.org/10.1088/0954-3899/38/1/015003}{\textbf{G38}
  (2011) 015003},
  \MYhref[eprintLinks]{http://arxiv.org/abs/1005.3482}{{\ttfamily
  arXiv:1005.3482 [hep-ph]}}.

\bibitem{Canales:2012dr}
F.~Gonzalez~Canales, A.~Mondragon and M.~Mondragon, \emph{{The $S_3$ Flavour
  Symmetry: Neutrino Masses and Mixings}},
  \MYhref[journalLinks]{http://dx.doi.org/10.1002/prop.201200121}{Fortsch.Phys.
  }\MYhref[journalLinks]{http://dx.doi.org/10.1002/prop.201200121}{\textbf{61}
  (2013) 546--570},
  \MYhref[eprintLinks]{http://arxiv.org/abs/1205.4755}{{\ttfamily
  arXiv:1205.4755 [hep-ph]}}.

\bibitem{Canales:2013cga}
F.~González~Canales et~al., \emph{{Quark sector of S3 models: classification
  and comparison with experimental data}},
  \MYhref[journalLinks]{http://dx.doi.org/10.1103/PhysRevD.88.096004}{Phys.Rev.
  }\MYhref[journalLinks]{http://dx.doi.org/10.1103/PhysRevD.88.096004}{\textbf{D88}
  (2013) 096004},
  \MYhref[eprintLinks]{http://arxiv.org/abs/1304.6644}{{\ttfamily
  arXiv:1304.6644 [hep-ph]}}.

\bibitem{Pakvasa:1977in}
S.~Pakvasa and H.~Sugawara, \emph{{Discrete Symmetry and Cabibbo Angle}},
  \MYhref[journalLinks]{http://dx.doi.org/10.1016/0370-2693(78)90172-7}{Phys.
  Lett.
  }\MYhref[journalLinks]{http://dx.doi.org/10.1016/0370-2693(78)90172-7}{\textbf{73B}
  (1978) 61--64}.

\bibitem{Kubo:2004ps}
J.~Kubo, H.~Okada and F.~Sakamaki, \emph{{Higgs potential in minimal S(3)
  invariant extension of the standard model}},
  \MYhref[journalLinks]{http://dx.doi.org/10.1103/PhysRevD.70.036007}{Phys.
  Rev.
  }\MYhref[journalLinks]{http://dx.doi.org/10.1103/PhysRevD.70.036007}{\textbf{D70}
  (2004) 036007},
  \MYhref[eprintLinks]{http://arxiv.org/abs/hep-ph/0402089}{{\ttfamily
  arXiv:hep-ph/0402089 [hep-ph]}}.

\bibitem{EmmanuelCosta:2007zz}
D.~Emmanuel-Costa, O.~Felix-Beltran, M.~Mondragon and E.~Rodriguez-Jauregui,
  \emph{{Stability of the tree-level vacuum in a minimal S(3) extension of the
  standard model}},
  \MYhref[journalLinks]{http://dx.doi.org/10.1063/1.2751981}{AIP Conf. Proc.
  }\MYhref[journalLinks]{http://dx.doi.org/10.1063/1.2751981}{\textbf{917}
  (2007) 390--393}, [,390(2007)].

\bibitem{Beltran:2009zz}
O.~F. Beltran, M.~Mondragon and E.~Rodriguez-Jauregui, \emph{{Conditions for
  vacuum stability in an S(3) extension of the standard model}},
  \MYhref[journalLinks]{http://dx.doi.org/10.1088/1742-6596/171/1/012028}{J.
  Phys. Conf. Ser.
  }\MYhref[journalLinks]{http://dx.doi.org/10.1088/1742-6596/171/1/012028}{\textbf{171}
  (2009) 012028}.

\bibitem{Teshima:2012cg}
T.~Teshima, \emph{{Higgs potential in S3 invariant model for quarklepton mass
  and mixing}},
  \MYhref[journalLinks]{http://dx.doi.org/10.1103/PhysRevD.85.105013}{Phys.
  Rev.
  }\MYhref[journalLinks]{http://dx.doi.org/10.1103/PhysRevD.85.105013}{\textbf{D85}
  (2012) 105013},
  \MYhref[eprintLinks]{http://arxiv.org/abs/1202.4528}{{\ttfamily
  arXiv:1202.4528 [hep-ph]}}.

\bibitem{Barradas-Guevara:2014yoa}
E.~Barradas-Guevara, O.~Félix-Beltrán and E.~Rodríguez-Jáuregui,
  \emph{{Trilinear self-couplings in an S(3) flavored Higgs model}},
  \MYhref[journalLinks]{http://dx.doi.org/10.1103/PhysRevD.90.095001}{Phys.
  Rev.
  }\MYhref[journalLinks]{http://dx.doi.org/10.1103/PhysRevD.90.095001}{\textbf{D90}
  (2014) 9 095001},
  \MYhref[eprintLinks]{http://arxiv.org/abs/1402.2244}{{\ttfamily
  arXiv:1402.2244 [hep-ph]}}.

\bibitem{Das:2014fea}
D.~Das and U.~K. Dey, \emph{{Analysis of an extended scalar sector with $S_3$
  symmetry}},
  \MYhref[journalLinks]{http://dx.doi.org/10.1103/PhysRevD.91.039905,
  10.1103/PhysRevD.89.095025}{Phys. Rev.
  }\MYhref[journalLinks]{http://dx.doi.org/10.1103/PhysRevD.91.039905,
  10.1103/PhysRevD.89.095025}{\textbf{D89} (2014) 9 095025}, [Erratum: Phys.
  Rev.D91,no.3,039905(2015)],
  \MYhref[eprintLinks]{http://arxiv.org/abs/1404.2491}{{\ttfamily
  arXiv:1404.2491 [hep-ph]}}.

\bibitem{Gomez-Bock:2021uyu}
M.~G\'omez-Bock, M.~Mondrag\'on and A.~P\'erez-Mart\'\i{}nez, \emph{{Scalar and
  gauge sectors in the 3-Higgs Doublet Model under the $S_3$ symmetry}},
  \MYhref[journalLinks]{http://dx.doi.org/10.1140/epjc/s10052-021-09731-3}{Eur.
  Phys. J. C
  }\MYhref[journalLinks]{http://dx.doi.org/10.1140/epjc/s10052-021-09731-3}{\textbf{81}
  (2021) 10 942},
  \MYhref[eprintLinks]{http://arxiv.org/abs/2102.02800}{{\ttfamily
  arXiv:2102.02800 [hep-ph]}}.

\bibitem{ATLAS:2012yve}
G.~Aad et~al. (ATLAS), \emph{{Observation of a new particle in the search for
  the Standard Model Higgs boson with the ATLAS detector at the LHC}},
  \MYhref[journalLinks]{http://dx.doi.org/10.1016/j.physletb.2012.08.020}{Phys.
  Lett. B
  }\MYhref[journalLinks]{http://dx.doi.org/10.1016/j.physletb.2012.08.020}{\textbf{716}
  (2012) 1--29},
  \MYhref[eprintLinks]{http://arxiv.org/abs/1207.7214}{{\ttfamily
  arXiv:1207.7214 [hep-ex]}}.

\bibitem{CMS:2012qbp}
S.~Chatrchyan et~al. (CMS), \emph{{Observation of a New Boson at a Mass of 125
  GeV with the CMS Experiment at the LHC}},
  \MYhref[journalLinks]{http://dx.doi.org/10.1016/j.physletb.2012.08.021}{Phys.
  Lett. B
  }\MYhref[journalLinks]{http://dx.doi.org/10.1016/j.physletb.2012.08.021}{\textbf{716}
  (2012) 30--61},
  \MYhref[eprintLinks]{http://arxiv.org/abs/1207.7235}{{\ttfamily
  arXiv:1207.7235 [hep-ex]}}.

\bibitem{Kuncinas:2020wrn}
A.~Kun\v{c}inas, O.~M. Ogreid, P.~Osland and M.~N. Rebelo, \emph{{S3 -inspired
  three-Higgs-doublet models: A class with a complex vacuum}},
  \MYhref[journalLinks]{http://dx.doi.org/10.1103/PhysRevD.101.075052}{Phys.
  Rev. D
  }\MYhref[journalLinks]{http://dx.doi.org/10.1103/PhysRevD.101.075052}{\textbf{101}
  (2020) 7 075052},
  \MYhref[eprintLinks]{http://arxiv.org/abs/2001.01994}{{\ttfamily
  arXiv:2001.01994 [hep-ph]}}.

\bibitem{Khater:2021wcx}
W.~Khater et~al., \emph{{Dark matter in three-Higgs-doublet models with S$_{3}$
  symmetry}},
  \MYhref[journalLinks]{http://dx.doi.org/10.1007/JHEP01(2022)120}{JHEP
  }\MYhref[journalLinks]{http://dx.doi.org/10.1007/JHEP01(2022)120}{\textbf{01}
  (2022) 120}, \MYhref[eprintLinks]{http://arxiv.org/abs/2108.07026}{{\ttfamily
  arXiv:2108.07026 [hep-ph]}}.

\bibitem{Kuncinas:2023ycz}
A.~Kun\v{c}inas, O.~M. Ogreid, P.~Osland and M.~N. Rebelo, \emph{{Complex
  S$_{3}$-symmetric 3HDM}},
  \MYhref[journalLinks]{http://dx.doi.org/10.1007/JHEP07(2023)013}{JHEP
  }\MYhref[journalLinks]{http://dx.doi.org/10.1007/JHEP07(2023)013}{\textbf{07}
  (2023) 013}, \MYhref[eprintLinks]{http://arxiv.org/abs/2302.07210}{{\ttfamily
  arXiv:2302.07210 [hep-ph]}}.

\bibitem{Espinoza:2018itz}
C.~Espinoza, E.~A. Garc\'es, M.~Mondrag\'on and H.~Reyes-Gonz\'alez, \emph{{The
  $S3$ Symmetric Model with a Dark Scalar}},
  \MYhref[journalLinks]{http://dx.doi.org/10.1016/j.physletb.2018.11.028}{Phys.
  Lett. B
  }\MYhref[journalLinks]{http://dx.doi.org/10.1016/j.physletb.2018.11.028}{\textbf{788}
  (2019) 185--191},
  \MYhref[eprintLinks]{http://arxiv.org/abs/1804.01879}{{\ttfamily
  arXiv:1804.01879 [hep-ph]}}.

\bibitem{Gehrlein:2022nss}
J.~Gehrlein, S.~Petcov, M.~Spinrath and A.~Titov, \emph{{Testing neutrino
  flavor models}}, in \emph{{Snowmass 2021}} (2022)
  \MYhref[eprintLinks]{http://arxiv.org/abs/2203.06219}{{\ttfamily
  arXiv:2203.06219 [hep-ph]}}.

\bibitem{Hirsch:2009mx}
M.~Hirsch, S.~Morisi and J.~W.~F. Valle, \emph{{A4-based tri-bimaximal mixing
  within inverse and linear seesaw schemes}},
  \MYhref[journalLinks]{http://dx.doi.org/10.1016/j.physletb.2009.08.003}{Phys.
  Lett. B
  }\MYhref[journalLinks]{http://dx.doi.org/10.1016/j.physletb.2009.08.003}{\textbf{679}
  (2009) 454--459},
  \MYhref[eprintLinks]{http://arxiv.org/abs/0905.3056}{{\ttfamily
  arXiv:0905.3056 [hep-ph]}}.

\bibitem{Karmakar:2016cvb}
B.~Karmakar and A.~Sil, \emph{{An $A_4$ realization of inverse seesaw: neutrino
  masses, $\theta_{13}$ and leptonic non-unitarity}},
  \MYhref[journalLinks]{http://dx.doi.org/10.1103/PhysRevD.96.015007}{Phys.
  Rev. D
  }\MYhref[journalLinks]{http://dx.doi.org/10.1103/PhysRevD.96.015007}{\textbf{96}
  (2017) 1 015007},
  \MYhref[eprintLinks]{http://arxiv.org/abs/1610.01909}{{\ttfamily
  arXiv:1610.01909 [hep-ph]}}.

\bibitem{Gautam:2019pce}
N.~Gautam and M.~K. Das, \emph{{Phenomenology of keV scale sterile neutrino
  dark matter with $S_{4}$ flavor symmetry}},
  \MYhref[journalLinks]{http://dx.doi.org/10.1007/JHEP01(2020)098}{JHEP
  }\MYhref[journalLinks]{http://dx.doi.org/10.1007/JHEP01(2020)098}{\textbf{01}
  (2020) 098}, \MYhref[eprintLinks]{http://arxiv.org/abs/1904.10662}{{\ttfamily
  arXiv:1904.10662 [hep-ph]}}.

\bibitem{Thapa:2023fxu}
B.~Thapa, S.~Barman, S.~Bora and N.~K. Francis, \emph{{A minimal inverse seesaw
  model with S$_{4}$ flavour symmetry}},
  \MYhref[journalLinks]{http://dx.doi.org/10.1007/JHEP11(2023)154}{JHEP
  }\MYhref[journalLinks]{http://dx.doi.org/10.1007/JHEP11(2023)154}{\textbf{11}
  (2023) 154}, \MYhref[eprintLinks]{http://arxiv.org/abs/2305.09306}{{\ttfamily
  arXiv:2305.09306 [hep-ph]}}.

\bibitem{Garnica:2023ccx}
J.~C. Garnica, G.~Hern\'andez-Tom\'e and E.~Peinado, \emph{{Charged
  lepton-flavor violating processes and suppression of nonunitary mixing
  effects in low-scale seesaw models}},
  \MYhref[journalLinks]{http://dx.doi.org/10.1103/PhysRevD.108.035033}{Phys.
  Rev. D
  }\MYhref[journalLinks]{http://dx.doi.org/10.1103/PhysRevD.108.035033}{\textbf{108}
  (2023) 3 035033},
  \MYhref[eprintLinks]{http://arxiv.org/abs/2302.07379}{{\ttfamily
  arXiv:2302.07379 [hep-ph]}}.

\bibitem{Duy:2024lbd}
N.~T. Duy, D.~T. Huong and A.~E. Carcamo~Hernandez, \emph{{Flavor phenomenology
  of an extended 2HDM with inverse seesaw mechanism}}  (2024),
  \MYhref[eprintLinks]{http://arxiv.org/abs/2404.15935}{{\ttfamily
  arXiv:2404.15935 [hep-ph]}}.

\bibitem{Emmanuel-Costa:2016vej}
D.~Emmanuel-Costa, O.~M. Ogreid, P.~Osland and M.~N. Rebelo, \emph{{Spontaneous
  symmetry breaking in the $S_3$-symmetric scalar sector}},
  \MYhref[journalLinks]{http://dx.doi.org/10.1007/JHEP08(2016)169}{JHEP
  }\MYhref[journalLinks]{http://dx.doi.org/10.1007/JHEP08(2016)169}{\textbf{02}
  (2016) 154}, [Erratum: JHEP 08, 169 (2016)],
  \MYhref[eprintLinks]{http://arxiv.org/abs/1601.04654}{{\ttfamily
  arXiv:1601.04654 [hep-ph]}}.

\bibitem{Fukuura:1999ze}
K.~Fukuura, T.~Miura, E.~Takasugi and M.~Yoshimura, \emph{{Maximal CP
  violation, large mixings of neutrinos and democratic type neutrino mass
  matrix}},
  \MYhref[journalLinks]{http://dx.doi.org/10.1103/PhysRevD.61.073002}{Phys.
  Rev. D
  }\MYhref[journalLinks]{http://dx.doi.org/10.1103/PhysRevD.61.073002}{\textbf{61}
  (2000) 073002},
  \MYhref[eprintLinks]{http://arxiv.org/abs/hep-ph/9909415}{{\ttfamily
  arXiv:hep-ph/9909415}}.

\bibitem{Miura:2000sx}
T.~Miura, E.~Takasugi and M.~Yoshimura, \emph{{Large CP violation, large
  mixings of neutrinos and the Z(3) symmetry}},
  \MYhref[journalLinks]{http://dx.doi.org/10.1103/PhysRevD.63.013001}{Phys.
  Rev. D
  }\MYhref[journalLinks]{http://dx.doi.org/10.1103/PhysRevD.63.013001}{\textbf{63}
  (2001) 013001},
  \MYhref[eprintLinks]{http://arxiv.org/abs/hep-ph/0003139}{{\ttfamily
  arXiv:hep-ph/0003139}}.

\bibitem{Ma:2002ce}
E.~Ma, \emph{{The All purpose neutrino mass matrix}},
  \MYhref[journalLinks]{http://dx.doi.org/10.1103/PhysRevD.66.117301}{Phys.
  Rev. D
  }\MYhref[journalLinks]{http://dx.doi.org/10.1103/PhysRevD.66.117301}{\textbf{66}
  (2002) 117301},
  \MYhref[eprintLinks]{http://arxiv.org/abs/hep-ph/0207352}{{\ttfamily
  arXiv:hep-ph/0207352}}.

\bibitem{Grimus:2003yn}
W.~Grimus and L.~Lavoura, \emph{{A Nonstandard CP transformation leading to
  maximal atmospheric neutrino mixing}},
  \MYhref[journalLinks]{http://dx.doi.org/10.1016/j.physletb.2003.10.075}{Phys.
  Lett. B
  }\MYhref[journalLinks]{http://dx.doi.org/10.1016/j.physletb.2003.10.075}{\textbf{579}
  (2004) 113--122},
  \MYhref[eprintLinks]{http://arxiv.org/abs/hep-ph/0305309}{{\ttfamily
  arXiv:hep-ph/0305309}}.

\bibitem{Chen:2014wxa}
P.~Chen, C.-C. Li and G.-J. Ding, \emph{{Lepton Flavor Mixing and CP
  Symmetry}},
  \MYhref[journalLinks]{http://dx.doi.org/10.1103/PhysRevD.91.033003}{Phys.
  Rev. D
  }\MYhref[journalLinks]{http://dx.doi.org/10.1103/PhysRevD.91.033003}{\textbf{91}
  (2015) 033003},
  \MYhref[eprintLinks]{http://arxiv.org/abs/1412.8352}{{\ttfamily
  arXiv:1412.8352 [hep-ph]}}.

\bibitem{Ma:2015fpa}
E.~Ma, \emph{{Neutrino mixing: $A_4$ variations}},
  \MYhref[journalLinks]{http://dx.doi.org/10.1016/j.physletb.2015.11.049}{Phys.
  Lett. B
  }\MYhref[journalLinks]{http://dx.doi.org/10.1016/j.physletb.2015.11.049}{\textbf{752}
  (2016) 198--200},
  \MYhref[eprintLinks]{http://arxiv.org/abs/1510.02501}{{\ttfamily
  arXiv:1510.02501 [hep-ph]}}.

\bibitem{Joshipura:2015dsa}
A.~S. Joshipura and K.~M. Patel, \emph{{Generalized $\mu$-$\tau$ symmetry and
  discrete subgroups of O(3)}},
  \MYhref[journalLinks]{http://dx.doi.org/10.1016/j.physletb.2015.07.062}{Phys.
  Lett. B
  }\MYhref[journalLinks]{http://dx.doi.org/10.1016/j.physletb.2015.07.062}{\textbf{749}
  (2015) 159--166},
  \MYhref[eprintLinks]{http://arxiv.org/abs/1507.01235}{{\ttfamily
  arXiv:1507.01235 [hep-ph]}}.

\bibitem{Li:2015rtz}
G.-N. Li and X.-G. He, \emph{{CP violation in neutrino mixing with $\delta =
  -\pi/2$ in $A_4$ Type-II seesaw model}},
  \MYhref[journalLinks]{http://dx.doi.org/10.1016/j.physletb.2015.09.061}{Phys.
  Lett. B
  }\MYhref[journalLinks]{http://dx.doi.org/10.1016/j.physletb.2015.09.061}{\textbf{750}
  (2015) 620--626},
  \MYhref[eprintLinks]{http://arxiv.org/abs/1505.01932}{{\ttfamily
  arXiv:1505.01932 [hep-ph]}}.

\bibitem{He:2015xha}
H.-J. He, W.~Rodejohann and X.-J. Xu, \emph{{Origin of Constrained Maximal CP
  Violation in Flavor Symmetry}},
  \MYhref[journalLinks]{http://dx.doi.org/10.1016/j.physletb.2015.10.066}{Phys.
  Lett. B
  }\MYhref[journalLinks]{http://dx.doi.org/10.1016/j.physletb.2015.10.066}{\textbf{751}
  (2015) 586--594},
  \MYhref[eprintLinks]{http://arxiv.org/abs/1507.03541}{{\ttfamily
  arXiv:1507.03541 [hep-ph]}}.

\bibitem{Chen:2015siy}
P.~Chen, G.-J. Ding, F.~Gonzalez-Canales and J.~W.~F. Valle, \emph{{Generalized
  $\mu-\tau$ reflection symmetry and leptonic CP violation}},
  \MYhref[journalLinks]{http://dx.doi.org/10.1016/j.physletb.2015.12.069}{Phys.
  Lett. B
  }\MYhref[journalLinks]{http://dx.doi.org/10.1016/j.physletb.2015.12.069}{\textbf{753}
  (2016) 644--652},
  \MYhref[eprintLinks]{http://arxiv.org/abs/1512.01551}{{\ttfamily
  arXiv:1512.01551 [hep-ph]}}.

\bibitem{Ma:2016nkf}
E.~Ma, \emph{{Soft $A_4 \to Z_3$ symmetry breaking and cobimaximal neutrino
  mixing}},
  \MYhref[journalLinks]{http://dx.doi.org/10.1016/j.physletb.2016.02.032}{Phys.
  Lett. B
  }\MYhref[journalLinks]{http://dx.doi.org/10.1016/j.physletb.2016.02.032}{\textbf{755}
  (2016) 348--350},
  \MYhref[eprintLinks]{http://arxiv.org/abs/1601.00138}{{\ttfamily
  arXiv:1601.00138 [hep-ph]}}.

\bibitem{Damanik:2017jar}
A.~Damanik, \emph{{Neutrino masses from a cobimaximal neutrino mixing matrix}}
  (2017), \MYhref[eprintLinks]{http://arxiv.org/abs/1702.03214}{{\ttfamily
  arXiv:1702.03214 [physics.gen-ph]}}.

\bibitem{Ma:2017trv}
E.~Ma, \emph{{Cobimaximal neutrino mixing from $S_3 \times Z_2$}},
  \MYhref[journalLinks]{http://dx.doi.org/10.1016/j.physletb.2017.12.049}{Phys.
  Lett. B
  }\MYhref[journalLinks]{http://dx.doi.org/10.1016/j.physletb.2017.12.049}{\textbf{777}
  (2018) 332--334},
  \MYhref[eprintLinks]{http://arxiv.org/abs/1707.03352}{{\ttfamily
  arXiv:1707.03352 [hep-ph]}}.

\bibitem{Grimus:2017itg}
W.~Grimus and L.~Lavoura, \emph{{Cobimaximal lepton mixing from soft symmetry
  breaking}},
  \MYhref[journalLinks]{http://dx.doi.org/10.1016/j.physletb.2017.09.082}{Phys.
  Lett. B
  }\MYhref[journalLinks]{http://dx.doi.org/10.1016/j.physletb.2017.09.082}{\textbf{774}
  (2017) 325--331},
  \MYhref[eprintLinks]{http://arxiv.org/abs/1708.09809}{{\ttfamily
  arXiv:1708.09809 [hep-ph]}}.

\bibitem{CarcamoHernandez:2017owh}
A.~E. C\'arcamo~Hern\'andez, S.~Kovalenko, J.~W.~F. Valle and C.~A.
  Vaquera-Araujo, \emph{{Predictive Pati-Salam theory of fermion masses and
  mixing}},
  \MYhref[journalLinks]{http://dx.doi.org/10.1007/JHEP07(2017)118}{JHEP
  }\MYhref[journalLinks]{http://dx.doi.org/10.1007/JHEP07(2017)118}{\textbf{07}
  (2017) 118}, \MYhref[eprintLinks]{http://arxiv.org/abs/1705.06320}{{\ttfamily
  arXiv:1705.06320 [hep-ph]}}.

\bibitem{CarcamoHernandez:2018hst}
A.~E. C\'arcamo~Hern\'andez, S.~Kovalenko, J.~W.~F. Valle and C.~A.
  Vaquera-Araujo, \emph{{Neutrino predictions from a left-right symmetric
  flavored extension of the standard model}},
  \MYhref[journalLinks]{http://dx.doi.org/10.1007/JHEP02(2019)065}{JHEP
  }\MYhref[journalLinks]{http://dx.doi.org/10.1007/JHEP02(2019)065}{\textbf{02}
  (2019) 065}, \MYhref[eprintLinks]{http://arxiv.org/abs/1811.03018}{{\ttfamily
  arXiv:1811.03018 [hep-ph]}}.

\bibitem{Ma:2019iwj}
E.~Ma, \emph{{Scotogenic cobimaximal Dirac neutrino mixing from $\Delta (27)$
  and $U(1)_\chi $}},
  \MYhref[journalLinks]{http://dx.doi.org/10.1140/epjc/s10052-019-7440-x}{Eur.
  Phys. J. C
  }\MYhref[journalLinks]{http://dx.doi.org/10.1140/epjc/s10052-019-7440-x}{\textbf{79}
  (2019) 11 903},
  \MYhref[eprintLinks]{http://arxiv.org/abs/1905.01535}{{\ttfamily
  arXiv:1905.01535 [hep-ph]}}.

\bibitem{Hernandez:2021kju}
A.~E.~C. Hern\'andez, C.~Espinoza, J.~C. G\'omez-Izquierdo and M.~Mondrag\'on,
  \emph{{Fermion masses and mixings, dark matter, leptogenesis and $g-2$ muon
  anomaly in an extended 2HDM with inverse seesaw}},
  \MYhref[journalLinks]{http://dx.doi.org/10.1140/epjp/s13360-022-03432-w}{Eur.
  Phys. J. Plus
  }\MYhref[journalLinks]{http://dx.doi.org/10.1140/epjp/s13360-022-03432-w}{\textbf{137}
  (2022) 11 1224},
  \MYhref[eprintLinks]{http://arxiv.org/abs/2104.02730}{{\ttfamily
  arXiv:2104.02730 [hep-ph]}}.

\bibitem{CarcamoHernandez:2024ycd}
A.~E. C\'arcamo~Hern\'andez, D.~Salinas-Arizmendi, J.~Vignatti and A.~Zerwekh,
  \emph{{Phenomenology of an Extended $1+2$ Higgs Doublet Model with $S_3$
  Family Symmetry}}  (2024),
  \MYhref[eprintLinks]{http://arxiv.org/abs/2408.01497}{{\ttfamily
  arXiv:2408.01497 [hep-ph]}}.

\bibitem{Gomez-Izquierdo:2023mph}
J.~C. G\'omez-Izquierdo and A.~E.~P. Ram\'\i{}rez, \emph{{A lepton model with
  nearly Cobimaximal mixing}},
  \MYhref[journalLinks]{http://dx.doi.org/10.31349/RevMexFis.70.040801}{Rev.
  Mex. Fis.
  }\MYhref[journalLinks]{http://dx.doi.org/10.31349/RevMexFis.70.040801}{\textbf{70}
  (2024) 4 040801},
  \MYhref[eprintLinks]{http://arxiv.org/abs/2310.03000}{{\ttfamily
  arXiv:2310.03000 [hep-ph]}}.

\bibitem{Khan:2012zw}
S.~Khan, S.~Goswami and S.~Roy, \emph{{Vacuum Stability constraints on the
  minimal singlet TeV Seesaw Model}},
  \MYhref[journalLinks]{http://dx.doi.org/10.1103/PhysRevD.89.073021}{Phys.
  Rev. D
  }\MYhref[journalLinks]{http://dx.doi.org/10.1103/PhysRevD.89.073021}{\textbf{89}
  (2014) 7 073021},
  \MYhref[eprintLinks]{http://arxiv.org/abs/1212.3694}{{\ttfamily
  arXiv:1212.3694 [hep-ph]}}.

\bibitem{Hettmansperger:2011bt}
H.~Hettmansperger, M.~Lindner and W.~Rodejohann, \emph{{Phenomenological
  Consequences of sub-leading Terms in See-Saw Formulas}},
  \MYhref[journalLinks]{http://dx.doi.org/10.1007/JHEP04(2011)123}{JHEP
  }\MYhref[journalLinks]{http://dx.doi.org/10.1007/JHEP04(2011)123}{\textbf{04}
  (2011) 123}, \MYhref[eprintLinks]{http://arxiv.org/abs/1102.3432}{{\ttfamily
  arXiv:1102.3432 [hep-ph]}}.

\bibitem{tHooft:1979rat}
G.~'t~Hooft, \emph{{Naturalness, chiral symmetry, and spontaneous chiral
  symmetry breaking}},
  \MYhref[journalLinks]{http://dx.doi.org/10.1007/978-1-4684-7571-5_9}{NATO
  Sci. Ser. B
  }\MYhref[journalLinks]{http://dx.doi.org/10.1007/978-1-4684-7571-5_9}{\textbf{59}
  (1980) 135--157}.

\bibitem{Xing:2015fdg}
Z.-z. Xing and Z.-h. Zhao, \emph{{A review of ?-? flavor symmetry in neutrino
  physics}},
  \MYhref[journalLinks]{http://dx.doi.org/10.1088/0034-4885/79/7/076201}{Rept.
  Prog. Phys.
  }\MYhref[journalLinks]{http://dx.doi.org/10.1088/0034-4885/79/7/076201}{\textbf{79}
  (2016) 7 076201},
  \MYhref[eprintLinks]{http://arxiv.org/abs/1512.04207}{{\ttfamily
  arXiv:1512.04207 [hep-ph]}}.

\bibitem{Derman:1978rx}
E.~Derman, \emph{{Flavor Unification, $\tau$ Decay and $b$ Decay Within the Six
  Quark Six Lepton {Weinberg-Salam} Model}},
  \MYhref[journalLinks]{http://dx.doi.org/10.1103/PhysRevD.19.317}{Phys. Rev. D
  }\MYhref[journalLinks]{http://dx.doi.org/10.1103/PhysRevD.19.317}{\textbf{19}
  (1979) 317--329}.

\bibitem{Pakvasa:1978tx}
S.~Pakvasa and H.~Sugawara, \emph{{Mass of the t Quark in SU(2) x U(1)}},
  \MYhref[journalLinks]{http://dx.doi.org/10.1016/0370-2693(79)90436-2}{Phys.
  Lett. B
  }\MYhref[journalLinks]{http://dx.doi.org/10.1016/0370-2693(79)90436-2}{\textbf{82}
  (1979) 105--107}.

\bibitem{Derman:1979nf}
E.~Derman and H.-S. Tsao, \emph{{SU(2) X U(1) X S($n$) Flavor Dynamics and a
  Bound on the Number of Flavors}},
  \MYhref[journalLinks]{http://dx.doi.org/10.1103/PhysRevD.20.1207}{Phys. Rev.
  D
  }\MYhref[journalLinks]{http://dx.doi.org/10.1103/PhysRevD.20.1207}{\textbf{20}
  (1979) 1207}.

\bibitem{Fritzsch:1999ee}
H.~Fritzsch and Z.-z. Xing, \emph{{Mass and flavor mixing schemes of quarks and
  leptons}},
  \MYhref[journalLinks]{http://dx.doi.org/10.1016/S0146-6410(00)00102-2}{Prog.
  Part. Nucl. Phys.
  }\MYhref[journalLinks]{http://dx.doi.org/10.1016/S0146-6410(00)00102-2}{\textbf{45}
  (2000) 1--81},
  \MYhref[eprintLinks]{http://arxiv.org/abs/hep-ph/9912358}{{\ttfamily
  arXiv:hep-ph/9912358}}.

\bibitem{Garcia-Aguilar:2022gfw}
J.~D. Garc\'\i{}a-Aguilar, A.~E.~P. Ram\'\i{}rez, M.~M.~S. Casta\~neda and
  J.~C. G\'omez-Izquierdo, \emph{{Soft breaking of the \textmu{}
  \ensuremath{\leftrightarrow} \ensuremath{\tau} symmetry by S4
  \ensuremath{\otimes} Z2}},
  \MYhref[journalLinks]{http://dx.doi.org/10.31349/RevMexFis.69.030802}{Rev.
  Mex. Fis.
  }\MYhref[journalLinks]{http://dx.doi.org/10.31349/RevMexFis.69.030802}{\textbf{69}
  (2023) 3 030802},
  \MYhref[eprintLinks]{http://arxiv.org/abs/2209.01316}{{\ttfamily
  arXiv:2209.01316 [hep-ph]}}.

\bibitem{Garces:2018nar}
E.~A. Garc\'es, J.~C. G\'omez-Izquierdo and F.~Gonzalez-Canales,
  \emph{{Flavored non-minimal left\textendash{}right symmetric model fermion
  masses and mixings}},
  \MYhref[journalLinks]{http://dx.doi.org/10.1140/epjc/s10052-018-6271-5}{Eur.
  Phys. J. C
  }\MYhref[journalLinks]{http://dx.doi.org/10.1140/epjc/s10052-018-6271-5}{\textbf{78}
  (2018) 10 812},
  \MYhref[eprintLinks]{http://arxiv.org/abs/1807.02727}{{\ttfamily
  arXiv:1807.02727 [hep-ph]}}.

\bibitem{ParticleDataGroup:2022pth}
R.~L. Workman et~al. (Particle Data Group), \emph{{Review of Particle
  Physics}}, \MYhref[journalLinks]{http://dx.doi.org/10.1093/ptep/ptac097}{PTEP
  }\MYhref[journalLinks]{http://dx.doi.org/10.1093/ptep/ptac097}{\textbf{2022}
  (2022) 083C01}.

\bibitem{Martinez:2017lzg}
G.~D. Martinez et~al. (GAMBIT), \emph{{Comparison of statistical sampling
  methods with ScannerBit, the GAMBIT scanning module}},
  \MYhref[journalLinks]{http://dx.doi.org/10.1140/epjc/s10052-017-5274-y}{Eur.
  Phys. J. C
  }\MYhref[journalLinks]{http://dx.doi.org/10.1140/epjc/s10052-017-5274-y}{\textbf{77}
  (2017) 11 761},
  \MYhref[eprintLinks]{http://arxiv.org/abs/1705.07959}{{\ttfamily
  arXiv:1705.07959 [hep-ph]}}.

\bibitem{DarkMachinesHighDimensionalSamplingGroup:2021wkt}
C.~Bal\'azs et~al. (DarkMachines High Dimensional Sampling Group), \emph{{A
  comparison of optimisation algorithms for high-dimensional particle and
  astrophysics applications}},
  \MYhref[journalLinks]{http://dx.doi.org/10.1007/JHEP05(2021)108}{JHEP
  }\MYhref[journalLinks]{http://dx.doi.org/10.1007/JHEP05(2021)108}{\textbf{05}
  (2021) 108}, \MYhref[eprintLinks]{http://arxiv.org/abs/2101.04525}{{\ttfamily
  arXiv:2101.04525 [hep-ph]}}.

\bibitem{Scott:2012qh}
P.~Scott, \emph{{Pippi - painless parsing, post-processing and plotting of
  posterior and likelihood samples}},
  \MYhref[journalLinks]{http://dx.doi.org/10.1140/epjp/i2012-12138-3}{Eur.
  Phys. J. Plus
  }\MYhref[journalLinks]{http://dx.doi.org/10.1140/epjp/i2012-12138-3}{\textbf{127}
  (2012) 138}, \MYhref[eprintLinks]{http://arxiv.org/abs/1206.2245}{{\ttfamily
  arXiv:1206.2245 [physics.data-an]}}.

\bibitem{Fernandez-Martinez:2016lgt}
E.~Fernandez-Martinez, J.~Hernandez-Garcia and J.~Lopez-Pavon, \emph{{Global
  constraints on heavy neutrino mixing}},
  \MYhref[journalLinks]{http://dx.doi.org/10.1007/JHEP08(2016)033}{JHEP
  }\MYhref[journalLinks]{http://dx.doi.org/10.1007/JHEP08(2016)033}{\textbf{08}
  (2016) 033}, \MYhref[eprintLinks]{http://arxiv.org/abs/1605.08774}{{\ttfamily
  arXiv:1605.08774 [hep-ph]}}.

\bibitem{Blennow:2023mqx}
M.~Blennow et~al., \emph{{Bounds on lepton non-unitarity and heavy neutrino
  mixing}},
  \MYhref[journalLinks]{http://dx.doi.org/10.1007/JHEP08(2023)030}{JHEP
  }\MYhref[journalLinks]{http://dx.doi.org/10.1007/JHEP08(2023)030}{\textbf{08}
  (2023) 030}, \MYhref[eprintLinks]{http://arxiv.org/abs/2306.01040}{{\ttfamily
  arXiv:2306.01040 [hep-ph]}}.

\bibitem{deSalas:2020pgw}
P.~F. de~Salas et~al., \emph{{2020 global reassessment of the neutrino
  oscillation picture}},
  \MYhref[journalLinks]{http://dx.doi.org/10.1007/JHEP02(2021)071}{JHEP
  }\MYhref[journalLinks]{http://dx.doi.org/10.1007/JHEP02(2021)071}{\textbf{02}
  (2021) 071}, \MYhref[eprintLinks]{http://arxiv.org/abs/2006.11237}{{\ttfamily
  arXiv:2006.11237 [hep-ph]}}.

\bibitem{Esteban:2020cvm}
I.~Esteban et~al., \emph{{The fate of hints: updated global analysis of
  three-flavor neutrino oscillations}},
  \MYhref[journalLinks]{http://dx.doi.org/10.1007/JHEP09(2020)178}{JHEP
  }\MYhref[journalLinks]{http://dx.doi.org/10.1007/JHEP09(2020)178}{\textbf{09}
  (2020) 178}, \MYhref[eprintLinks]{http://arxiv.org/abs/2007.14792}{{\ttfamily
  arXiv:2007.14792 [hep-ph]}}.

\bibitem{Gonzalez-Garcia:2021dve}
M.~C. Gonzalez-Garcia, M.~Maltoni and T.~Schwetz, \emph{{NuFIT: Three-Flavour
  Global Analyses of Neutrino Oscillation Experiments}},
  \MYhref[journalLinks]{http://dx.doi.org/10.3390/universe7120459}{Universe
  }\MYhref[journalLinks]{http://dx.doi.org/10.3390/universe7120459}{\textbf{7}
  (2021) 12 459},
  \MYhref[eprintLinks]{http://arxiv.org/abs/2111.03086}{{\ttfamily
  arXiv:2111.03086 [hep-ph]}}.

\bibitem{Xing:2022uax}
Z.-z. Xing, \emph{{The \textmu{}\textendash{}\ensuremath{\tau} reflection
  symmetry of Majorana neutrinos $^{*}$}},
  \MYhref[journalLinks]{http://dx.doi.org/10.1088/1361-6633/acd8ce}{Rept. Prog.
  Phys.
  }\MYhref[journalLinks]{http://dx.doi.org/10.1088/1361-6633/acd8ce}{\textbf{86}
  (2023) 7 076201},
  \MYhref[eprintLinks]{http://arxiv.org/abs/2210.11922}{{\ttfamily
  arXiv:2210.11922 [hep-ph]}}.

\bibitem{Petcov:1976ff}
S.~T. Petcov, \emph{{The Processes $\mu \rightarrow e + \gamma, \mu \rightarrow
  e + \overline{e}, \nu' \rightarrow \nu + \gamma$ in the Weinberg-Salam Model
  with Neutrino Mixing}}, Sov. J. Nucl. Phys. \textbf{25} (1977) 340, [Erratum:
  Sov.J.Nucl.Phys. 25, 698 (1977), Erratum: Yad.Fiz. 25, 1336 (1977)].

\bibitem{Bilenky:1977du}
S.~M. Bilenky, S.~T. Petcov and B.~Pontecorvo, \emph{{Lepton Mixing, mu
  --\ensuremath{>} e + gamma Decay and Neutrino Oscillations}},
  \MYhref[journalLinks]{http://dx.doi.org/10.1016/0370-2693(77)90379-3}{Phys.
  Lett. B
  }\MYhref[journalLinks]{http://dx.doi.org/10.1016/0370-2693(77)90379-3}{\textbf{67}
  (1977) 309}.

\bibitem{Cheng:1976uq}
T.~P. Cheng and L.-F. Li, \emph{{Nonconservation of Separate mu - Lepton and e
  - Lepton Numbers in Gauge Theories with v+a Currents}},
  \MYhref[journalLinks]{http://dx.doi.org/10.1103/PhysRevLett.38.381}{Phys.
  Rev. Lett.
  }\MYhref[journalLinks]{http://dx.doi.org/10.1103/PhysRevLett.38.381}{\textbf{38}
  (1977) 381}.

\bibitem{Cheng:1980tp}
T.~P. Cheng and L.-F. Li, \emph{{$\mu \to e \gamma$ in Theories With Dirac and
  Majorana Neutrino Mass Terms}},
  \MYhref[journalLinks]{http://dx.doi.org/10.1103/PhysRevLett.45.1908}{Phys.
  Rev. Lett.
  }\MYhref[journalLinks]{http://dx.doi.org/10.1103/PhysRevLett.45.1908}{\textbf{45}
  (1980) 1908}.

\bibitem{Ilakovac:1994kj}
A.~Ilakovac and A.~Pilaftsis, \emph{{Flavor violating charged lepton decays in
  seesaw-type models}},
  \MYhref[journalLinks]{http://dx.doi.org/10.1016/0550-3213(94)00567-X}{Nucl.
  Phys. B
  }\MYhref[journalLinks]{http://dx.doi.org/10.1016/0550-3213(94)00567-X}{\textbf{437}
  (1995) 491},
  \MYhref[eprintLinks]{http://arxiv.org/abs/hep-ph/9403398}{{\ttfamily
  arXiv:hep-ph/9403398}}.

\bibitem{Deppisch:2004fa}
F.~Deppisch and J.~W.~F. Valle, \emph{{Enhanced lepton flavor violation in the
  supersymmetric inverse seesaw model}},
  \MYhref[journalLinks]{http://dx.doi.org/10.1103/PhysRevD.72.036001}{Phys.
  Rev. D
  }\MYhref[journalLinks]{http://dx.doi.org/10.1103/PhysRevD.72.036001}{\textbf{72}
  (2005) 036001},
  \MYhref[eprintLinks]{http://arxiv.org/abs/hep-ph/0406040}{{\ttfamily
  arXiv:hep-ph/0406040}}.

\bibitem{Bernstein:2013hba}
R.~H. Bernstein and P.~S. Cooper, \emph{{Charged Lepton Flavor Violation: An
  Experimenter's Guide}},
  \MYhref[journalLinks]{http://dx.doi.org/10.1016/j.physrep.2013.07.002}{Phys.
  Rept.
  }\MYhref[journalLinks]{http://dx.doi.org/10.1016/j.physrep.2013.07.002}{\textbf{532}
  (2013) 27--64},
  \MYhref[eprintLinks]{http://arxiv.org/abs/1307.5787}{{\ttfamily
  arXiv:1307.5787 [hep-ex]}}.

\bibitem{Vergados:2002pv}
J.~D. Vergados, \emph{{The Neutrinoless double beta decay from a modern
  perspective}},
  \MYhref[journalLinks]{http://dx.doi.org/10.1016/S0370-1573(01)00068-0}{Phys.
  Rept.
  }\MYhref[journalLinks]{http://dx.doi.org/10.1016/S0370-1573(01)00068-0}{\textbf{361}
  (2002) 1--56},
  \MYhref[eprintLinks]{http://arxiv.org/abs/hep-ph/0209347}{{\ttfamily
  arXiv:hep-ph/0209347}}.

\bibitem{Blennow:2010th}
M.~Blennow, E.~Fernandez-Martinez, J.~Lopez-Pavon and J.~Menendez,
  \emph{{Neutrinoless double beta decay in seesaw models}},
  \MYhref[journalLinks]{http://dx.doi.org/10.1007/JHEP07(2010)096}{JHEP
  }\MYhref[journalLinks]{http://dx.doi.org/10.1007/JHEP07(2010)096}{\textbf{07}
  (2010) 096}, \MYhref[eprintLinks]{http://arxiv.org/abs/1005.3240}{{\ttfamily
  arXiv:1005.3240 [hep-ph]}}.

\bibitem{Rodejohann:2011mu}
W.~Rodejohann, \emph{{Neutrino-less Double Beta Decay and Particle Physics}},
  \MYhref[journalLinks]{http://dx.doi.org/10.1142/S0218301311020186}{Int. J.
  Mod. Phys. E
  }\MYhref[journalLinks]{http://dx.doi.org/10.1142/S0218301311020186}{\textbf{20}
  (2011) 1833--1930},
  \MYhref[eprintLinks]{http://arxiv.org/abs/1106.1334}{{\ttfamily
  arXiv:1106.1334 [hep-ph]}}.

\bibitem{Vergados:2012xy}
J.~D. Vergados, H.~Ejiri and F.~Simkovic, \emph{{Theory of Neutrinoless Double
  Beta Decay}},
  \MYhref[journalLinks]{http://dx.doi.org/10.1088/0034-4885/75/10/106301}{Rept.
  Prog. Phys.
  }\MYhref[journalLinks]{http://dx.doi.org/10.1088/0034-4885/75/10/106301}{\textbf{75}
  (2012) 106301},
  \MYhref[eprintLinks]{http://arxiv.org/abs/1205.0649}{{\ttfamily
  arXiv:1205.0649 [hep-ph]}}.

\bibitem{Pas:2015eia}
H.~P\"as and W.~Rodejohann, \emph{{Neutrinoless Double Beta Decay}},
  \MYhref[journalLinks]{http://dx.doi.org/10.1088/1367-2630/17/11/115010}{New
  J. Phys.
  }\MYhref[journalLinks]{http://dx.doi.org/10.1088/1367-2630/17/11/115010}{\textbf{17}
  (2015) 11 115010},
  \MYhref[eprintLinks]{http://arxiv.org/abs/1507.00170}{{\ttfamily
  arXiv:1507.00170 [hep-ph]}}.

\bibitem{Agostini:2013mzu}
M.~Agostini et~al. (GERDA), \emph{{Results on Neutrinoless Double-$\beta$ Decay
  of $^{76}$Ge from Phase I of the GERDA Experiment}},
  \MYhref[journalLinks]{http://dx.doi.org/10.1103/PhysRevLett.111.122503}{Phys.
  Rev. Lett.
  }\MYhref[journalLinks]{http://dx.doi.org/10.1103/PhysRevLett.111.122503}{\textbf{111}
  (2013) 12 122503},
  \MYhref[eprintLinks]{http://arxiv.org/abs/1307.4720}{{\ttfamily
  arXiv:1307.4720 [nucl-ex]}}.

\end{thebibliography}
\end{document}